\documentclass[12pt]{article}
\usepackage{amsfonts,amssymb,graphicx}
\usepackage{fullpage}
\usepackage{subfig}
\usepackage{rotating,cancel}
\usepackage{sidecap}
\pdfoutput=1
\usepackage{ifpdf}
\ifpdf
 \usepackage{color}
 \definecolor{darkblue}{rgb}{0.3,0.3,0.6}
 \definecolor{darkgreen}{rgb}{0,0.6,0}
 \usepackage[pdftitle={Towards Geometric D6-Brane Model Building on non-Factorisable Toroidal Orbifolds},pdfauthor={Mikel Berasaluce-Gonz\'alez, Gabriele Honecker, Alexander Seifert},pdfsubject={String theory},pdfkeywords={string theory},colorlinks=true,linkcolor=black,urlcolor=darkblue,filecolor=darkblue,citecolor=darkblue,menucolor=darkblue,breaklinks=true,bookmarks=true,anchorcolor=black,unicode=true]{hyperref}
\else
 
\fi
\usepackage[numbers,compress]{natbib}
\usepackage{a4wide}
\usepackage{longtable}
\usepackage[table]{xcolor}
\usepackage{graphicx}
\usepackage[centertags]{amsmath}
\usepackage{multirow}
\usepackage{slashed}
\usepackage{pdflscape}
\usepackage{caption} 
\usepackage{MnSymbol}
\usepackage{marginnote}

\usepackage{mathtools}
\usepackage{bbm}
\usepackage{dsfont}
\usepackage{hhline}

\newcommand{\bCentering}{\centering}

\def\muc{\multicolumn}

\def\Z{\mathbb{Z}}

\def\N{\mathbf{N}}

\def\1{{\bf 1}}
\def\2{{\bf 2}}
\def\3{{\bf 3}}
\def\4{{\bf 4}}
\def\6{{\bf 6}}
\def\8{{\bf 8}}
\def\OR{\Omega\mathcal{R}}

\def\targ#1#2{\genfrac{[}{]}{0pt}{}{#1}{#2}}

\def\targ2#1#2{\genfrac{}{}{0pt}{}{#1}{#2}}

\definecolor{blus}{rgb}{0.1,0.1,0.8}
\definecolor{GreenYellow}{cmyk}{0.15,0,0.69,0}
\definecolor{Yellow}{cmyk}{0,0,1,0}
\definecolor{Goldenrod}{cmyk}{0,0.10,0.84,0}
\definecolor{Dandelion}{cmyk}{0,0.29,0.84,0}
\definecolor{Apricot}{cmyk}{0,0.32,0.52,0}
\definecolor{Peach}{cmyk}{0,0.50,0.70,0}
\definecolor{Melon}{cmyk}{0,0.46,0.50,0}
\definecolor{YellowOrange}{cmyk}{0,0.42,1,0}
\definecolor{Orange}{cmyk}{0,0.61,0.87,0}
\definecolor{BurntOrange}{cmyk}{0,0.51,1,0}
\definecolor{Bittersweet}{cmyk}{0,0.75,1,0.24}
\definecolor{RedOrange}{cmyk}{0,0.77,0.87,0}
\definecolor{Mahogany}{cmyk}{0,0.85,0.87,0.35}
\definecolor{Maroon}{cmyk}{0,0.87,0.68,0.32}
\definecolor{BrickRed}{cmyk}{0,0.89,0.94,0.28}
\definecolor{Red}{cmyk}{0,1,1,0}
\definecolor{OrangeRed}{cmyk}{0,1,0.50,0}
\definecolor{RubineRed}{cmyk}{0,1,0.13,0}
\definecolor{WildStrawberry}{cmyk}{0,0.96,0.39,0}
\definecolor{Salmon}{cmyk}{0,0.53,0.38,0}
\definecolor{CarnationPink}{cmyk}{0,0.63,0,0}
\definecolor{Magenta}{cmyk}{0,1,0,0}
\definecolor{VioletRed}{cmyk}{0,0.81,0,0}
\definecolor{Rhodamine}{cmyk}{0,0.82,0,0}
\definecolor{Mulberry}{cmyk}{0.34,0.90,0,0.02}
\definecolor{RedViolet}{cmyk}{0.07,0.90,0,0.34}
\definecolor{Fuchsia}{cmyk}{0.47,0.91,0,0.08}
\definecolor{Lavender}{cmyk}{0,0.48,0,0}
\definecolor{Thistle}{cmyk}{0.12,0.59,0,0}
\definecolor{Orchid}{cmyk}{0.32,0.64,0,0}
\definecolor{DarkOrchid}{cmyk}{0.40,0.80,0.20,0}
\definecolor{Purple}{cmyk}{0.45,0.86,0,0}
\definecolor{Plum}{cmyk}{0.50,1,0,0}
\definecolor{Violet}{cmyk}{0.79,0.88,0,0}
\definecolor{RoyalPurple}{cmyk}{0.75,0.90,0,0}
\definecolor{BlueViolet}{cmyk}{0.86,0.91,0,0.04}
\definecolor{Periwinkle}{cmyk}{0.57,0.55,0,0}
\definecolor{CadetBlue}{cmyk}{0.62,0.57,0.23,0}
\definecolor{CornflowerBlue}{cmyk}{0.65,0.13,0,0}
\definecolor{MidnightBlue}{cmyk}{0.98,0.13,0,0.43}
\definecolor{NavyBlue}{cmyk}{0.94,0.54,0,0}
\definecolor{RoyalBlue}{cmyk}{1,0.50,0,0}
\definecolor{Blue}{cmyk}{1,1,0,0}
\definecolor{Cerulean}{cmyk}{0.94,0.11,0,0}
\definecolor{Cyan}{cmyk}{1,0,0,0}
\definecolor{ProcessBlue}{cmyk}{0.96,0,0,0}
\definecolor{SkyBlue}{cmyk}{0.62,0,0.12,0}
\definecolor{Turquoise}{cmyk}{0.85,0,0.20,0}
\definecolor{TealBlue}{cmyk}{0.86,0,0.34,0.02}
\definecolor{Aquamarine}{cmyk}{0.82,0,0.30,0}
\definecolor{BlueGreen}{cmyk}{0.85,0,0.33,0}
\definecolor{Emerald}{cmyk}{1,0,0.50,0}
\definecolor{JungleGreen}{cmyk}{0.99,0,0.52,0}
\definecolor{SeaGreen}{cmyk}{0.69,0,0.50,0}
\definecolor{Green}{cmyk}{1,0,1,0}
\definecolor{ForestGreen}{cmyk}{0.91,0,0.88,0.12}
\definecolor{PineGreen}{cmyk}{0.92,0,0.59,0.25}
\definecolor{LimeGreen}{cmyk}{0.50,0,1,0}
\definecolor{YellowGreen}{cmyk}{0.44,0,0.74,0}
\definecolor{SpringGreen}{cmyk}{0.26,0,0.76,0}
\definecolor{OliveGreen}{cmyk}{0.64,0,0.95,0.40}
\definecolor{RawSienna}{cmyk}{0,0.72,1,0.45}
\definecolor{Sepia}{cmyk}{0,0.83,1,0.70}
\definecolor{Brown}{cmyk}{0,0.81,1,0.60}
\definecolor{Tan}{cmyk}{0.14,0.42,0.56,0}
\definecolor{Gray}{cmyk}{0,0,0,0.50}
\definecolor{Black}{cmyk}{0,0,0,1}
\definecolor{White}{cmyk}{0,0,0,0}
\definecolor{LightGray}{gray}{0.8}

\definecolor{mygr}{rgb}{0,0.6,0}
\definecolor{mygrey}{rgb}{0,0.1,0.2}
\definecolor{myblue}{rgb}{0,0.5,0.9}
\definecolor{myblue2}{rgb}{0,0.5,0.5}
\definecolor{myorange}{rgb}{1,0.5,0}
\definecolor{mypurple}{rgb}{0.6,0,1}
\definecolor{mygolden}{rgb}{1,0.8,0.2}

\newcommand{\bCaptionfonts}{\small}
\makeatletter
\long\def\@makecaption#1#2{%
 \vskip\abovecaptionskip
 \sbox\@tempboxa{{\bCaptionfonts #1: #2}}%
 \ifdim \wd\@tempboxa >\hsize
 {\bCaptionfonts #1: #2\par}
 \else
 \hbox to\hsize{\hfil\box\@tempboxa\hfil}%
 \fi
 \vskip\belowcaptionskip}
\makeatother

\makeatletter
\let\ORIGINALlatex@openbib@code=\@openbib@code
\renewcommand{\@openbib@code}{\ORIGINALlatex@openbib@code\setlength{\itemsep}{1ex plus.5ex minus.5ex}\setlength{\parsep}{0pt}}
\makeatother

\allowdisplaybreaks[4]

\begin{document}
\begin{center}
\begin{flushright}
{\small MITP/16-052\\ 
\today}

\end{flushright}

\vspace{25mm}
{\Large\bf Towards Geometric D6-Brane Model Building on non-Factorisable Toroidal $\Z_4$-Orbifolds}
\vspace{12mm}

{\large Mikel Berasaluce-Gonz\'alez${}^{\clubsuit}$, Gabriele Honecker${}^{\heartsuit}$ and Alexander Seifert${}^{\spadesuit}$
}

\vspace{8mm}
{
\it PRISMA Cluster of Excellence \& Institut f\"ur Physik (WA THEP), \\Johannes-Gutenberg-Universit\"at, D-55099 Mainz, Germany
\;$^{\clubsuit}${\tt mberasal@uni-mainz.de},~$^{\heartsuit}${\tt Gabriele.Honecker@uni-mainz.de},~$^{\spadesuit}${\tt alseifer@uni-mainz.de}}

\vspace{15mm}{\bf Abstract}\\[2ex]\parbox{140mm}{We present a geometric approach to D-brane model building on the non-factorisable torus backgrounds of $T^6/\Z_4$, which are $A_3 \times A_3$ and 
$A_3 \times A_1 \times B_2$. Based on the counting of `short' supersymmetric three-cycles per complex structure {\it vev}, the number of physically inequivalent lattice orientations with respect to the anti-holomorphic
involution ${\cal R}$ of the Type IIA/$\OR$ orientifold can be reduced to three for the $A_3 \times A_3$ lattice and four for the $A_3 \times A_1 \times B_2$ lattice. While four independent three-cycles on $A_3 \times A_3$
cannot accommodate phenomenologically interesting global models with a chiral spectrum, the eight-dimensional space of three-cycles on $A_3 \times A_1 \times B_2$ is rich enough to provide for particle physics models, with several globally consistent two- and four-generation Pati-Salam models presented here. 
\\
We further show that for fractional {\it sLag} three-cycles, the compact geometry can be rewritten in a $(T^2)^3$ factorised form, paving the way for a generalisation of known CFT methods to determine the vector-like spectrum and to derive the low-energy effective action for open string states.
}
\end{center}

\thispagestyle{empty}
\clearpage 

\tableofcontents
\setlength{\parskip}{1em plus1ex minus.5ex}
%

\section{Introduction}\label{S:intro}

Ever since the finding in 1985 that string theory constitutes a framework for unifying quantum field theory and gravity, which - when compactified on a Calabi-Yau-threefold or some singular limit thereof such as a toroidal orbifold - 
leads to ${\cal N}=1$ supersymmetry in four dimensions~\cite{Candelas:1985en}, the search for vacuum configurations with not only the Standard Model spectrum but also its interactions has been intensively pursued.
Starting from sporadic models of the $E_8 \times E_8$ heterotic string such as~\cite{Witten:1985xc,Ibanez:1987sn}, by implementing systematic computer searches, large classes of vacua with particle physics spectra on $T^6/\Z_N$ and $T^6/\Z_N \times \Z_M$ orbifolds~\cite{Lebedev:2006kn,Lebedev:2008un,Nilles:2011aj} (see also~\cite{Choi:2004wn} for the heterotic $SO(32)$ string theory) and Calabi-Yau manifolds~\cite{Bak:2008ey,Anderson:2011ns,Anderson:2012yf,Anderson:2013xka,Nibbelink:2015ixa} could be constructed.

With the identification of D-branes as dynamical objects in Type I and II string theory~\cite{Polchinski:1995mt} in 1995, model building also opened up in these theories, which are conjectured to be related by S-duality~\cite{Vafa:1994tf,Polchinski:1995df} and M-/F-theory duality~\cite{Schwarz:1995jq,Vafa:1996xn} to the vacua of the heterotic string. The virtue of D-brane model building lies in the fact that physical quantities, like e.g. the number of particle generations, are expressed in terms of topology and geometry of the compact space, which is particularly intuitive for the case of D6-branes on three-cycles in Type IIA orientifold compactifications, see e.g.~\cite{Ibanez:2012zz} for a broad recent overview.
This intuition, however, comes at the cost of the relatively little explored symplectic structure of generic Calabi-Yau threefolds. 
Since supersymmetric - or mathematically expressed {\it special Lagrangian} ({\it sLag}) - three-cycles constitute a largely unexplored area (except for~\cite{Joyce:2001xt,Joyce:2001nm} and~\cite{Palti:2009bt}), 
intersecting D6-brane models in Type IIA orientifolds have focussed on tori and toroidal orbifolds, and more specifically on backgrounds which are factorised into two-tori, $T^6=(T^2)^3$, see e.g.~\cite{Ibanez:2001nd,Cremades:2002te,Cremades:2002cs}, and $(T^2)^3/\Gamma$ with Abelian point groups $\Gamma= \Z_N$ or $\Z_N \times \Z_M$. In the latter case, models with all three-cycles inherited from the underlying torus have been constructed for $\Gamma=\Z_2 \times \Z_2$~\cite{Cvetic:2001tj,Cvetic:2001nr,Blumenhagen:2005mu,Gmeiner:2005vz,Blumenhagen:2006ci,Gmeiner:2006vb} and $\Z_2 \times \Z_4$~\cite{Honecker:2003vq,Honecker:2003vw,Honecker:2004np} without discrete torsion, while fractional three-cycles consisting of components inherited from the torus plus exceptional divisors at orbifold singularities have been employed for $\Gamma=\Z_4$~\cite{Blumenhagen:2002gw}, $\Z_6$~\cite{Honecker:2004kb,Honecker:2004np,Gmeiner:2007we,Gmeiner:2009fb}, $\Z_6'$~\cite{Bailin:2006zf,Bailin:2007va,Gmeiner:2007zz,Bailin:2008xx,Gmeiner:2008xq,Gmeiner:2009fb} and $\Z_{12-II}$~\cite{Bailin:2013sya} with a single $\Z_2$ sector and $\Gamma=\Z_2 \times \Z_2$~\cite{Blumenhagen:2005tn,Forste:2010gw}, $\Z_2 \times \Z_6'$~\cite{Forste:2010gw,Honecker:2012qr,Honecker:2013kda} and $\Z_2 \times \Z_6$~\cite{Forste:2010gw,Ecker:2014hma,Ecker:2015vea,Honecker:2015qba} with discrete torsion with two $\Z_2$ subgroups.

 All of the above mentioned types of string vacua contain a plethora of scalar fields with flat directions. While the above models with a $\Z_2 \times \Z_2$ subsymmetry allow for {\it rigid} fractional three-cycles which, in the open string spectrum, provide gauge groups without brane recombination/splitting moduli in the adjoint representation, it has recently been noticed that (most) twisted complex structure moduli associated to deformations of singularities are in fact stabilised by the existence of D-branes with $U(1)$ symmetries~\cite{Blaszczyk:2014xla,Blaszczyk:2015oia,Honecker:2015qba,Koltermann:2015oyv}.
 To further stabilise the dilaton and untwisted complex structure moduli, one usually argues that closed string background NS-NS fluxes (see~\cite{Grana:2005jc,Koerber:2010bx,Larfors:2015umb} for reviews) provide a non-trivial scalar potential, see also~\cite{Ihl:2006pp,Ihl:2007ah} for attempts to incorporate NS-NS fluxes on the factorisable $T^6/\Z_4$ orbifold and~\cite{Bailin:2011am} for the factorisable $T^6/\Z_6'$ orbifold. However, incorporating 
 a non-trivial NS-NS flux $H_3$ will in general violate the factorisation into two-tori and instead lead to so-called non-factorisable torus backgrounds~\cite{Camara:2005dc,Aldazabal:2006up,Marchesano:2006ns}.

Orbifolds of non-factorisable tori have, to our best knowledge, scarcely been considered in the literature. 
Within Type IIA orientifolds, one of the first studies of non-factorisable $\mathbb{Z}_N$ orbifolds can be found in~\cite{Blumenhagen:2004di}, where special configurations of D6-branes on top of the O6-planes lead to a {\it local} cancellation of the RR tadpoles within the compact space. In~\cite{Kimura:2007ey} a similar analysis for $\mathbb{Z}_N\times\mathbb{Z}_M$ orbifolds was performed. In~\cite{Forste:2007zb,Forste:2008ex} orientifolds of $T^6/(\mathbb{Z}_2\times\mathbb{Z}_2)$ with (non-)factorisable lattices were considered, including D6-branes which are not parallel to the O6-planes, and in~\cite{Bailin:2013sya}, three-cycles on the $D_4 \times A_2$ and $D_4 \times A_1 \times A_1$ lattices with $\Z_{12-I}$ and $\Z_{12-II}$ orbifold symmetry, respectively, were studied. Finally, in \cite{Forste:2014bfa} the Yukawa couplings for a torus generated by a $D_6$ lattice were computed.
Here, we will for the first time perform a thorough study of all possible {\it sLag} three-cycles on the two different non-factorisable lattice backgrounds $A_3 \times A_3$ and $A_3 \times A_1 \times B_2$ of $T^6/\Z_4$, for which we briefly provided some preliminary results in~\cite{Seifert:2015fwr}.
 
 Besides from being able to classify {\it sLag} cycles on toroidal orbifolds, these geometrically simple backgrounds are equipped with the non-negligible power of allowing for an explicit string quantisation and thus Conformal Field Theory (CFT) techniques, which do not only reproduce the RR tadpole cancellation conditions and chiral spectrum, but are indispensable for distinguishing gauge group enhancements $U(N) \hookrightarrow SO(2N)$ or $USp(2N)$ for D-branes wrapped on orientifold invariant three-cycles and for deriving the low-energy effective action (for particle physics models based on powerful RCFT techniques see e.g. also~\cite{Dijkstra:2004ym,Dijkstra:2004cc,Anastasopoulos:2006da,Anastasopoulos:2010hu}). 
 So far, CFT results within the geometrically intuitive approach to Type II string model building with D-branes have been obtained only for gauge couplings at one-loop and $n$-point couplings at tree-level using
{\it bulk} cycles on the factorisable six-torus $(T^2)^3$~\cite{Lust:2003ky,Abel:2003vv,Cvetic:2003ch,Abel:2003yx,Lust:2004cx,Akerblom:2007uc}. For {\it fractional} cycles on $(T^2)^3/\Z_N$ and $\Z_N \times \Z_M$ orbifolds~\cite{Blumenhagen:2007ip,Gmeiner:2009fb,Honecker:2011sm,Honecker:2011hm}, the one-loop corrections to gauge couplings and the K\"ahler potential at leading order could be derived, while the one-loop corrections to the open string K\"ahler potential are only known for {\it bulk} cycles on such orbifolds~\cite{Berg:2011ij}, 
see also~\cite{Forste:2010gw,Honecker:2012qr,Honecker:2013sww,Ecker:2014hma} for the distinction of $U(N) \hookrightarrow SO(2N)$ versus $USp(2N)$ gauge group enhancement using one-loop gauge threshold computations. 
 The aim of the present work is not only to generalise the geometric methods of deriving the chiral spectrum from topological intersection numbers, but also to initiate the generalisation of CFT techniques to so-called non-factorisable lattice backgrounds.

The outline of this article is as follows: in section~\ref{S:T6Z4-nonFact} we first study the geometry of $T^6/\Z_4$ on non-factorisable tori, then implement anti-holomorphic involutions on the lattices to study
orientifolds of Type IIA string theory in section~\ref{S:IBWs}, after which we proceed to discuss first hints on physical equivalences among different choices, and finally we study supersymmetric D6-branes wrapped on (fractional) three-cycles. In section~\ref{S:Factorisation}, we argue that any supersymmetric fractional three-cycle on $T^6/\Z_4$ can be written in a factorised form, which paves the way for implementing CFT methods in order to distinguish gauge group enhancements $U(N) \hookrightarrow SO(2N)$ versus $USp(2N)$ and to derive the vector-like matter spectrum.
We then proceed to provide some explicit Pati-Salam models with two and four particle generations in section~\ref{S:ConcretePSModels}. Section~\ref{S:Conclusions} contains our conclusions and outlook, and appendix~\ref{A:MoreModels}
contains some further explicit examples of globally consistent D6-brane configurations with chiral matter on the $A_3 \times A_1 \times B_2$ lattice.

\section{Non-factorisable $T^6/\Z_4$ orbifold geometries}\label{S:T6Z4-nonFact}

In this section, we discuss the three-cycle geometry on the two non-factorisable background lattices, $A_3\times A_3$ and $A_3\times A_1\times B_2$, of $T^6/\Z_4$.
The $\Z_4$-action is usually encoded in the shift vector $\vec{\zeta}=\frac{1}{4}(1,-2,1)$ if the six-torus is parameterized by three complex coordinates, and more generally
the $\Z_4$-action is generated by the Coxeter element $Q$ which acts on the root lattice of the corresponding orbifold spanned by the simple roots $\{e_i\}_{i=1,\dots,6}$. 
Furthermore, we denote the six toroidal one-cycles along the directions $\{e_i\}$ by $\pi_i$ and toroidal two- and three-cycles by $\pi_{ij}:=\pi_i\wedge\pi_j$ and $\pi_{ijk}:=\pi_i\wedge\pi_j\wedge\pi_k$,
respectively. The Hodge numbers of all three possible lattice backgrounds - one factorisable and two non-factorisable ones - are summarized in table~\ref{Tab:HodgeNumbers-Z4}
(cf. e.g. \cite{Lust:2006zh}). 
\begin{table}
\begin{equation*}
\begin{array}{|c||c|c|c||c|c|c|}\hline
\muc{7}{|c|}{\text{\bf Hodge numbers of $T^6/\Z_4$ orbifolds}}
\\\hline\hline
\text{Lattice} & h_{11}^{\text{bulk}} & h_{11}^{\Z_2} & h_{11}^{\Z_4} & h_{21}^{\text{bulk}} & h_{21}^{\Z_2} & h_{21}^{\Z_4}
\\\hline\hline
B_2 \times (A_1)^2 \times B_2 & 5 & 10 & 16 & 1 & 6 & 0
\\\hline
A_3 \times A_3 & 5 & 4 & 16 & 1 & 0 & 0
\\\hline
A_3 \times A_1 \times B_2 & 5 & 6 & 16 & 1 & 2 & 0
\\\hline
\end{array}
\end{equation*}
\caption{Summary of the Hodge numbers per untwisted and twisted sector of the $T^6/\Z_4$ orbifolds on factorisable and non-factorisable tori.
\label{Tab:HodgeNumbers-Z4}}
\end{table}
We are in particular interested in characterization of the three-homology $H_3(T^6/\Z_4,\,\Z)$ of each background lattice. 
This homology class contains in general the $\Z_4$-invariant bulk $\pi^\text{bulk}$ and exceptional three-cycles $\pi^\text{exc}$ as well as fractional linear combinations thereof, the so-called fractional three-cycles $\pi^\text{frac}$. The bulk three-cycles are inherited from the underlying torus and can be computed by taking the $\Z_4$-orbits thereof:
\begin{equation}\label{1.bulk_cycle}
\pi^\text{bulk}:=\sum_{i=0}^{i=3}Q^i\pi^\text{torus}\,.
\end{equation}
The exceptional three-cycles arise for the $T^6/\Z_4$ orbifolds only in the $\Z_2$-twisted sector. They stem from the resolution of the $\Z_2$-invariant two-tori and can be written as a product of an exceptional two-cycle $\mathbf{e}_{\alpha\beta}$ (with $\alpha\beta$ labelling the location of the cycle)
 and a $\Z_2$-invariant toroidal one-cycle (plus some $\Z_4$-image).
 Finally, fractional three-cycles are either one-half of a bulk cycle or linear combinations of one-half of some bulk and exceptional three-cycles with the combinatorics depending on the corresponding singularities traversed by the bulk cycles as well as sign factors associated to the choice of some discrete Wilson line, or geometrically speaking the orientation the corresponding exceptional three-cycle is wrapped.

Based on the discussion in this section, we will proceed to discuss (supersymmetric) Type IIA orientifolds with O6-planes and D6-branes and the associated anti-holomorphic involution
on the background geometry in section~\ref{S:IBWs}.

\subsection{$B_2 \times (A_1)^2 \times B_2$}\label{Ss:B2xA1-2xB2}

\begin{figure}[h]
 	\centering
 		\includegraphics[width=15.5cm]{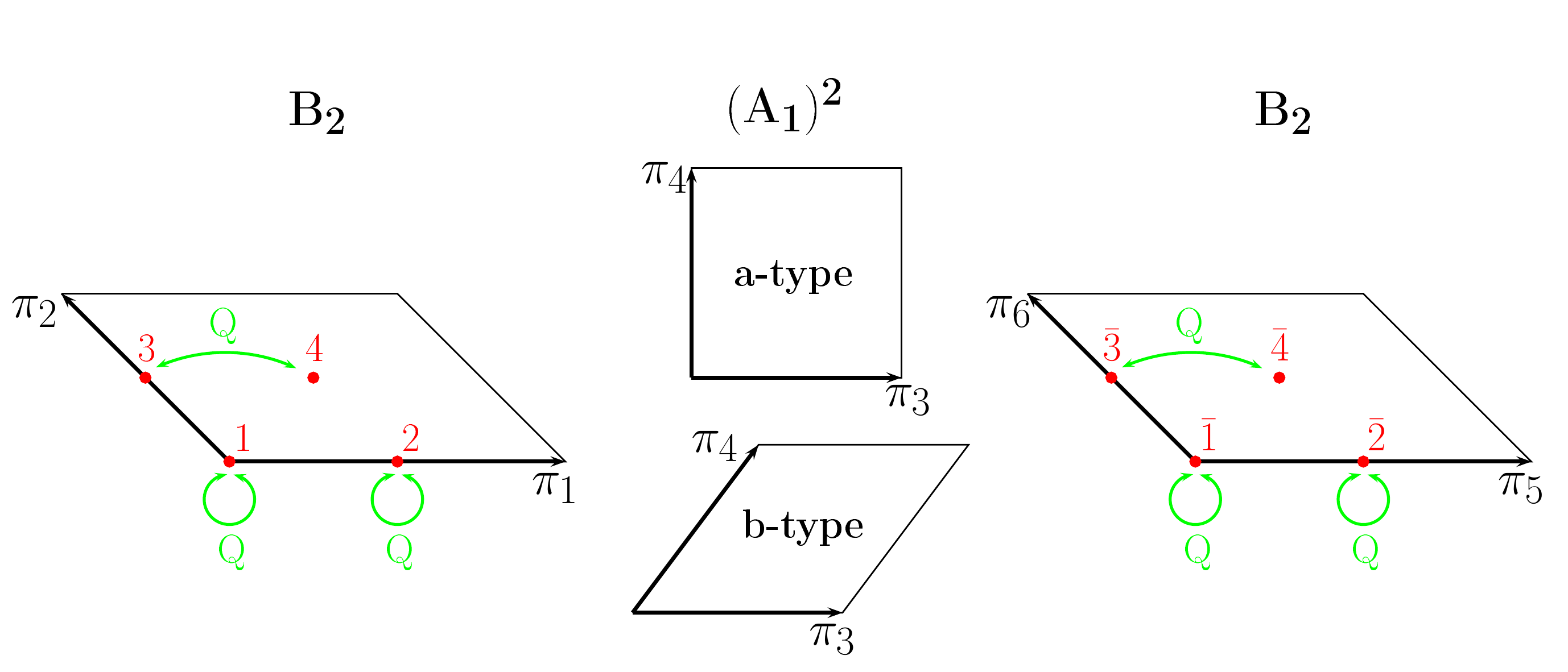}
 		\caption{$T^6/\mathbb{Z}_4$-orbifold on the $B_2\times (A_1)^2\times B_2$-lattice and the $\mathbb{Z}_2$-invariant points (\textcolor{red}{ in red}) on $T^2_{(1)}\times T^2_{(3)}$
		with both possible shapes {\bf a} and {\bf b} displayed for the $(A_1)^2$ torus.}
 		\label{fig:Fig0a}
\end{figure}
Before investigating the three-cycle geometry of the non-factorisable lattices, we briefly review three-cycles on the factorisable background on the group lattice $B_2 \times (A_1)^2 \times B_2$
as first discussed in~\cite{Blumenhagen:2002gw}, see e.g. also the appendix of~\cite{Forste:2010gw} for the Hodge numbers per twist sector displayed in table~\ref{Tab:HodgeNumbers-Z4}. 
Although the $B_2$-torus is a square-torus, we take the positive simple roots of the $B_2$-Lie algebra as basis of the torus lattice (see figure \ref{fig:Fig0a}). The $\Z_4$-action is then generated by the Coxeter element $Q$:
\begin{equation}
Q:=\begin{pmatrix}
1&-1\\2&-1
\end{pmatrix}
\oplus
\begin{pmatrix}
-1&0\\0&-1
\end{pmatrix}
\oplus\begin{pmatrix}
1&-1\\2&-1
\end{pmatrix}\,.
\end{equation}
acting on three two-tori $T^2_{(i)}$. A basis of bulk three-cycles on the factorisable lattice is given by 
\begin{equation}
\begin{aligned}
\gamma_1 := \sum_{k=0}^3 Q^k (\pi_{235}+\pi_{236})= 2 \, (\pi_{136} +\pi_{235}+2\pi_{236}),
\quad
\gamma_2 := -\sum_{k=0}^3 Q^k \pi_{236} = 2 \, (\pi_{135}+\pi_{136}+ \pi_{235}),
\\
\bar{\gamma}_1 := \sum_{k=0}^3 Q^k (\pi_{245}+\pi_{246})= 2 \, (\pi_{146} +\pi_{245}+2\pi_{246}),
\quad
\bar{\gamma}_2 := -\sum_{k=0}^3 Q^k \pi_{246} =2 \, (\pi_{145}+\pi_{146}+ \pi_{245}).
\end{aligned}
\end{equation}
An arbitrary bulk three-cycle can be represented by the (pairwise co-prime) toroidal wrapping numbers $(n^i,m^i)_{i=1,2,3}$, 
\begin{equation}\label{Eq:Pi-Bulk-Factoriable}
\begin{aligned}
\pi^{\text{bulk}} =& \sum_{k=0}^3 Q^k \left(\bigwedge_{i=1}^3 (n^i \pi_{2i-1} + m^i \pi_{2i} )\right)
\\
=& A n^2 \, \gamma_1 + A m^2 \, \bar{\gamma}_1 + B n^2 \, \gamma_2 + B m^2 \, \bar{\gamma}_2,
\end{aligned}
\end{equation}
where on the second line the $\Z_4$ invariant bulk wrapping numbers with 
\begin{equation}
A := n^1m^3+m^1n^3-2n^1n^3,
\qquad
B := n^1 m^3 + m^1n^3-m^1m^3,
\end{equation}
have been used.
The basic non-vanishing bulk intersection numbers are computed from 
\begin{equation}\label{1.intersection_numbers}
\pi_a^{\text{bulk}} \circ \pi_b^{\text{bulk}} \equiv \frac{1}{4} \left(\sum_{i=0}^3 Q^i \pi^{\text{torus}}_a \right) \circ \left(\sum_{i=0}^3 Q^i \pi^{\text{torus}}_b \right)
= \pi^{\text{torus}}_a \circ \left(\sum_{i=0}^3 Q^i \pi^{\text{torus}}_b \right)
\end{equation}
and read
\begin{equation}
\gamma_i \circ \bar{\gamma}_j = 2 \, \delta_{ij}.
\end{equation}
The bulk three-cycles thus do not form an unimodular basis.
In addition to the bulk three-cycles, there exist twelve exceptional cycles with basis
\begin{equation}
\varepsilon_i = (e_{\alpha\beta} - e_{Q(\alpha)Q(\beta)}) \wedge \pi_3,
\qquad
\bar{\varepsilon}_i = (e_{\alpha\beta} - e_{Q(\alpha)Q(\beta)}) \wedge \pi_4
\end{equation}
(where $\alpha$ and $\beta$ denote the $\Z_2$-invariant points on the four-torus $T^2_{(1)}\times T^2_{(3)}$
and $Q(\alpha)$ and $Q(\beta)$ their $\Z_4$ images), and with non-vanishing intersection numbers
\begin{equation}
\varepsilon_i \circ \bar{\varepsilon}_j = - 2 \, \delta_{ij}.
\end{equation}
Fractional three-cycles of the form
\begin{equation}\label{Eq:factorisable-3-cycle}
\pi^{\text{frac}} = \frac{1}{2} \left(\pi^{\text{bulk}} +(-1)^{\tau^{\Z_2}} \sum_{\text{fixed set of } i} [\pm (n^2 \varepsilon_i + m^2 \bar{\varepsilon}_i)]
\right),
\end{equation}
with the $\Z_2$ eigenvalue $(-1)^{\tau^{\Z_2}}$
parametrised by $\tau^{\Z_2} \in \{0,1\}$ and the sum over four $i$ such that the product of the signs $\pm$ gives +1, then generate the unimodular sixteen dimensional basis of three-cycles.

\subsection{$A_3 \times A_3$}\label{Ss:A3xA3}

\begin{figure}[h]
 	\centering
 		\includegraphics[width=15.5cm]{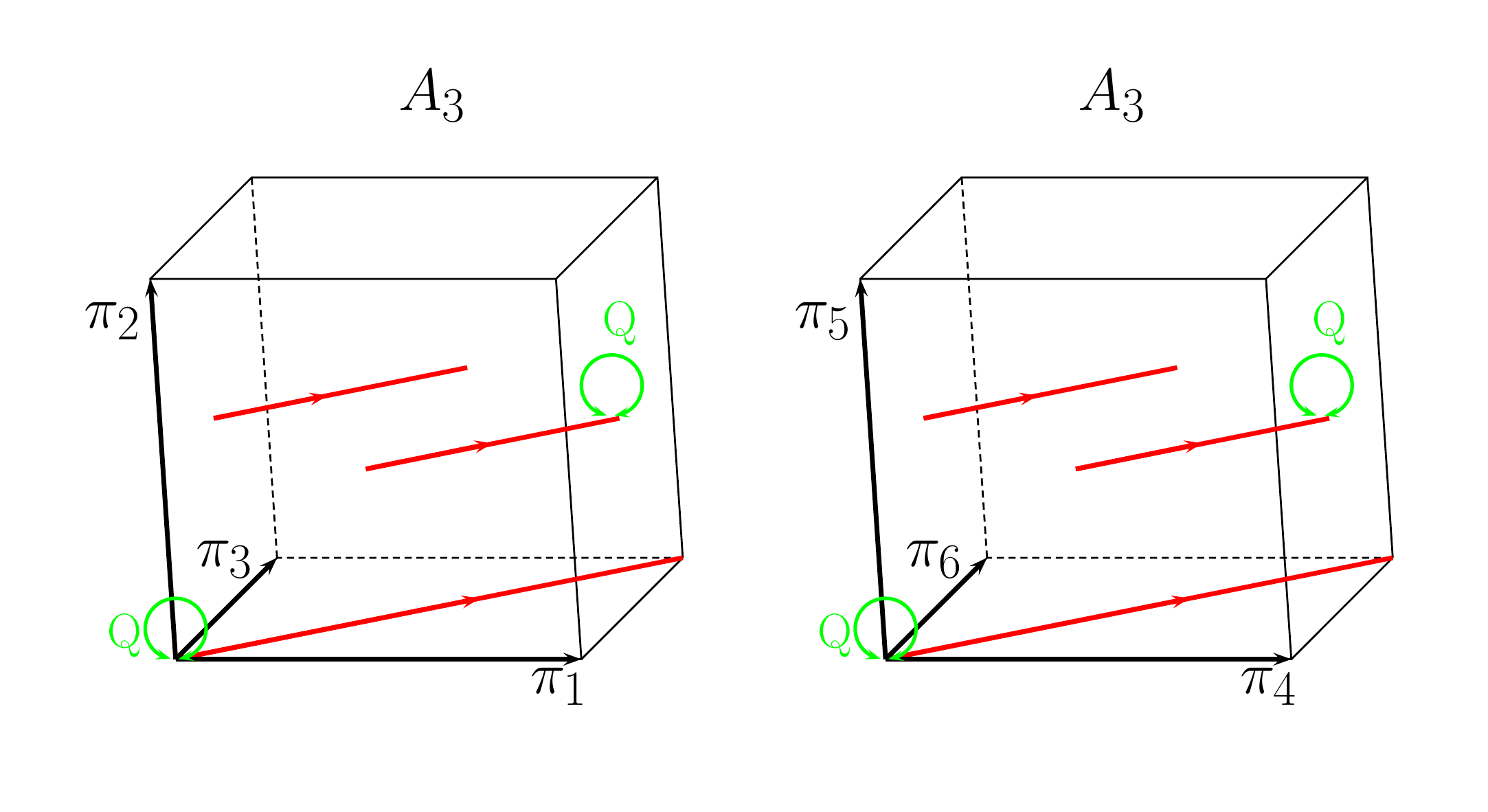}
 		\caption{$T^6/\mathbb{Z}_4$-orbifold on the $A_3\times A_3$-lattice and its $\mathbb{Z}_2$-fixed lines (\textcolor{red}{ in red}).}
 		\label{fig:Fig1a}
 \end{figure}

We start the discussion of non-factorisable $\mathbb{Z}_4$-orbifolds with the lattice of the type $A_3\times A_3$ (see figure \ref{fig:Fig1a}). 
The $\mathbb{Z}_4$ group acts by the Coxeter element $Q$ on the vectors $\{e_i\}_{i=1,\dots,6}$, which span the six-torus:
\begin{equation}\label{2.coxeter_1} 
\begin{array}{cccc}
&Qe_1=e_2\,,&Qe_2=e_3\,,&Qe_3=-e_1-e_2-e_3\,,\\
&Qe_4=e_5\,,&Qe_5=e_6\,,&Qe_6=-e_4-e_5-e_6\,,
\end{array}
\end{equation}
which can be written in the matrix form
\begin{equation}\label{3a} 
Q:=\begin{pmatrix}
0&0&-1&0&0&0\\
1&0&-1&0&0&0\\
0&1&-1&0&0&0\\
0&0&0&0&0&-1\\
0&0&0&1&0&-1\\
0&0&0&0&1&-1
\end{pmatrix}\,.
\end{equation}
This action forms a discrete subgroup of $SU(3)$ and thus preserves $\mathcal{N}=2$ supersymmetry in four dimensions when compactifying Type II string theory. 
The Hodge numbers of this orbifold are (see e.g.~\cite{Lust:2006zh}) $h_{21}=h^\text{untw}_{21}+h_{21}^{\Z_2}=1+0$ and $h_{11}=h^\text{untw}_{11}+h_{11}^{\Z_4}+h_{11}^{\Z_2}=5+16+4$ as displayed in the middle row in table~\ref{Tab:HodgeNumbers-Z4}. 
Due to $h_{21}^{\Z_2}=0$, the three-homology of this orbifold contains only (fractions of) bulk three-cycles and is four-dimensional. 

For the $Q$-action to be an isometry of the lattice, one has to require the invariance of the scalar product $Q^tgQ=g$ \cite{Spalinski:1991yw},
which restricts the shape of the metric of the six-torus to
\begin{equation}\label{3b}
g=\left(
\begin{array}{cccccc}
 R_1^2 & a R_1^2 & -(2 a+1) R_1^2 & e R_1 R_2 & s R_1 R_2 & c R_1
 R_2 \\
 a R_1^2 & R_1^2 & a R_1^2 & d R_1 R_2 & e R_1 R_2 & s R_1 R_2 \\
 -(2 a+1) R_1^2 & a R_1^2 & R_1^2 & c R_1 R_2 & d R_1 R_2 & e R_1 R_2 \\
 e R_1 R_2 & d R_1 R_2 & c R_1 R_2 & R_2^2 & b R_2^2 & -(2 b+1) R_2^2 \\
 s R_1 R_2 & e R_1 R_2 & d R_1 R_2 & b R_2^2 & R_2^2 & b R_2^2 \\
 c R_1 R_2 & s R_1 R_2 & e R_1 R_2 & -(2 b+1) R_2^2 & b R_2^2 &
 R_2^2 \\
\end{array}
\right)
\end{equation}
with $s:=-(c+d+e)$. The moduli $R_1$ and $R_2$ describe the radii of $A_3\times A_3$ whereas $a$ $(b)$ specifies the cosine of the angle between the vectors $e_1$ and $e_2$ ($e_4$ and $e_5$) and $c,d$ and $e$ specify the cosines of the angles between $e_4$ and the vectors $e_3$, $e_2$ and $e_1$, respectively.

To describe the three-cycles on the six-torus $T^6$ we use the usual notation $\pi_{ikl}=\pi_i\wedge\pi_k\wedge\pi_l$. Due to the action of $Q$, it suffices to consider the three-cycles which wrap a two-cycle on one $A_3$-torus and a one-cycle on the other one. Such cycles can {\it a priori} be described by twelve wrapping numbers $(m^i,n^i,p^i,q^i,r^i,s^i)^{i=1,2}$:
\begin{equation}\label{2.toroidal_3-cycle} 
\pi^\text{torus}:=(m^1\pi_1+n^1\pi_2+p^1\pi_3)
\wedge(m^2\pi_1+n^2\pi_2+p^2\pi_3+q^2\pi_4+r^2\pi_5+s^2\pi_6)\wedge(q^1\pi_4+r^1\pi_5+s^1\pi_6)\,.
\end{equation} 
By taking orbits of the $Q$-action, the basis of $\mathbb{Z}_4$-invariant three-cycles is given by
\begin{equation}\label{2.bulk_basis1}
\begin{array}{cll}
\gamma_1&:=&\sum_{i=0}^3Q^i\pi_{124}=-\pi_{125}-\pi_{126}-\pi_{134}-\pi_{135}
+\pi_{235}+\pi_{236}\,,\\
\gamma_2&:=&\sum_{i=0}^3Q^i\pi_{125}=\pi_{124}+\pi_{125}-\pi_{135}-\pi_{136}
-\pi_{234}-\pi_{235}\,,\\
\bar{\gamma}_1&:=&\sum_{i=0}^3Q^i\pi_{145}=-\pi_{146}-\pi_{245}-\pi_{246}
+\pi_{256}-\pi_{345}+\pi_{356}\,,\\
\bar{\gamma}_2&:=&\sum_{i=0}^3Q^i\pi_{245}=\pi_{145}-\pi_{156}+\pi_{245}
-\pi_{246}-\pi_{256}-\pi_{346}\,.
\end{array}
\end{equation} 
Using the ansatz $\eqref{2.toroidal_3-cycle}$ for a toroidal three-cycle, we can compute the corresponding bulk cycle with the orbifold map $\sum_{i=0}^3Q^i$ and decompose it in the basis $\eqref{2.bulk_basis1}$:
\begin{equation}\label{1f}
\begin{split}
\pi^\text{bulk}&=\left(A_1(q^1-s^1)-A_2(q^1+r^1-s^1)+A_3r^1\right)\gamma_1\\
 &+\left(A_1r^1+A_2(q^1-r^1-s^1)-A_3(q^1-s^1)\right)\gamma_2\\
 &+\left(B_1(m^1-p^1)-B_2(m^1+n^1-p^1)+B_3n^1\right)\bar{\gamma}_1\\
 &+\left(B_1n^1+B_2(m^1-n^1-p^1)-B_3(m^1-p^1)\right)\bar{\gamma}_2
\end{split}
\end{equation}
with
\begin{equation*}
\begin{array}{ccc}
A_1:=m^1n^2-n^1m^2\,,\quad &A_2:=m^1p^2-p^1m^2\,,\quad &A_3:=n^1p^2-p^1n^2\,,\\
B_1:=q^2r^1-r^2q^1\,,\quad &B_2:=q^2s^1-s^2q^1\,,\quad &B_3:=r^2s^1-s^2r^1\,.
\end{array}
\end{equation*}
It is easy to verify that the non-vanishing intersection numbers between the basis elements \eqref{2.bulk_basis1} are given by
\begin{equation}\label{2.intersection1}
\gamma_i\circ\bar{\gamma}_j=-\delta_{ij},
\end{equation} 
where we used analogously to the factorised lattice background in section~\ref{Ss:B2xA1-2xB2} that we can define the intersection number between two toroidal three-cycles $
\pi^{\text{torus}}_a$ and $\pi^{\text{torus}}_b$ on $T^6$ as $\pi^{\text{torus}}_a\circ\pi^{\text{torus}}_b=\pi^{\text{torus}}_a \wedge\pi^{\text{torus}}_b/\text{Vol}(T^6)$
to arrive at the intersection number between two $\Z_4$-invariant bulk three-cycles 
\begin{equation}\label{Eq:IntersectionNumber-with-Q}
\pi_a\circ\pi_b=\frac{1}{4}\bigl(\sum_{i=0}^{3}Q^i\pi^{\text{torus}}_a\bigr)
\circ\bigl(\sum_{i=0}^{3}Q^i\pi^{\text{torus}}_b\bigr).
\end{equation}
Therefore, $\{\gamma_1,\,\gamma_2,\,\bar{\gamma}_1,\,\bar{\gamma}_2\}$ already builds the unimodular basis. 

Despite this and the fact that this orbifold does not have any exceptional three-cycles,
in order to compare with the other non-factorisable lattice in section~\ref{Ss:A3A1B2} below, 
we can consider the special class of three-cycles, for which the toroidal building blocks are $\Z_2$-invariant, i.e. fractional three-cycles\footnote{This construction will be used in the next sections. We will see that if both $A_3$-tori are orthogonal to each other, any fractional three-cycle is Lagrangian.\\
 Note also that for $T^6/(\Z_2 \times \Z_2)$ without discrete torsion, the unimodular basis is constructed analogously using $\Z_2 \times \Z_2$-invariant three-cycles $\pi^{\text{torus}}=\frac{1}{4}\pi^{\text{bulk}}$, cf. e.g.~\cite{Blumenhagen:2002wn}. However, since $Q$ as defined in~\eqref{3a} permutes toroidal one-cycles on $T^6/\Z_4$ non-trivially, here we have to restrict to a special subclass of all {\it a priori} allowed bulk three-cycles. }
 \begin{equation}\label{2.fact2}
\pi^\text{frac}=\frac{1}{2}\pi^\text{bulk}.
\end{equation}
Using the $Q$-transformation of the wrapping numbers
\begin{equation}\label{1g}
\left(\begin{array}{ccc}
m^i\,,&n^i\,,&p^i\\
 q^i\,,&r^i\,,&s^i
\end{array}\right)^{i=1,2}\quad \stackrel{Q}{\longrightarrow} \quad
\left(\begin{array}{ccc}
-p^i\,,&m^i-p^i\,,&n^i-p^i\\
 -s^i\,,&q^i-s^i\,,&r^i-s^i
\end{array}\right)^{i=1,2},
\end{equation}
we obtain the following condition. Any toroidal three-cycle $\pi^\text{torus}$ of the form $\eqref{2.toroidal_3-cycle}$ is $\mathbb{Z}_2$-invariant if and only if
\begin{equation}\label{1h}
\begin{array}{ccc}
r^1(A_1+A_3)=0\,,\quad&n^1(B_1+B_3)=0\,,\\
(s^1-r^1)A_3=q^1A_1\,,\quad&(p^1-n^1)B_3=m^1B_1\,,\\
(A_1+A_3)(s^1-q^1)=0\,,\quad&(B_1+B_3)(m^1-p^1)=0\,,\\
(A_1+A_3)s^1=(q^1-r^1+s^1)A_2\,,\quad&(B_1+B_3)p^1=(m^1-n^1+p^1)B_2.
\end{array}
\end{equation}

Among this special class of three-cycles from $\Z_2$-invariant toroidal cycles, there are also those that satisfy $Q\pi^\text{torus}=-\pi^\text{torus}$. They do not contribute to the bulk cycles, or in other words are trivial in the $H_3(T^6/\Z_4,\Z)$ homology.
Taking into account this fact, we can reduce the conditions \eqref{1h} to 
\begin{equation}\label{2.frac.cond}
\begin{array}{cccc}
A_1+A_3=0\,,\quad&B_1+B_3=0\,,&\\
(q^1-r^1+s^1)A_i=0\,,\quad&(m^1-n^1+p^1)B_i=0& \text{for all }i\,.
\end{array}
\end{equation}

Note that due to the above $\Z_2$-invariance constraints on the wrapping numbers, the basis of these fractional three-cycles coincides with the basis $\{\gamma_1,\,\gamma_2,\,\bar{\gamma}_1,\,\bar{\gamma}_2\}$.

Let us for example consider the toroidal three-cycle $\pi=(\pi_1+\pi_2)\wedge(\pi_1+\pi_3)\wedge(\pi_4-\pi_6)$. Obviously, it is $\Z_2$-invariant and gives rise to the fractional three-cycle $\pi^\text{frac}=\pi+Q\pi=-2\gamma_1$ on $T^6/\Z_4$. Although the original cycle has coprime wrapping numbers and no other toroidal three-cycle $\pi^\prime$ exists such that $\pi=2\pi^\prime$, the corresponding bulk cycle $-2\gamma_1$ is non-coprime. This will play a role for computing the gauge group in the section \ref{Sss:Crosscheck}. 
\\
\subsection{$A_3 \times A_1\times B_2$}\label{Ss:A3A1B2}

Now we consider three-cycles on $T^6/\mathbbm{Z}_4$ with the lattice of type $A_3\times A_1\times B_2$. The $\Z_4$-action is generated by the Coxeter element which acts on the root lattice spanned by the simple roots $\{e_i\}_{i=1,\dots,6}$ in the following way 
\begin{equation}\label{2.coxeter_2} 
\begin{array}{llll}
&Qe_1=e_2\,,&Qe_2=e_3\,,&Qe_3=-e_1-e_2-e_3\,,\\
&Qe_4=-e_4\,,&Qe_5=e_5+2e_6\,,&Qe_6=-e_5-e_6\,,
\end{array}
\end{equation}
which can be cast in the matrix form
\begin{equation}\label{1a}
Q:=\left(
\begin{array}{cccccc}
 0 & 0 & -1 & 0 & 0 & 0 \\
 1 & 0 & -1 & 0 & 0 & 0 \\
 0 & 1 & -1 & 0 & 0 & 0 \\
 0 & 0 & 0 & -1 & 0 & 0 \\
 0 & 0 & 0 & 0 & 1 & -1 \\
 0 & 0 & 0 & 0 & 2 & -1 \\
\end{array}
\right)\,.
\end{equation}
Again, this action corresponds to a discrete subgroup of $SU(3)$, and thus preserves $\mathcal{N}=2$ supersymmetry in four dimensions when considering Type II string theory compactifications. 
The Hodge numbers of this orbifold are \cite{Lust:2006zh} $h_{21}=h^\text{untw}_{21}+h_{21}^{\Z_2}=1+2$ and $h_{11}=h^\text{untw}_{11}+h_{11}^{\Z_4}+h_{11}^{\Z_2}=5+16+6$ (see table \ref{Tab:HodgeNumbers-Z4}). Thus, we expect four bulk and four exceptional three-cycles on this orbifold.

From solving the equation $Q^tgQ=g$ we obtain
\begin{equation}\label{1b}
g:=e_i\cdot e_j=\begin{pmatrix}
R_3^2&aR_3^2&-(1+2a)R_3^2&dR_3R_1&bR_3R_2&cR_3R_2\\
aR_3^2&R_3^2&aR_3^2&-dR_3R_1&-(b+2c)R_3R_2&(b+c)R_3R_2\\
-(1+2a)R_3^2&aR_3^2&R_3^2&dR_3R_1&-bR_3R_2&-cR_3R_2\\
dR_3R_1&-dR_3R_1&dR_3R_1&R_1^2&0&0\\
bR_3R_2&-(b+2c)R_3R_2&-bR_3R_2&0&2R_2^2&-R_2^2\\
cR_3R_2&(b+c)R_3R_2&-cR_3R_2&0&-R_2^2&R_2^2\end{pmatrix}
.
\end{equation}
The real positive moduli $R_3$, $R_1$ and $R_2$ describe the radii of $A_3\times A_1 \times B_2$, respectively, and $a$, $b$, $c$ and $d$ specify the cosines of angles between the vectors of the lattice. More precisely, $a$ is the cosine of the angle between the vectors $e_1$ and $e_2$, $d$ the cosine of the angle between $e_1$ and $e_4$, $b$ ($c$) is the cosine of the angle between $e_1$ and $e_5$ ($e_6$). \par\noindent
\textit{\textbf{Bulk three-cycles}}\par
We make the ansatz that any toroidal three-cycle 
is factorisable in the sense that it can be characterised {\it a priori} by ten wrapping numbers $(m^1,n^1,p^1,q^1)\times(m^2,n^2,p^2,q^2)\times(m^3,n^3)$ and written as
\begin{equation}\label{2.torus_cycle} 
\pi^{\text{torus}}:=\bigwedge_{i=1}^2(m^i\pi_1+n^i\pi_2+p^i\pi_3+q^i\pi_4) \wedge(m^3\pi_5+n^3\pi_6)\,.
\end{equation}
The last doublet $(m^3,n^3)$ gives us the one-cycle on the $B_2$-torus. The two quadruplets $(m^i,n^i,p^i,q^i)$
parametrise the two-cycle on $A_3\times A_1$. It is easy to see that the representation of three-cycles by this ansatz is not unique, i.e.\;the same three-cycle can be described by different wrapping numbers, e.g. when permuting the indices $i=1$ and $2$.
 
By taking orbits of the $Q$-action, we can define a basis of the $\Z_4$-invariant {\it bulk} three-cycles
\begin{equation}\label{2.bulk_basis} 
\begin{split}
\gamma_1&:=-\sum_{i=0}^3 Q^i\pi_{136}=
2(\pi_{125}+\pi_{126}-\pi_{136}-\pi_{235}-\pi_{236})\,,\\
\gamma_2&:=-\sum_{i=0}^3 Q^i\pi_{125}=
2(\pi_{126}+\pi_{135}+\pi_{136}-\pi_{236})\,,\\
\bar{\gamma}_1&:=\sum_{i=0}^3 Q^i\pi_{146}=
\pi_{145}+2\pi_{146}+2\pi_{245}+2\pi_{246}+\pi_{345}\,,\\
\bar{\gamma}_2&:=\sum_{i=0}^3 Q^i\pi_{246}=
-\pi_{145}+2\pi_{246}+\pi_{345}+2\pi_{346}\,.
\end{split}
\end{equation} 
Note that here the linear combinations $\frac{1}{2}(\gamma_1\pm\gamma_2)$ are also bulk cycles. \\
The decomposition in the basis $\{\gamma_{1,2},\,\bar{\gamma}_{1,2}\}$ of any bulk three-cycle inherited from a toroidal one of the type $\eqref{2.torus_cycle}$ is given by
\begin{equation}\label{2.bulk_decomposition} 
\begin{split}
\pi^{\text{bulk}}=& P\gamma_1+Q\gamma_2+\bar{P}\bar{\gamma}_1 +\bar{Q}\bar\gamma_2\\
=&[(A_2-A_3)m^3+\frac{1}{2}(A_1-2A_2+A_3)n^3]\gamma_1+[(A_2-A_1)m^3+\frac{1}{2}(A_1-A_3)n^3]\gamma_2+\\
&[(-B_1+B_2+B_3)m^3+(B_1-B_3)n^3]\bar{\gamma}_1+[(-B_1-B_2+B_3)m^3+B_2n^3]\bar{\gamma}_2
\, ,
\end{split}
\end{equation}
with
\begin{equation}\label{2.definition_of_A_and_B}
\begin{split}
A_1&:=m^1n^2-n^1m^2,\quad B_1:=m^1q^2-q^1m^2,\\
A_2&:=m^1p^2-p^1m^2,\quad\: B_2:=n^1q^2-q^1n^2,\\
A_3&:=n^1p^2-p^1n^2,\quad\;\;\: B_3:=p^1q^2-q^1p^2\,.
\end{split}
\end{equation}
Using the formula for bulk intersection numbers~\eqref{Eq:IntersectionNumber-with-Q}, we obtain for the bulk basis of $\eqref{2.bulk_basis}$ 
\begin{equation}\label{2.intersection_number_bulk} 
\gamma_i\circ\bar{\gamma}_j=-2\delta_{ij},\quad\gamma_i\circ\gamma_j=\bar{\gamma}_i\circ\bar{\gamma}_j=0\,,
\end{equation}
which shows that the integral basis of bulk three-cycles is not unimodular.
\par
\textbf{\textit{Exceptional three-cycles}}\par
\begin{figure}[h]
 	\centering
 		\includegraphics[width=15.5cm]{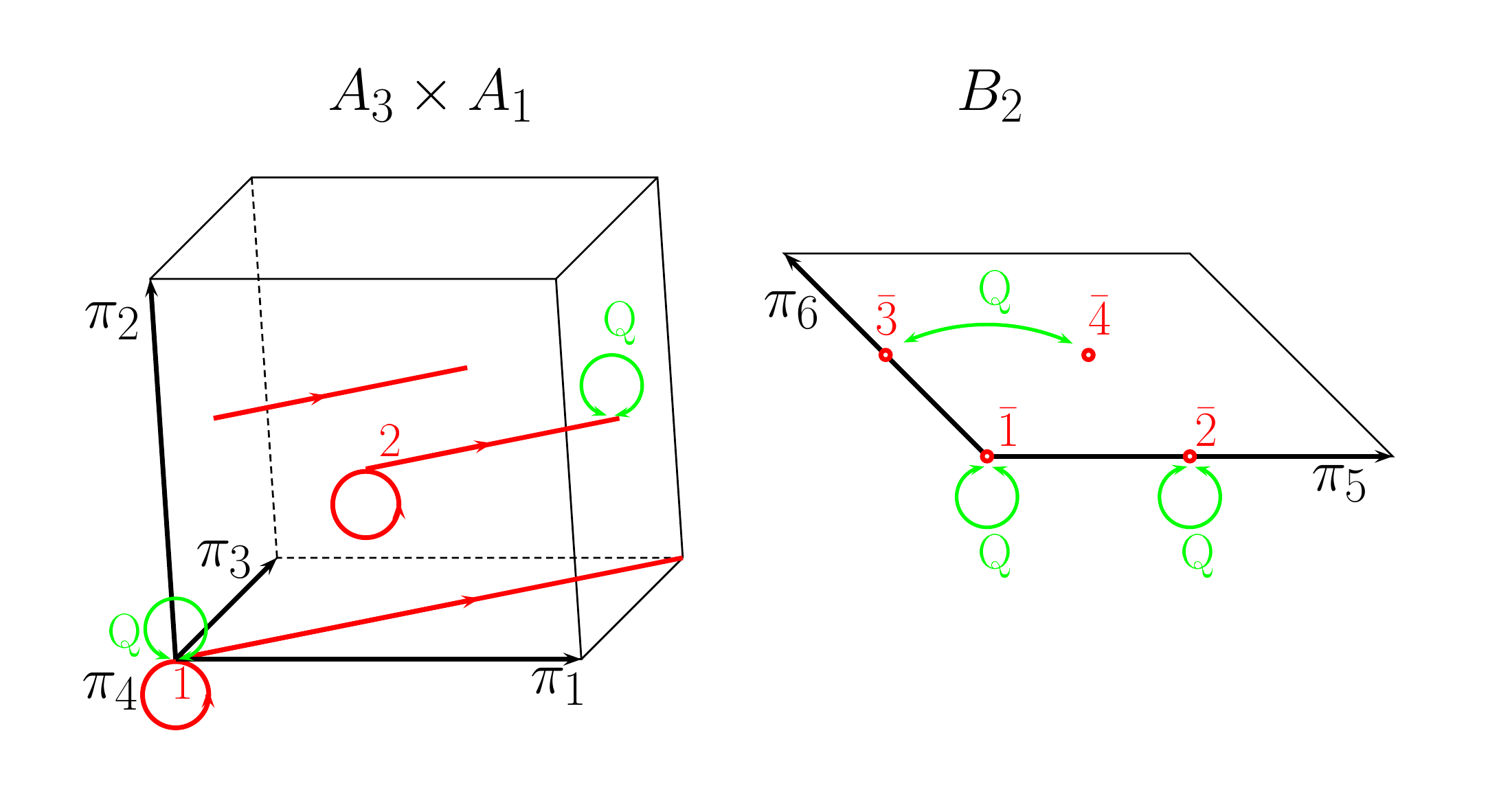}
 		\caption{$T^6/\mathbb{Z}_4$-orbifold on the $A_3\times A_1\times B_2$-lattice and its $\mathbb{Z}_2$-fixed lines (\textcolor{red}{in red}).}
 		\label{fig:Fig2}
 \end{figure}

Besides the four bulk three-cycles $\gamma_i,\bar{\gamma}_i$ (with $i=1,2$), there are also 
four
 exceptional three-cycles appearing in the $\mathbbm{Z}_2$ twisted sector of the orbifold. It is easy to see that $Q^2$ acts trivially on the sub-manifold $(\pi_1+\pi_3)\wedge\pi_4$. One can show that there are eight such $Q^2$-invariant sub-manifolds, which are indicated in red in figure $\ref{fig:Fig2}$. We numerate them by ${i \bar{j}}$ where the first index denotes the $\Z_2$-invariant two-tori ({\color{red}1,\;2}) on the $A_3\times A_1$-torus and the second one the $\Z_2$-invariant points ({$\color{red}\bar{1},\,\bar{2},\,\bar{3},\,\bar{4}$}) on the $B_2$-torus. These $Q^2$-invariant sub-manifolds can be arranged in six congruence classes (under the $Q$-action): $\{1\bar{1}\},\,\{1\bar{2}\},\,\{1\bar{3},\,1\bar{4}\},\,\{2\bar{1}\},\,\{2\bar{2}\},\,\{2\bar{3},\,2\bar{4}\}. $ 
 \\

The resolution of these six $\Z_2$-singular sub-manifolds gives rise to six four-dimensional sub-manifolds with the topology $S^2\times T^2$ in accordance with $h_{22}^{\Z_2}\stackrel{\text{Poincar\'e}}{=}h_{11}^{\Z_2}=6$ in table~\ref{Tab:HodgeNumbers-Z4}, where the exceptional two-cycle $\mathbf{e}_{i\bar{j}}$ describes the $S^2$-part, and the two one-cycles $\pi_1+\pi_3$ and $\pi_4$ span a two-torus. The index of $\mathbf{e}_{i\bar{j}}$ is inherited from the numeration of the $\Z_2$-invariant two-tori. Finally by splitting $T^2$ into one-cycles $\pi_1+\pi_3$ and $\pi_4$, we construct $Q$-invariant exceptional three-cycles. Due to the anti-symmetric action of $Q$ on the one-cycles, only the exceptional two-cycles $\mathbf{e}_{1\bar{3}},\, \mathbf{e}_{1\bar{4}},\,\mathbf{e}_{2\bar{3}},\,\mathbf{e}_{2\bar{4}}$ provide non-trivial results in the construction. Thus, the exceptional three-cycles are: 
\begin{equation}\label{2.except_cycles} 
\begin{split}
\gamma_3&:=(\mathbf{e}_{1\bar{3}}-\mathbf{e}_{1\bar{4}})\wedge(\pi_1+\pi_3)\,,\qquad \bar{\gamma}_3:=(\mathbf{e}_{1\bar{3}}-\mathbf{e}_{1\bar{4}})\wedge\pi_4\,,\\
\gamma_4&:=(\mathbf{e}_{2\bar{3}}-\mathbf{e}_{2\bar{4}})\wedge(\pi_1+\pi_3)\,,\qquad \bar{\gamma}_4:=(\mathbf{e}_{2\bar{3}}-\mathbf{e}_{2\bar{4}})\wedge\pi_4,
\end{split}
\end{equation} 
with the intersection numbers
\begin{equation}\label{2.intersection_number_exc_cycles} 
\gamma_i\circ\bar{\gamma}_j=2\delta_{ij}\,,\quad\gamma_i\circ\gamma_j=\bar{\gamma}_i\circ\bar{\gamma}_j=0\qquad i=3,4\,.
\end{equation}
Since the intersection form of the $\gamma_i$'s, and $\bar{\gamma}_i$'s $(i=1,2,3,4)$ is not unimodular, these three-cycles do not form the minimal integral basis. We thus have to consider fractional three-cycles, which can consist of half a bulk cycle and simultaneously of half an exceptional cycle.

\textbf{\textit{Fractional three-cycles and their integral basis}}

In order to write down an integral basis for the three-cycles such that the intersection form is unimodular, we start by specifying the construction of fractional cycles, which are $\mathbbm{Z}_2$-invariant,
analogously to the factorisable case reviewed in section~\ref{Ss:B2xA1-2xB2}. The fractional cycles can wrap half a bulk cycle and half an exceptional one
\begin{equation*}
\pi^{\text{frac}}=\frac{1}{2}\pi^{\text{bulk}}+\frac{1}{2}\pi^{\text{exc}}\,.
\end{equation*}
Motivated by the factorisable orbifold, we apply the well-known techniques of construction of the fractional cycles to our case. The non-factorisable structure of the lattice gives rise to some differences and therefore some modification of these techniques is needed.

In the case of the factorisable torus $T^2_1\times T^2_2\times T^2_3$, every $\Z_2$-invariant three-cycle passes through two $\mathbbm{Z}_2$-invariant points per $T^2$ (here $T^2_1 \times T^2_3$) and contains a $\Z_2$-invariant one-cycle on the remaining two-torus (here $T^2_2$). In the $A_3\times A_3$-case we saw that this does not hold true and we had to generalise this condition to $Q^2\pi^\text{torus}=\pi^\text{torus}$. The same happens in the present case of the $A_3 \times A_1 \times B_2$-lattice. Let us for example consider the toroidal cycle $(1,0,0,0)\times(0,0,0,1)\times(1,0)\equiv\pi_{145}$ through the origin. Despite the fact that this three-cycle contains the $\mathbbm{Z}_2$-invariant one-cycle $\pi_4$ and passes through a $\mathbbm{Z}_2$-invariant point, $\pi_{145}$ is not $\mathbbm{Z}_2$-invariant and therefore cannot be fractional. Indeed, one can check that $Q^2\pi_{145}=-\pi_{345}$. 
To impose the additional constraint on the numbers $(m^i,n^i,p^i,q^i)_{i=1,2}\times(m^3,n^3)$ ensuring the $Q^2$-invariance of the toroidal three-cycle we use the $Q$-action on the wrapping numbers
\begin{equation}\label{2.Q-action_on_wrap_numbers}
\begin{pmatrix}
m^i,&n^i,&p^i,&q^i\\
&m^3,&n^3&
\end{pmatrix}\;\;\stackrel{Q}{\longrightarrow}\;\;
\begin{pmatrix}
-p^i,&m^i-p^i,&n^i-p^i,&-q^i\\
&m^3-n^3,&2m^3-n^3&
\end{pmatrix}
.
\end{equation}

For any toroidal three-cycle $\pi^{\text{torus}}$ of the form $\eqref{2.torus_cycle}$, it can be shown that the following statements are equivalent:
\begin{equation}\label{2.fractional_condition}
\pi^{\text{torus}} \text{ is $\Z_2$-invariant}\Longleftrightarrow\;Q^2\pi^{\text{torus}}=\pi^{\text{torus}}\;\Longleftrightarrow\; A_1+A_3=0\;\;\text{and }\;B_1-B_2+B_3=0.
\end{equation}
Only such $\Z_2$-invariant toroidal three-cycles can be used for the construction of the fractional cycles.\\ 
From the factorisable case we know that the toroidal three-cycles giving rise to the fractional cycles have to contain 
$\Z_2$-invariant one-cycles. Indeed, it can be verified that any cycle which satisfies the conditions $\eqref{2.fractional_condition}$ can be written as
\begin{equation}\label{2.Z2-inv_cycles}
(m^1,n^1,p^1,q^1)\times(m^2,n^2,p^2,q^2)\times(m^3,n^3)=(m^1,n^1,p^1,0)\times(\tilde{p}^2,0,\tilde{p}^2,\tilde{q}^2)\times(m^3,n^3)+\text{R}
,
\end{equation} 
where $\tilde{p}^2:=p^2-\frac{n^2}{n^1}p^1$ and $\tilde{q}^2:=q^2-\frac{n^2}{n^1}q^1$ for $n^1\neq0$.\footnote{{} $n^1=0$ gives rise to a similar result.} The remaining term
\begin{equation*}\label{a7d2}
\text{R}:=(0,0,0,q^1)\times(\tilde{p}^2,0,\tilde{p}^2,\tilde{q}^2)\times(m^3,n^3)
\end{equation*}
does not contribute to the bulk three-cycle and can be neglected. Therefore, any $\mathbbm{Z}_2$-invariant three-cycle contains a linear combination of the $\mathbbm{Z}_2$-invariant one-cycles $\pi_1+\pi_3$ and $\pi_4$.

The next step to construct the fractional cycles is to determine the exceptional part. In the factorisable case we identify which $\Z_2$-invariant points the toroidal three-cycle passes through. On the $A_3\times A_1\times B_2$-orbifold also the $\Z_2$-invariant lines are involved and so we have to calculate their intersection with the toroidal three-cycle.
Another important difference is that for the factorisable lattice $(T^2)^3$, the co-prime condition on the wrapping numbers $(n^i,m^i)$ per $T^2_i$ is necessary and sufficient for the fractional cycle to have a co-prime decomposition in the unimodular basis. On the contrary, in the non-factorisable case the sufficiency condition gets lost and some further restrictions on the wrapping numbers $(m^i,n^i,p^i,q^i)$ are needed. It turns out that it is convenient to neglect these constraints at first and better to select the co-prime cycles after they are expressed in the unimodular basis. Thus, in this case, for the computation of the exceptional part we have to know not just which $\Z_2$-invariant lines the toroidal three-cycle intersects but also how many times. That requires some combinatorics.

Using the property \eqref{2.Z2-inv_cycles} for the $\Z_2$-invariant three-cycles, we have only to compute how many times the cycle with the wrapping numbers $(m^1,n^1,p^1,0)$ intersects the one-cycle $\pi_1+\pi_3$. We find that the number $N$ of intersection points is
\begin{equation}\label{2. number_of_intersections}
N=\text{g.c.d.}(m^1-p^1,n^1)\,.
\end{equation}
Furthermore, we make the following considerations and set several notations:
\begin{itemize}
\item The three-cycle can pass through both fixed lines 1 and 2 on the $A_3$-torus if and only if $\frac{m^1-p^1}{N}$ and $\frac{n^1}{N}$ are odd. We introduce three new parameters counting the different fixed lines traversed by the three-cycle
\begin{equation}\label{2.tau_parameter} 
\tau:=\begin{cases}
\displaystyle 2 &\text{if } \frac{m^1-p^1}{N} \text{ and } \frac{n^1}{N} \text{ are odd },\\
1& \text{ otherwise } 
\end{cases}
\end{equation}
and 
\begin{equation}\label{2.taus_12_parameter}
\tau^1,\tau^2\in\{0,1\}\quad\text{with }\;\tau^1+\tau^2=\tau.
\end{equation}
\item On the $B_2$-torus, a one-cycle with even wrapping number $n^3$ passes through the fixed points $\bar{1}$ and $\bar{2}$ (both $\Z_4$-invariant), or $\bar{3}$ and $\bar{4}$ (both $\Z_2$-invariant). In this case, the fractional three-cycle contains either none or two exceptional three-cycles. If $n^3$ is odd, the one-cycle on $B_2$ intersects one $\Z_4$- and one $\Z_2$-invariant fixed point, which in any case gives rise to one corresponding exceptional cycle. Thus, we define the parameter $\sigma$, which counts the number of exceptional three-cycles contributing to a fractional cycle for given bulk part, by
\begin{equation}\label{2.sigma_parameter}
\sigma:=\begin{cases}\displaystyle 0,2&\quad\text{if }n^3\;\;\text{is even,}\\
1&\quad\text{if }n^3\;\;\text{is odd}.
\end{cases}
\end{equation} 
\item Furthermore, we introduce two parameters $s^1,s^2=\pm1$ which describe the winding directions of the exceptional cycles. 
\end{itemize}

Now we are able to write down the final form of the fractional three-cycle which is stemmed from the $\Z_2$-invariant toroidal three-cycle with the wrapping numbers $(m^1,n^1,p^1,0)\times(\tilde{p}^2,0,\tilde{p}^2,\tilde{q}^2)\times(m^3,n^3)$. It has the following form:
\begin{equation}\label{a7j}
\pi^{\text{frac}}=\frac{1}{2}\pi^{\text{bulk}}+N\frac{\tilde{p}^2\sigma}{2}(s^1\tau^1\gamma_3+s^2\tau^2\gamma_4)+N\frac{\tilde{q}^2\sigma}{2}(s^1\tau^1\bar{\gamma}_3+s^2\tau^2\bar{\gamma}_4).
\end{equation}
It is easy to check that the coefficients of $\pi^{\text{bulk}}$, stemming from the toroidal cycle with the wrapping numbers $(m^1,n^1,p^1,0)\times(\tilde{p}^2,0,\tilde{p}^2,\tilde{q}^2)\times(m^3,n^3)$, have $N$ as defined in~\eqref{2. number_of_intersections}
as a common divisor too. Naively one can expect the restriction to $N=1$ for the co-prime condition. However, this is not necessarily true. Let us for example consider the toroidal cycle with wrapping numbers $(1,0,-1,0)\times(0,0,0,1)\times(0,1)$ passing through the fixed point 1. It gives rise to the fractional cycle $\frac{1}{2}(2\bar{\gamma}_1\pm 2\bar{\gamma}_3)$. Although $N=2$ this cycle is co-prime because the cycle $\frac{1}{2}(\bar{\gamma}_1\pm \bar{\gamma}_3)$ does not exist due to the condition $\eqref{2.fractional_condition}$.

As a consequence of the considerations above we can summarise: 
\begin{itemize}
\item Any fractional cycle containing only the one-cycle factor $\pi_1+\pi_3$ ($\tilde{q}^2=0$) within the exceptional part
can be expressed as a sum over the unbarred cycles $\gamma_i$, 
\begin{equation}\label{a7e} 
\pi:=v^1\gamma_1+v^2\gamma_2+v^3\gamma_3+v^4\gamma_4,
\end{equation}
with an even number of the coefficients $v^i$ with half-integer values.
\item 
Any fractional cycle containing only the one-cycle factor $\pi_4$ ($\tilde{p}^2=0$) within the exceptional part
 can be expanded as a sum over the barred cycles $\bar{\gamma}_i$,
\begin{equation}\label{a7f} 
\bar{\pi}:=\bar{v}^1\bar{\gamma}_1+\bar{v}^2\bar{\gamma}_2-\bar{v}^3\bar{\gamma}_3
-\bar{v}^4\bar{\gamma}_4,
\end{equation}
where $\bar{v}^i$ are either all half-integer or all integer.
\end{itemize} 

In order to determine an integral symplectic basis for the three-cycles, we search for pairs of cycles $\pi$ and $\bar{\pi}$ such that $\pi\circ\bar{\pi}=-2$, with the coefficients $v^i$ and $\bar{v}^i$ satisfying\footnote{The signum function is given by sgn$(x):=\begin{cases}\displaystyle-1&\text{if }x<0\,,\\0&\text{if }x=0\,,\\1&\text{if }x>0\,.\end{cases}$} 
sgn\:$v^i=\,$sgn\:$\bar{v}^i$ for all $i$.
All possible cycles can be combined in three sets:
\begin{enumerate} 
\item $(v^1,v^2,v^3,v^4)=(\underline{\pm\frac{1}{2},\pm\frac{1}{2},0,0})$ and $(\bar{v}^1,\bar{v}^2,\bar{v}^3,\bar{v}^4)=(\underline{\pm 1,\pm 1,0,0})$ where the underlying denotes all permutations of entries. 
\item $(v^1,v^2,v^3,v^4)=(\underline{\pm 1,0,0,0})$ for the (un-)barred cycles. 
\item $(v^1,v^2,v^3,v^4)=(\pm\frac{1}{2},\pm\frac{1}{2},\pm\frac{1}{2},\pm\frac{1}{2})$ for the (un-)barred cycles. 
\end{enumerate}

Altogether there are 48 unbarred and 48 barred cycles. It is not difficult to specify a basis of these cycles. For the unbarred part we obtain
\begin{equation}\label{2.basis_unbarred_cycles} 
\begin{split}
\alpha_1:=&(0,\frac{1}{2},-\frac{1}{2},0)\,,\\
\alpha_2:=&(0,0,\frac{1}{2},-\frac{1}{2})\,,\\
\alpha_3:=&(0,0,0,1)\,,\\
\alpha_4:=&(\frac{1}{2},-\frac{1}{2},-\frac{1}{2},-\frac{1}{2}),
\end{split}
\end{equation} 
and for the barred one,
\begin{equation}\label{2.basis_barred_cycles} 
\begin{split}
\bar{\alpha}_1:=&(0,1,1,0)\,,\\
\bar{\alpha}_2:=&(0,0,-1,1)\,,\\
\bar{\alpha}_3:=&(0,0,0,-1)\,,\\
\bar{\alpha}_4:=&(\frac{1}{2},-\frac{1}{2},\frac{1}{2},\frac{1}{2})\,.
\end{split}
\end{equation} 
Therefore, the fractional cycles form the $F_4\oplus F_4$-lattice, and the intersection matrix takes the form
\begin{equation}\label{2.unimodular_matrix} 
\alpha_i\circ\bar{\alpha}_j=\begin{pmatrix}
-2&1&0&0\\
1&-2&2&0\\
0&1&-2&1\\
0&0&1&-2
\end{pmatrix}\,.
\end{equation}
This is the Cartan matrix for $F_4$, and it is unimodular. Indeed, it is easy to verify that any bulk, exceptional or fractional cycle can be expanded in the basis $\eqref{2.basis_unbarred_cycles}$ and $\eqref{2.basis_barred_cycles}$ with integer coefficients. In other words, since the determinant of the intersection matrix \eqref{2.unimodular_matrix} is 1, the $\alpha_i$'s and $\bar{\alpha}_i$'s form an integral basis of the homology lattice $H_3(M,\mathbbm{Z})$.

\section{Intersecting brane worlds}\label{S:IBWs}

The aim of this article is not only the study of non-factorisable $\Z_4$-orbifolds but also model building with $\Z_4$-orientifolds of Type IIA superstring theory. We are interested in global supersymmetric models with a semi-realistic chiral spectrum, in particular Pati-Salam-models as D6-brane realisations with only three visible stacks, which have on other lattices been the most simple kind of global GUT model to be found, see e.g.~\cite{Blumenhagen:2002gw,Honecker:2003vq,Cvetic:2003xs,Cvetic:2004ui,Chen:2006gd,Gmeiner:2007zz,Honecker:2012qr,Ecker:2014hma}.
The introduction of an anti-holomorphic involution $\mathcal{R}$ on the $\Z_4$-orbifold gives rise to orientifold six-planes (O6-planes), which wrap the fixed loci of $\mathcal{R}Q^k$ which together form some
element of $H_3(T^6/\Z_4,\Z)$. In the following, we denote this homological cycle by $\pi_\text{O6}$. The ($\Z_4$-orbits of) O6-planes have negative RR charge, which has to be canceled by introducing stacks of $N_a$ space-time filling D6-branes which wrap a three-cycle $\pi_a$ on the orbifold. The RR tadpole cancellation condition is given by~\cite{Blumenhagen:2002wn} 
\begin{equation}\label{tadpole}
\sum_aN_a(\pi_a+\pi_a^\prime)-4\pi_\text{O6}=0\,,
\end{equation}
where $\pi_a^\prime$ is the $\mathcal{R}$-image of the three-cycle $\pi_a$ with in general $\pi_a^{\prime} \neq \pi_a$. The resulting gauge group is then generically $\prod_aU(N_a)$. The case $\pi_a^\prime=\pi_a$ gives rise to the 
rank-preserving symmetry enhancement: $U(N_a)\,\hookrightarrow\,USp(2N_a)$ or $SO(2N_a)$.
We call a model with the RR tadpole condition~\eqref{tadpole} implemented {\it global}, otherwise {\it local}. \\
The chiral massless spectrum can be computed from topological intersection numbers~\cite{Blumenhagen:2002wn}. For the gauge group $\prod_aU(N_a)$ it is given in the table \ref{Tab:chiral}.

\begin{table}
\centering
\begin{tabular}{|c|c|}
\hline 
\muc{2}{|c|}{\text{\bf Chiral spectrum}}
\\\hline\hline
Representation& Multiplicity\\
\hline
\textbf{(Sym)}$_a$ & $ \frac{1}{2}(\pi_a\circ\pi_a^\prime-\pi_a\circ\pi_\text{O6})$ \\ 
\hline 
\textbf{(Anti)}$_a$ & $ \frac{1}{2}(\pi_a\circ\pi_a^\prime+\pi_a\circ\pi_\text{O6})$ \\ 
\hline 
$(\N_a,\,\bar{\N}_b)$ & $\pi_a\circ\pi_b$ \\ 
\hline 
$(\N_a,\,\N_b)$ & $\pi_a\circ\pi_b^\prime$\\ 
\hline 
\end{tabular} 
\caption{Chiral spectrum for intersecting D6-branes with gauge group $\prod_a U(N_a)$.}
\label{Tab:chiral}
\end{table}
Furthermore, the ($\Z_4$-orbits of the) O6-planes preserve $\mathcal{N}=1$. For semi-realistic models to be supersymmetric, we have to require that all D6-branes preserve the same
supersymmetry, i.e., that they are wrapped on {\it special Lagrangian} three-cycles with the same calibration as the O6-planes.
These additional constraints on the wrapped three-cycles will be considered in more details in the section~\ref{Sss:Susy}.

\subsection{Anti-holomorphic involutions $\mathcal{R}$ of $T^6/\Z_4$}\label{Ss:Orientifold}

Before we start with the construction of the ($\Z_4$-orbit of) O6-plane(s), we have to calculate how the anti-holomorphic involution acts on the real lattice in each case. 
It is known \cite{Dixon:1985jw, Dixon:1986jc} 
that there exist a set of complex coordinates $\{z^i\}_{i=1,2,3}$ on which the twist $Q$ $\eqref{2.coxeter_2}$ acts diagonally
\begin{equation}\label{2.twist} 
Q^tz^i=e^{2\pi i\zeta_i}z^i\;
\end{equation}
with the eigenvalues $(i,-1,i)$ for the shift vector $\vec{\zeta}=\frac{1}{4}(1,-2,1)$. In these coordinates, the anti-holomorphic involution (including $\Z_4$-twists)
 is simply given by complex conjugation
\begin{equation}\label{2.involution_def} 
\mathcal{R}(Q^t)^n: 
\;\;z^i\rightarrow e^{i\theta_{n_i}}\bar{z}^i
\end{equation}
for some real parameter $\theta_{n_i}$. We will now discuss the {\it a priori} different possible choices of $(\theta_{n_1},\theta_{n_2},\theta_{n_3})$ for each $\Z_4$-invariant background lattice and argue that
some of the different choices lead to physically equivalent vacua, at least based on the allowed ranks of gauge groups in the RR tadpole cancellation conditions
as well as on the counting of supersymmetric bulk cycles per given length.

\subsubsection{$B_2 \times (A_1)^2 \times B_2$}\label{Sss:B2A1-2B2-antihol}

In the factorisable case, there exist {\it a priori} two choices of orientifold axes per two-torus~\cite{Blumenhagen:1999md,Blumenhagen:1999ev,Forste:2000hx} denoted by {\bf A} (reflection along the short one-cycle, here $\pi_{2/6}$ or some $\Z_4$-images thereof, cf. figure~\ref{fig:Fig0a}) and {\bf B} (reflection along the long one-cycle, here $\pi_{1/5}$) for the $B_2$-tori and two lattice orientations {\bf a} (rectangular lattice) and {\bf b} (tilted lattice) for the $(A_1)^2$ torus.
Due to the permutation symmetry $T^2_1 \leftrightarrow T^2_3$, here the combinatorics provides at most six inequivalent choices of phases~\eqref{2.involution_def}, denoted by
 {\bf AaA}, {\bf AaB}, {\bf BaB} and {\bf AbA}, {\bf AbB}, {\bf BbB}, for which the O6-planes are displayed 
in table~\ref{Tab:B2A1-2B2-RRtcc}.

\begin{table}[th]
\bCentering
\resizebox{\linewidth}{!}{
$\begin{array}{|c||c|c|c|c|c|c|}\hline
\muc{4}{|c|}{\text{\bf O6-planes on the $B_2 \times (A_1)^2 \times B_2$ lattice}}
\\\hline\hline
\text{lattice} & {\bf Aa/bA} & {\bf Aa/bB} & {\bf Ba/bB}
\\\hline\hline
\mathcal{R} & 8(1-b)\pi_{236} & 4(1-b)\pi_{235} & 2(1-b)\pi_{135}
\\\hline
\mathcal{R}Q & 2(\pi_{1}+2\pi_2)\wedge (\pi_4-b\pi_3)\wedge(\pi_5+2\pi_6) & 4(\pi_1+2\pi_2)\wedge (b\pi_3-\pi_4)\wedge \pi_6 & 8\pi_2\wedge (\pi_4-b\pi_3)\wedge\pi_6
\\\hline
\mathcal{R}Q^2 & -8(1-b)(\pi_1+\pi_2)\wedge\pi_3\wedge(\pi_5+\pi_6) & 4(1-b)(\pi_1+\pi_2)\wedge\pi_3\wedge(\pi_5+2\pi_6) & -2(1-b)(\pi_1+2\pi_2)\wedge\pi_3\wedge(\pi_5+2\pi_6)
\\\hline
\mathcal{R}Q^3 & 2\pi_{1}\wedge (b\pi_3-\pi_4)\wedge\pi_5 & 4\pi_1\wedge (\pi_4-b\pi_3)\wedge (\pi_5+\pi_6) & 8(\pi_1+\pi_2)\wedge (b\pi_3-\pi_4)\wedge(\pi_5+\pi_6)
\\\hline
\end{array}$}
\caption{
Fixed planes for the different lattice orientations of $B_2 \times (A_1)^2 \times B_2$, weighted with the number $N_{O6}=2(1-b)$ of parallel O6-planes along $T^2_{2}$ with
$b =0, \frac{1}{2}$ for {\bf a} and {\bf b}, respectively.
 \label{Tab:B2A1-2B2-RRtcc}}
\end{table}
However, the massless closed string spectrum - encoded in the orientifold-even and -odd Hodge numbers $(h_{11}^+,h_{11}^-,h_{21})$ counting vectors, K\"ahler and complex structure moduli as derived for generic Calabi-Yau backgrounds in~\cite{Grimm:2004ua} -, 
which was derived using CFT techniques in~\cite{Blumenhagen:2002gw,Forste:2010gw} suggests that there are pairwise relations
${\bf AaA} \leftrightarrow {\bf BaB}$ and ${\bf AbA} \leftrightarrow {\bf BbB}$, reducing the number of physically inequivalent backgrounds to four. This assumption is supported 
by the analogy to the relations among different lattice backgrounds derived for $\Z_6^{(\prime)}$ and $\Z_2 \times \Z_6^{(\prime)}$ in~\cite{Gmeiner:2008xq,Honecker:2012qr,Ecker:2014hma}.

Since D6-branes on the factorisable $T^6/(\Z_4 \times \OR)$ orientifold have been considered at length in~\cite{Blumenhagen:2002gw} with the result that at most two generations
of chiral particles can be engineered by two supersymmetric intersecting D6-brane orbits, we will from now on concentrate on backgrounds with non-factorisable tori, where we will determine all {\it a priori} different choices of 
lattice orientations and then search for physical equivalences.

\subsubsection{$A_3 \times A_3$}\label{Sss:A3A3-antihol}

The eigenvectors of $Q^t$ with $Q$ defined in \eqref{3a} give rise to the complex coordinates \cite{Lust:2006zh}

\begin{equation}\label{A3A3compl_coord.}
\begin{split}
z^1:=&\frac{1}{\sqrt{2}}\left(x^1+ix^2-x^3\right)\,,\\
z^2:=&\frac{1}{\sqrt{8u_2}}\left((x^1-x^2+x^3+\mathcal{U}(x^4-x^5+x^6)\right)\,,\\
z^3:=&\frac{1}{\sqrt{2}}\left(x^4+ix^5-x^6\right) \, ,
\end{split}
\end{equation}
where the complex structure is $\mathcal{U}:=u_1+iu_2=-\frac{R_2}{2aR_1}\left(c+e+i\sqrt{-(c+e)^2+4ab}\right)$.

With the transformation from the complex coordinates to the real ones, we are able to write down the action of $\mathcal{R}$
 on the real lattice. There are {\it a priori} four possible choices of angles in the anti-holomorphic involution: 
\begin{equation}\label{2.r1} 
\vec{\theta}=(0,0,0)\Rightarrow\left\{\begin{array}{cll}
&\mathcal{R}_1e_1=e_1\,, &\mathcal{R}_1e_2=-e_1-e_2-e_3\,,\\
&\mathcal{R}_1e_3=e_3\,,&\mathcal{R}_1e_4=u_1(e_1+e_3)-e_6\,,\\
&\mathcal{R}_1e_5=-u_1(e_1+e_3)-e_5\,,&\mathcal{R}_1e_6=u_1(e_1+e_3)-e_4\,,
\end{array}\right\}\quad \textbf{AAA}
\end{equation}
\begin{equation}\label{2.r2} 
\vec{\theta}=(\frac{\pi}{2},0,0)\Rightarrow\left\{\begin{array}{cll}
&\mathcal{R}_2e_1=e_1+e_2+e_3\,,&\mathcal{R}_2e_2=-e_3\,,\\
&\mathcal{R}_2e_3=-e_2\,,&\mathcal{R}_1e_4=u_1(e_1+e_3)-e_6\,,\\
&\mathcal{R}_1e_5=-u_1(e_1+e_3)-e_5\,,&\mathcal{R}_1e_6=u_1(e_1+e_3)-e_4\,,
\end{array}\right\}\quad \textbf{BAA}
\end{equation}
\begin{equation}\label{2.r3} 
\vec{\theta}=(0,0,\frac{\pi}{2})\Rightarrow\left\{\begin{array}{cll}
&\mathcal{R}_3e_1=e_1\,,&\mathcal{R}_3e_2=-e_1-e_2-e_3\,,\\
&\mathcal{R}_3e_3=e_3\,,&\mathcal{R}_3e_4=u_1(e_1+e_3)+e_5\,,\\
&\mathcal{R}_3e_5=u_1(e_1+e_3)+e_4\,,&\mathcal{R}_3e_6=u_1(e_1+e_3)-e_4-e_5-e_6\,,
\end{array}\right\}\quad \textbf{AAB}
\end{equation}
\begin{equation}\label{2.r4} 
\vec{\theta}=(\frac{\pi}{2},0,\frac{\pi}{2})\Rightarrow\left\{\begin{array}{cll}
&\mathcal{R}_4e_1=e_1+e_2+e_3\,,&\mathcal{R}_4e_2=-e_3\,,\\
&\mathcal{R}_4e_3=-e_2\,,&\mathcal{R}_4e_4=u_1(e_1+e_3)+e_5\,,\\
&\mathcal{R}_4e_5=u_1(e_1+e_3)+e_4\,,&\mathcal{R}_4e_6=u_1(e_1+e_3)-e_4-e_5-e_6.
\end{array}\right\}\quad \textbf{BAB}
\end{equation}
Here we took the notation for the lattices from \cite{Blumenhagen:2004di}, which we will now explain.
The complex coordinates (\ref{A3A3compl_coord.}) parametrise the three-planes\footnote{These three-planes are closely related to the ``factorisation'' of the non-factorisable torus we will discuss in the next section.} where $Q$ acts as a rotation. {\bf A} means that, in their choice of basis, the orientifold plane lies along the horizontal axis in the corresponding plane, while {\bf B} corresponds to the angle of the orientifold plane with respect to the horizontal axis being $\pi/4$.

Note also that in contrast to the $B_2\times (A_1)^2\times B_2$-orbifold, the {\bf BAA}- and {\bf AAB}-lattices are really different geometrically and cannot be related by exchanging $T^3_1 \leftrightarrow T^3_2$. This will be further specified in the sections \ref{Sss:Relations} and \ref{S:Factorisation}.

Since only a crystallographic action on the lattice is allowed 
$(\mathcal{R}_ir=r+\Lambda$ for arbitrary lattice vectors $r\in\Lambda)$, the real part of the complex structure can take only the value $u_1=0$ ($\Rightarrow\;c+e=0$ in \eqref{3b} 
and only {\bf A} as middle entry of the lattice orientation). 
The orientifold projections act then on the homology 
classes of three-cycles \eqref{2.bulk_basis1} in the following way:
\begin{equation}\label{2.r1_homol} 
\mathcal{R}_1:\qquad
\gamma_1\leftrightarrow-\gamma_2\,,\qquad\bar{\gamma}_1\leftrightarrow\bar{\gamma}_2\,.
\end{equation}
\begin{equation}\label{2.r23_homol} 
\mathcal{R}_{2/3}:\qquad\begin{cases}
\gamma_1\rightarrow\gamma_1\,,\quad &\quad\bar{\gamma}_1\rightarrow-\bar{\gamma}_1\,,\\
\gamma_2\rightarrow-\gamma_2\,,\quad &\quad\bar{\gamma}_2\rightarrow\bar{\gamma}_2\,.
\end{cases}
\end{equation}
\begin{equation}\label{2.r4_homol} 
\mathcal{R}_4:\qquad
\gamma_1\leftrightarrow\gamma_2\,,\quad \quad\bar{\gamma}_1\leftrightarrow-\bar{\gamma}_2\,.
\end{equation}
Furthermore, we calculate the fixed point set for the orientifold involutions $ \{\mathcal{R}_i\,,\, \mathcal{R}_iQ\,,\,\mathcal{R}_iQ^2\,,$ $\mathcal{R}_iQ^3\,\}$.
The results are listed in tables \ref{tab:t1}-\ref{tab:t4} for the respective $\mathcal{R}_i$. 
\begin{table}[h!] 
\centering
\begin{tabular}{|c|c|c|}
\hline 
\muc{2}{|c|}{\bf O6-planes for AAA-lattice of $A_3\times A_3$-orbifold}\\
\hline\hline
Projection & Fixed point set \\ 
\hline 
$\mathcal{R}_1$ & $2(\pi_{134}-\pi_{136})$ \\ 
\hline 
$\mathcal{R}_1Q$ & $2(\pi_{245}+\pi_{246}-\pi_{256}+\pi_{345}+\pi_{346}-\pi_{356}$) \\ 
\hline 
$\mathcal{R}_1Q^2$ & $2(\pi_{124}+2\pi_{125}+\pi_{126}-\pi_{234}-2\pi_{235}-\pi_{236})$ \\ 
\hline 
$\mathcal{R}_1Q^3$ &$2(-\pi_{145}+\pi_{146}+\pi_{156}-\pi_{245}+\pi_{246}+\pi_{256})$ \\ 
\hline 
\end{tabular}
\caption{Toroidal cycles wrapped by the O6-planes on the \textbf{AAA} lattice orientation of $A_3 \times A_3$.}
\label{tab:t1}
\end{table}

The fixed point set under $\mathcal{R}_iQ^n$ is computed in the following way, e.g.\;for $\mathcal{R}_1$: First, we calculate the three eigenvectors of $\mathcal{R}_1$ corresponding to the eigenvalue 1, in this case $e_1\,,\,e_3$ and $e_4-e_6$. The to these vectors corresponding one-cycles $\pi_i$ span a three-cycle $\pi_{134}-\pi_{136}$. The next step is to determine how many three-cycles of the same homology class we have which are point-wise $\mathcal{R}_1$-invariant. In the case at hand, we obtain two such submanifolds, which go through the $\mathcal{R}_1$
 fixed points $(0,0,0,0,\frac{m}{2},0)$ with $m\in\{0,1\}$. The projection $\mathcal{R}_1Q$ leads to similar results. The fixed point sets of $\mathcal{R}Q^2$ and $\mathcal{R}Q^3$ can be calculated by acting with $Q$ on the $\mathcal{R}$- and $\mathcal{R}Q$-fixed point sets, respectively. After all fixed point sets have been computed, from purely geometric considerations there still remains an ambiguity in both the global sign of all sets and the relative sign between the $\mathcal{R}$- and $\mathcal{R}Q$-sets which needs to be fixed, i.e., the three-cycle $\pm(\text{Fix}(\mathcal{R}+\mathcal{R}Q^2)\pm \text{Fix}(\mathcal{R}Q+\mathcal{R}Q^3))$ is invariant for any choice of the signs. Each of the two sets of fixed points corresponds to a different bulk cycle. Asking both of them to have the same calibration will fix the relative sign between them. The global sign is fixed once we choose one of the two possible calibration conditions (either $\int (e^{i\varphi}\Omega_3) >0$ or $\int (e^{i\varphi+i\pi}\Omega_3) >0$); alternatively, choosing the global sign will fix the calibration condition that has to be used. This calibration will be explained in more detail in section \ref{Sss:Susy}.

\begin{table}[h!] 
\centering
\begin{tabular}{|c|c|c|}
\hline 
\muc{2}{|c|}{\bf O6-planes for BAA-lattice of $A_3\times A_3$-orbifold}\\
\hline\hline
Projection & Fixed point set \\ 
\hline 
$\mathcal{R}_2$ & $2(-\pi_{124}+\pi_{126}+\pi_{134}-\pi_{136}+\pi_{234}-\pi_{236})$ \\ 
\hline 
$\mathcal{R}_2Q$ & $2(-\pi_{145}-\pi_{146}+\pi_{156} +\pi_{345}+\pi_{346}-\pi_{356}$) \\ 
\hline 
$\mathcal{R}_2Q^2$ & $2(\pi_{124}+2\pi_{125}+\pi_{126}+\pi_{134}+2\pi_{135}+\pi_{136}
-\pi_{234}-2\pi_{235}-\pi_{236})$ \\ 
\hline 
$\mathcal{R}_2Q^3$ &$2(-\pi_{145}+\pi_{146}+\pi_{156}-2\pi_{245}+2\pi_{246}+2\pi_{256}-\pi_{345}+\pi_{346}+\pi_{356})$ \\ 
\hline 
\end{tabular}
\caption{Toroidal cycles wrapped by the O6-planes for the \textbf{BAA} lattice orientation of $A_3 \times A_3$.}
\label{tab:t2}
\end{table}

\begin{table}[h!] 
\centering
\begin{tabular}{|c|c|c|}
\hline 
\muc{2}{|c|}{\bf O6-planes for AAB-lattice of $A_3\times A_3$-orbifold}\\
\hline\hline
Projection & Fixed point set \\ 
\hline 
$\mathcal{R}_3$ & $2(\pi_{134}+\pi_{135})$ \\ 
\hline 
$\mathcal{R}_3Q$ & $2(\pi_{246}+\pi_{346}$) \\ 
\hline 
$\mathcal{R}_3Q^2$ & $2(\pi_{125}+\pi_{126}-\pi_{235}-\pi_{236})$ \\ 
\hline 
$\mathcal{R}_3Q^3$ &$2(-\pi_{145}+\pi_{156}-\pi_{245}+\pi_{256})$ \\ 
\hline 
\end{tabular}
\caption{Toroidal cycles wrapped by the O6-planes for the \textbf{AAB} lattice orientation of $A_3 \times A_3$.}
\label{tab:t3}
\end{table}

\begin{table}[h!] 
\centering
\begin{tabular}{|c|c|c|}
\hline 
\muc{2}{|c|}{\bf O6-planes for BAB-lattice of $A_3\times A_3$-orbifold}\\
\hline\hline
Projection & Fixed point set \\ 
\hline 
$\mathcal{R}_4$ & $2(-\pi_{124}-\pi_{125}+\pi_{134}+\pi_{135}+\pi_{234}+\pi_{235})$ \\ 
\hline 
$\mathcal{R}_4Q$ & $2(\pi_{146}-\pi_{346}$) \\ 
\hline 
$\mathcal{R}_4Q^2$ & $2(\pi_{125}+\pi_{126}+\pi_{135}+\pi_{136}
-\pi_{235}-\pi_{236})$ \\ 
\hline 
$\mathcal{R}_4Q^3$ &$2(\pi_{145}-\pi_{156}+2\pi_{245}-2\pi_{256}+\pi_{345}-\pi_{356})$ \\ 
\hline 
\end{tabular}
\caption{Toroidal cycles wrapped by the O6-planes for the \textbf{BAB} lattice orientation of $A_3 \times A_3$.}
\label{tab:t4}
\end{table}

Adding all contributions, we can express the corresponding O6-planes as elements of $H_3(T^6/\Z_{4},\Z)$:
\begin{subequations}\label{2.O6-planes}
\begin{align}
\pi_{\text{O}6_1}&:=2(\gamma_2-\gamma_1)-2(\bar{\gamma}_1+\bar{\gamma}_2)\,,\\
\pi_{\text{O}6_2}&:=-4\gamma_1-4\bar{\gamma}_2\,,\\
\pi_{\text{O}6_3}&:=-2\gamma_1-2\bar{\gamma}_2\,,\\
\pi_{\text{O}6_4}&:=2(\gamma_1+\gamma_2)+2(\bar{\gamma}_2-\bar{\gamma}_1)\,.
\end{align}
\end{subequations}

Obviously, the O6-planes \eqref{2.O6-planes} are invariant under the corresponding orientifold projections given in equations~\eqref{2.r1_homol}, \eqref{2.r23_homol} and~\eqref{2.r4_homol}. 

\subsubsection{$A_3 \times A_1\times B_2$}
We can approach the $A_3\times A_1\times B_2$-case in an analogous way. Due to \eqref{2.twist}, the complex coordinates are:
\begin{equation}\label{A3A1B2complex_coord} 
\begin{split}
z^1&=\frac{1}{\sqrt{2}}(x^1+ix^2-x^3)\,,\\
z^2&=\frac{1}{2\sqrt{2\text{Im}(\mathcal{U})}}(x^1-x^2+x^3+2\mathcal{U}x^4)\,,\\
z^3&=x^5-\frac{x^6}{2}+i\frac{x^6}{2} \, ,
\end{split}
\end{equation}
with the complex structure
\begin{equation}\label{2.complex_structure} 
\mathcal{U}:=u_1+iu_2:=-\frac{R_1}{2aR_3}(d+i\sqrt{-a-d^2})\,.
\end{equation}
For this non-factorisable lattice, there are four possible orientations:
\begin{equation}\label{2.R1} 
\vec{\theta}=(0,0,0)\Rightarrow\left\{\begin{array}{cll}
&\mathcal{R}_1e_1=e_1\,, &\mathcal{R}_1e_2=-e_1-e_2-e_3\,,\\
&\mathcal{R}_1e_3=e_3\,,&\mathcal{R}_1e_4=2u_1(e_1+e_3)-e_4\,,\\
&\mathcal{R}_1e_5=e_5\,,&\mathcal{R}_1e_6=-e_5-e_6\,,
\end{array}\right\}\quad \textbf{AAB}
\end{equation}
\begin{equation}\label{2.R2} 
\vec{\theta}=(0,0,-\frac{\pi}{2})\Rightarrow\left\{\begin{array}{cll}
&\mathcal{R}_2e_1=e_1\,,&\mathcal{R}_2e_2=-e_1-e_2-e_3\,,\\
&\mathcal{R}_2e_3=e_3\,,&\mathcal{R}_2e_4=2u_1(e_1+e_3)-e_4\,,\\
&\mathcal{R}_2e_5=-e_5-2e_6\,,&\mathcal{R}_2e_6=e_6\,,
\end{array}\right\}\quad \textbf{AAA}
\end{equation}
\begin{equation}\label{2.R3} 
\vec{\theta}=(-\frac{\pi}{2},0,-\frac{\pi}{2})\Rightarrow\left\{\begin{array}{cll}
&\mathcal{R}_3e_1=-e_2\,,&\mathcal{R}_3e_2=-e_1\,,\\
&\mathcal{R}_3e_3=e_1+e_2+e_3\,,&\mathcal{R}_3e_4=2u_1(e_1+e_3)-e_4\,,\\
&\mathcal{R}_3e_5=-e_5-2e_6\,,&\mathcal{R}_3e_6=e_6\,,
\end{array}\right\}\quad \textbf{ABA}
\end{equation}
and
\begin{equation}\label{2.R4} 
\vec{\theta}=(-\frac{\pi}{2},0,0)\Rightarrow\left\{\begin{array}{cll}
&\mathcal{R}_4e_1=-e_2\,,&\mathcal{R}_4e_2=-e_1\,,\\
&\mathcal{R}_4e_3=e_1+e_2+e_3\,,&\mathcal{R}_4e_4=2u_1(e_1+e_3)-e_4\,,\\
&\mathcal{R}_4e_5=e_5\,,&\mathcal{R}_4e_6=-e_5-e_6.
\end{array}\right\}\quad \textbf{ABB}
\end{equation}

Here we also took again the notation for the lattices from \cite{Blumenhagen:2004di}. Since only the crystallographic action on the lattice is allowed $(\mathcal{R}_ir=r+\Lambda$ for arbitrary lattice vectors $r\in\Lambda)$, the real part of the complex structure can take only two values $u_1=0,\frac{1}{2}$. We adapt the above lattice notation for different values of $u_1$ in the following way: We write a small \textbf{a} and \textbf{b} as a subscript after the first \textbf{A} for $u_1=0$ and $u_1=\frac{1}{2}$, respectively. The {\bf b}-type lattice is to our best knowledge investigated here for the first time, while the D6-brane configurations, which cancel the bulk RR tadpoles locally on top of the O6-planes, in~ \cite{Blumenhagen:2004di} correspond to the {\bf a}-type choice $u_1=0$.

\begin{figure}[h!]
 	\centering
 		\includegraphics[width=14cm]{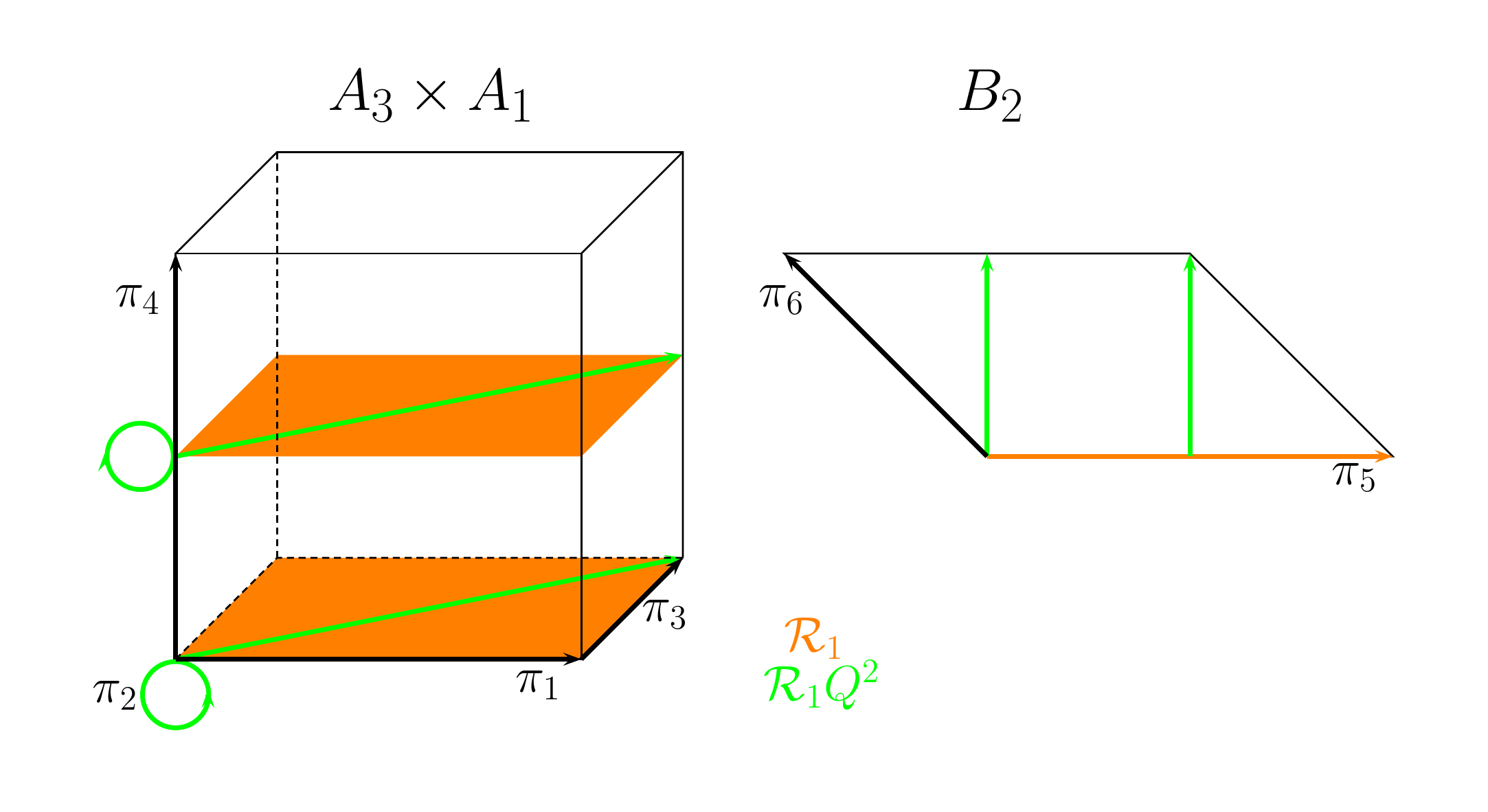}
 		\caption{$\mathcal{R}_1$- and $\mathcal{R}_1Q^2$-contributions to the O6-planes of the $A_3\times A_1\times B_2$-orientifold with \textbf{A${}_{\bf a}$AB}-lattice ($u_1=0$).}
 		\label{fig:Fig5}
 \end{figure}

\begin{figure}[h!]
 	\centering
 		\includegraphics[width=14cm]{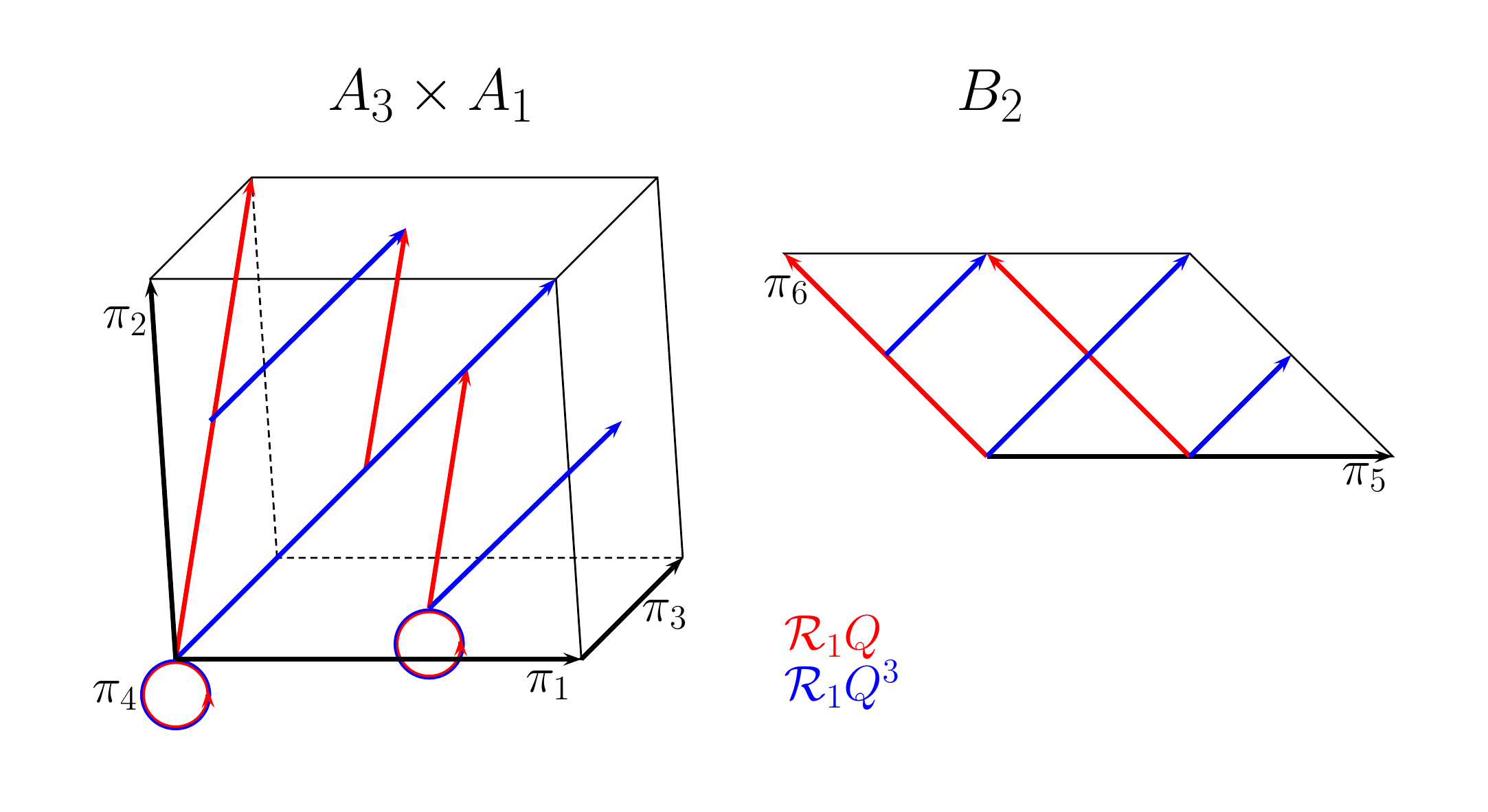}
 		\caption{$\mathcal{R}_1Q$- and $\mathcal{R}_1Q^3$-contributions to the O6-plane of the $A_3\times A_1\times B_2$-orientifold with \textbf{A${}_{\bf a}$AB}-lattice ($u_1=0$).}
 		\label{fig:Fig6}
 \end{figure}

\begin{figure}[h!]
 	\centering
 		\includegraphics[width=14cm]{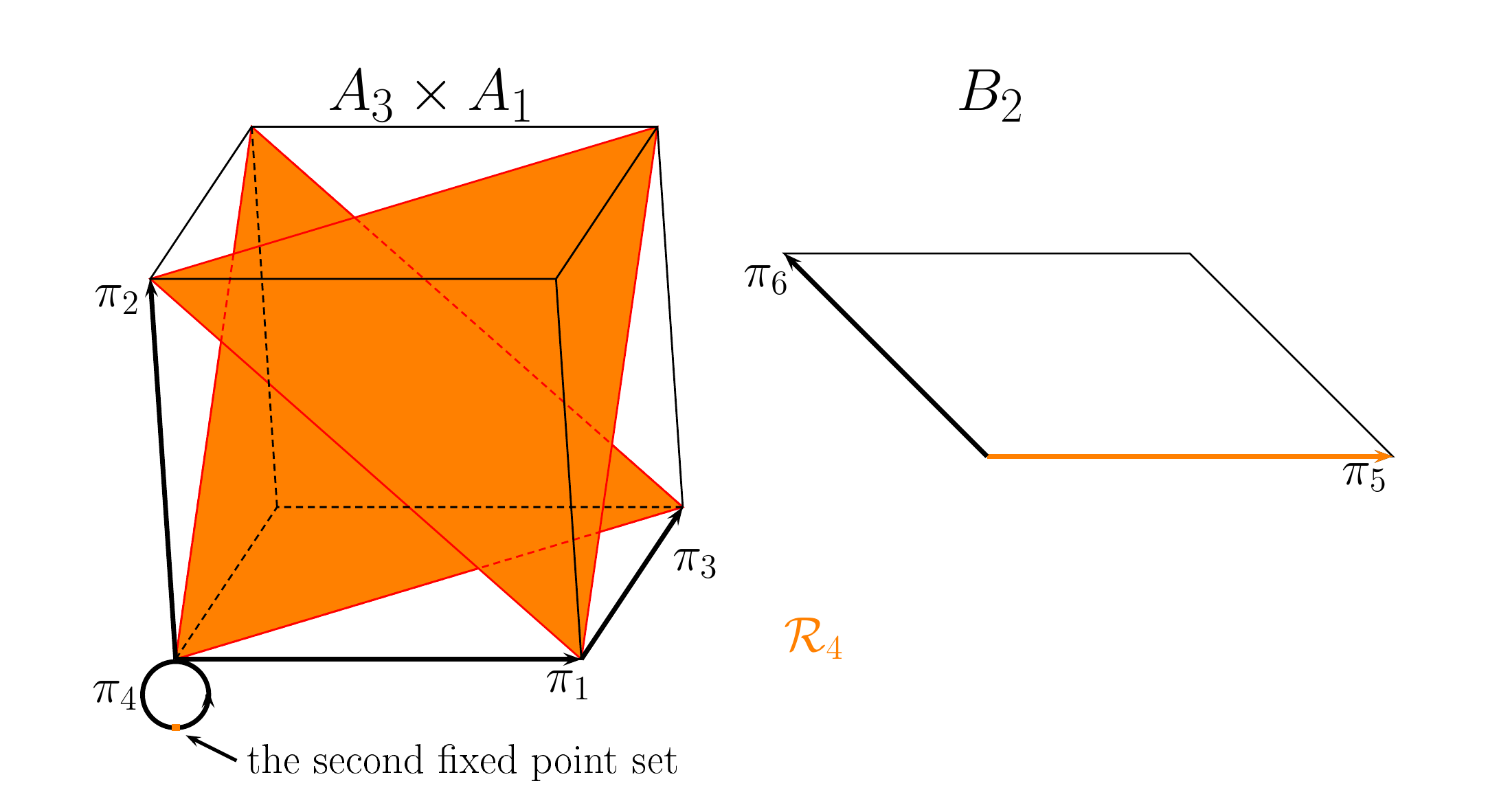}
 		\caption{$\mathcal{R}_4$-contribution to the O6-plane of the $A_3\times A_1\times B_2$-orientifold with \textbf{A${}_{\bf a}$BB}-lattice ($u_1=0$). The second fixed point set is located at $\pi_4=\frac{1}{2}$ and denoted by the orange point in the picture.}
 		\label{fig:Fig7}
 \end{figure}

\begin{figure}[h!]
 	\centering
 		\includegraphics[width=14cm]{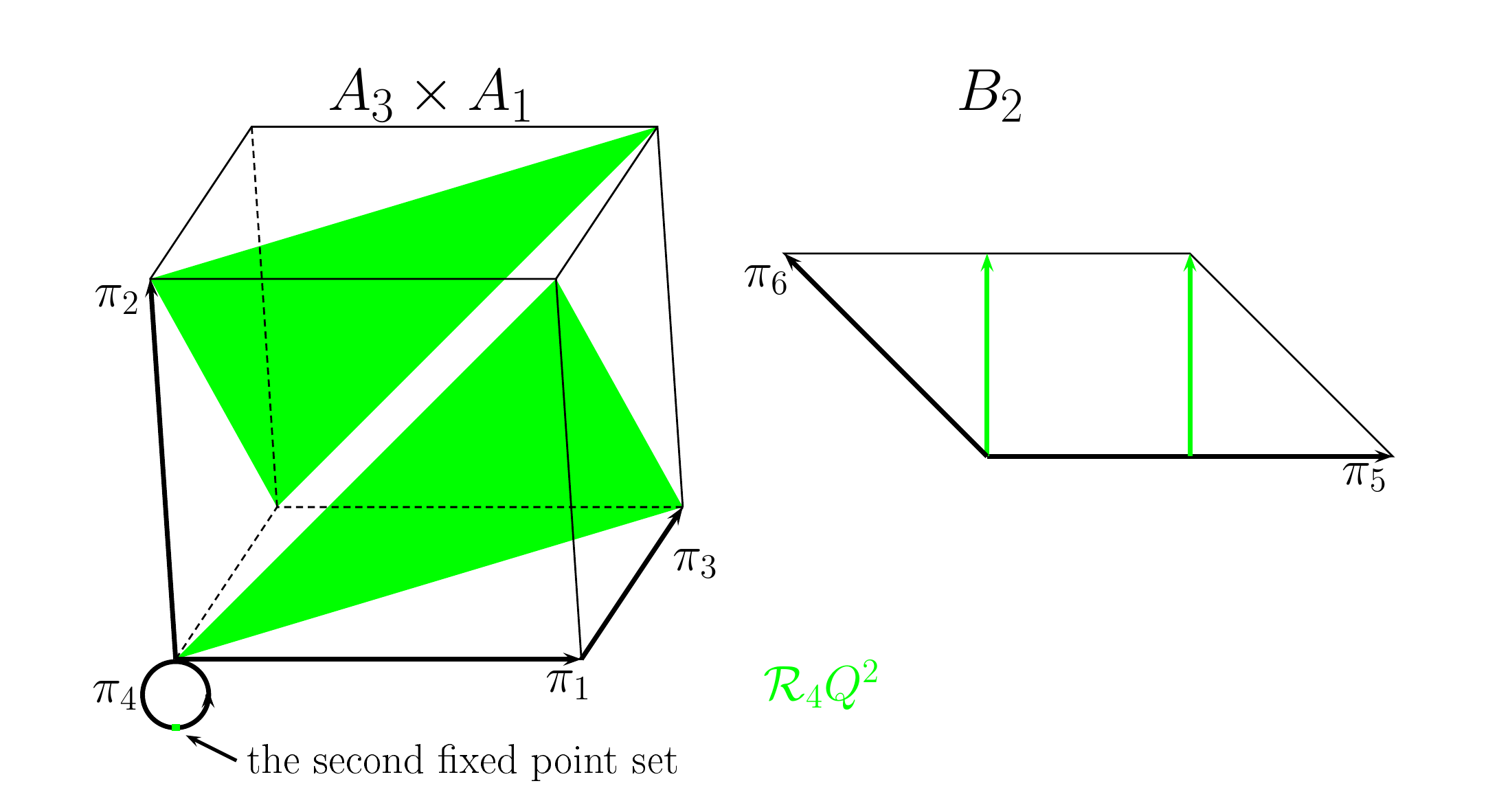}
 		\caption{$\mathcal{R}_4Q^2$-contributions to the O6-plane of the $A_3\times A_1\times B_2$-orientifold with \textbf{A${}_{\bf a}$BB}-lattice ($u_1=0$). The second fixed point set is located at $\pi_4=\frac{1}{2}$ and denoted by the green point in the picture.}
 		\label{fig:Fig8}
 \end{figure}

\begin{figure}[h!]
 	\centering
 		\includegraphics[width=14cm]{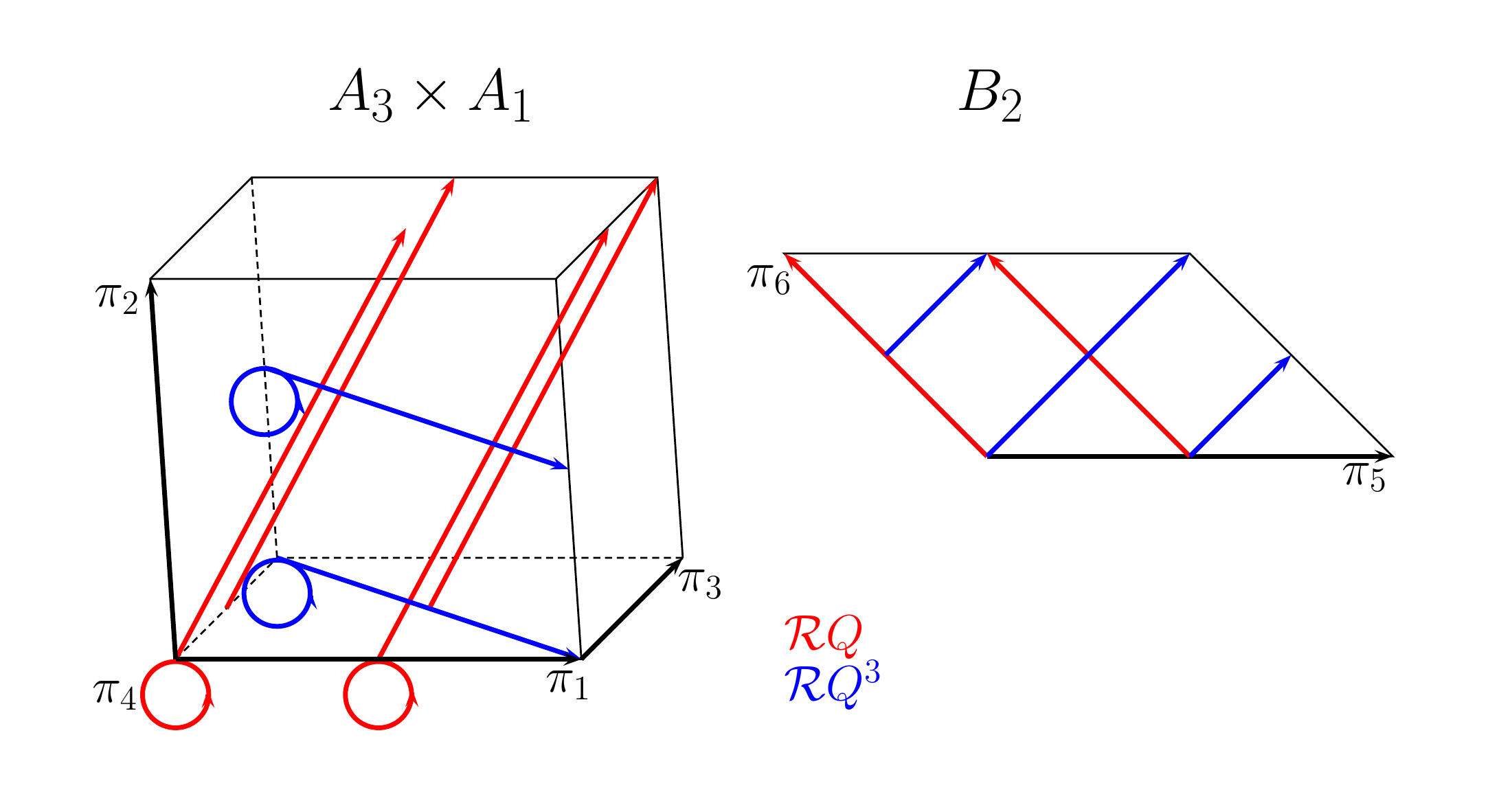}
 		\caption{$\mathcal{R}_4Q$- and $\mathcal{R}_4Q^3$-contributions to the O6-plane of the $A_3\times A_1\times B_2$-orientifold with \textbf{A${}_{\bf a}$BB}-lattice ($u_1=0$).}
 		\label{fig:Fig9}
 \end{figure}

The orientifold action on the homological three-cycles is given by 
\begin{equation}\label{2.R1_homol} 
\mathcal{R}_1:\qquad\begin{cases}
\gamma_1\rightarrow\gamma_2\,,\qquad\bar{\gamma}_1\rightarrow-\bar{\gamma}_2-2u_1\gamma_2\,,\\
\gamma_2\rightarrow\gamma_1\,,\qquad\bar{\gamma}_2\rightarrow-\bar{\gamma}_1-2u_1\gamma_1\,,\\
\gamma_3\rightarrow\gamma_3\,,\qquad\bar{\gamma}_3\rightarrow-\bar{\gamma}_3+2u_1\gamma_3\,,\\
\gamma_4\rightarrow\gamma_4\,,\qquad\bar{\gamma}_4\rightarrow-\bar{\gamma}_4+2u_1\gamma_4\,.
\end{cases}
\end{equation}
\begin{equation}\label{2.R2_homol} 
\mathcal{R}_2:\qquad\begin{cases}
\gamma_1\rightarrow\gamma_1\,,\quad &\quad\bar{\gamma}_1\rightarrow-\bar{\gamma}_1-2u_1\gamma_1\,,\\
\gamma_2\rightarrow-\gamma_2\,,\quad &\quad\bar{\gamma}_2\rightarrow\bar{\gamma}_2+2u_1\gamma_2\,,\\
\gamma_3\rightarrow-\gamma_3\,,\quad &\quad\bar{\gamma}_3\rightarrow\bar{\gamma}_3-2u_1\gamma_3\,,\\
\gamma_4\rightarrow-\gamma_4\,,\quad &\quad\bar{\gamma}_4\rightarrow\bar{\gamma}_4-2u_1\gamma_4\,.
\end{cases}
\end{equation}
\begin{equation}\label{2.R3_homol} 
\mathcal{R}_3:\qquad\begin{cases}
\gamma_1\rightarrow-\gamma_2\,,\qquad\bar{\gamma}_1\rightarrow\bar{\gamma}_2+2u_1\gamma_2\,,\\
\gamma_2\rightarrow-\gamma_1\,,\qquad\bar{\gamma}_2\rightarrow\bar{\gamma}_1+2u_1\gamma_1\,,\\
\gamma_3\rightarrow-\gamma_3\,,\qquad\bar{\gamma}_3\rightarrow\bar{\gamma}_3-2u_1\gamma_3\,,\\
\gamma_4\rightarrow-\gamma_4\,,\qquad\bar{\gamma}_4\rightarrow\bar{\gamma}_4-2u_1\gamma_4\,.
\end{cases}
\end{equation}

\begin{equation}\label{2.R4_homol} 
\mathcal{R}_4:\qquad\begin{cases}
\gamma_1\rightarrow\gamma_1\,,\quad &\quad\bar{\gamma}_1\rightarrow-\bar{\gamma}_1-2u_1\gamma_1\,,\\
\gamma_2\rightarrow-\gamma_2\,,\quad &\quad\bar{\gamma}_2\rightarrow\bar{\gamma}_2+2u_1\gamma_2\,,\\
\gamma_3\rightarrow\gamma_3\,,\quad &\quad\bar{\gamma}_3\rightarrow-\bar{\gamma}_3+2u_1\gamma_3\,,\\
\gamma_4\rightarrow\gamma_4\,,\quad &\quad\bar{\gamma}_4\rightarrow-\bar{\gamma}_4+2u_1\gamma_4\,.
\end{cases}
\end{equation}
Furthermore, we can calculate the fixed point sets for the orientifold projections $\{\mathcal{R}_i\,,\; \mathcal{R}_iQ\,,$ $\mathcal{R}_iQ^2\,,\;\mathcal{R}_iQ^3\,\}$. The results are listed in tables \ref{tab:T2}-\ref{tab:T4a} for the respective $\mathcal{R}_i$. In figures \ref{fig:Fig5}-\ref{fig:Fig9} we illustrate the O6-planes for the lattices \textbf{A}$_\textbf{a}$\textbf{AB} and \textbf{A}$_\textbf{a}$\textbf{BB}.\footnote{We will see in the next section that the other \textbf{a}-type lattices are related to these.}

\begin{table}[!h] 
\centering
\begin{tabular}{|c|c|c|}
\hline 
\muc{2}{|c|}{\bf O6-planes for AAB-lattice of $A_3\times A_1\times B_2$-orbifold}\\
\hline\hline
Projection & Fixed point set \\ 
\hline 
$\mathcal{R}_1$ & $2\pi_{135}$ \\ 
\hline 
$\mathcal{R}_1Q$ & $4(u_1(\pi_{126}+\pi_{136}-\pi_{236})+\pi_{246}+\pi_{346}$) \\ 
\hline 
$\mathcal{R}_1Q^2$ & $2\pi_{125}+4\pi_{126}-2\pi_{235}-4\pi_{236}$ \\ 
\hline 
$\mathcal{R}_1Q^3$ &$4(u_1(\pi_{135}+\pi_{136}-\pi_{125}-\pi_{126}+\pi_{235}+\pi_{236})-\pi_{145}-\pi_{146}-\pi_{245}-\pi_{246})$ \\ 
\hline 
\end{tabular}
\caption{Toroidal cycles wrapped by the O6-planes for the \textbf{AAB} orientation of $A_3 \times A_1 \times B_2$.}
\label{tab:T2}
\end{table}

\begin{table}[!h] 
\centering
\begin{tabular}{|c|c|c|}
\hline 
\muc{2}{|c|}{\bf O6-planes for AAA-lattice of $A_3\times A_1\times B_2$-orbifold}\\
\hline\hline
Projection & Fixed point set \\ 
\hline 
$\mathcal{R}_2$ & $4\pi_{136}$ \\ 
\hline 
$\mathcal{R}_2Q$ & $2(-u_1 (\pi_{125}+\pi_{135}-\pi_{235}+2(\pi_{126}+\pi_{136}-\pi_{236}))-\pi_{245}-\pi_{345}-2(\pi_{246}+\pi_{346})) $\\ 
\hline 
$\mathcal{R}_2Q^2$ & $4(-\pi_{125}-\pi_{126}+\pi_{235}+\pi_{236})$ \\ 
\hline 
$\mathcal{R}_2Q^3$ &$2(-u_1(\pi_{135}-\pi_{125}+\pi_{235})-\pi_{145}-\pi_{245})$ \\ 
\hline 
\end{tabular}
\caption{Toroidal cycles wrapped by the O6-planes for the \textbf{AAA} orientation of $A_3 \times A_1 \times B_2$.}
\label{tab:T3}
\end{table}
\begin{table}[!h] 
\centering
\begin{tabular}{|c|c|c|}
\hline 
\muc{2}{|c|}{\bf O6-planes for ABA-lattice of $A_3\times A_1\times B_2$-orbifold}\\
\hline\hline
Projection & Fixed point set \\ 
\hline 
$\mathcal{R}_3$ & $(4-4u_1)(\pi_{126}+\pi_{136}-\pi_{236})$ \\ 
\hline 
$\mathcal{R}_3Q$ & $2(-2u_1 (\pi_{125}-\pi_{235}+2\pi_{126}-2\pi_{236})-\pi_{145}-2\pi_{245}-\pi_{345}-2(\pi_{146}+2\pi_{246}+\pi_{346})) $\\ 
\hline 
$\mathcal{R}_3Q^2$ & $(4-4u_1)(-\pi_{125}-\pi_{126}+\pi_{135}+\pi_{136}+\pi_{235}+\pi_{236})$ \\ 
\hline 
$\mathcal{R}_3Q^3$ &$2(-2u_1\pi_{135}+\pi_{145}-\pi_{345})$ \\ 
\hline 
\end{tabular}
\caption{Toroidal cycles wrapped by the O6-planes for the \textbf{ABA} orientation of $A_3 \times A_1 \times B_2$.}
\label{tab:T4}
\end{table}
\begin{table}[!h] 
\centering
\begin{tabular}{|c|c|c|}
\hline 
\muc{2}{|c|}{\bf O6-planes for ABB-lattice of $A_3\times A_1\times B_2$-orbifold}\\
\hline\hline
Projection & Fixed point set \\ 
\hline 
$\mathcal{R}_4$ & $(2-2u_1)(\pi_{125}+\pi_{135}-\pi_{235})$ \\ 
\hline 
$\mathcal{R}_4Q$ & $4(2u_1 (\pi_{126}-\pi_{236})+\pi_{146}+2\pi_{246}+\pi_{346}) $\\ 
\hline 
$\mathcal{R}_4Q^2$ & $(2-2u_1)(\pi_{125}+2\pi_{126}-\pi_{135}-2\pi_{136}-\pi_{235}-2\pi_{236})$ \\ 
\hline 
$\mathcal{R}_4Q^3$ &$4(2u_1(\pi_{135}+\pi_{136})-\pi_{145}-\pi_{146}+\pi_{345}+\pi_{346})$ \\ 
\hline 
\end{tabular}
\caption{Toroidal cycles wrapped by the O6-planes for the \textbf{ABB} orientation of $A_3 \times A_1 \times B_2$.}
\label{tab:T4a}
\end{table} 

 
Adding up all contributions we obtain the corresponding O6-planes: 
\begin{subequations}\label{2.O6-Planes}
\begin{align}
\pi_{\text{O}6_1}&:=(1-2u_1)\gamma_1+(1+2u_1)\gamma_2-2(\bar{\gamma}_1-\bar{\gamma}_2)\,,\\
\pi_{\text{O}6_2}&:=-2\gamma_1-2u_1\gamma_2-2\bar{\gamma}_2\,,\\
\pi_{\text{O}6_3}&:=2(\gamma_2-\gamma_1)-4u_1\gamma_2-2(\bar{\gamma}_1+\bar{\gamma}_2)\,,\\
\pi_{\text{O}6_4}&:=(2-2u_1)\gamma_1+4u_1\gamma_2+4\bar{\gamma}_2\,.
\end{align}
\end{subequations}
It is easy to check that the resulting O6-planes are invariant under corresponding orientifold projections defined in equations~\eqref{2.R1_homol} to ~\eqref{2.R4_homol}.

\clearpage

\subsection{Supersymmetric cycles}\label{Sss:Susy}

From the phenomenological point of view (like chirality and stability at low energies) $\mathcal{N}=1$ supersymmetric models are particularly interesting. This requires the D6-branes to preserve some supersymmetry, which leads to additional geometrical conditions on the allowed three-cycles; namely, the three-cycles have to be special Lagrangian. A three-cycle $\pi$ is called special Lagrangian (\textit{sLag}) if it satisfies
\begin{subequations}\label{2.sLag} 
\begin{align}
J\mid_{\pi}=0\,,\\
\text{Im}(e^{i\varphi}\Omega_3)\mid_{\pi}=0\,,\\
\text{Re}(e^{i\varphi}\Omega_3)\mid_{\pi}>0\,,
\end{align}
\end{subequations}
where $\varphi$ is an arbitrary constant phase, and $J$ and $\Omega_3$ are the covariantly constant K\"ahler two-form and holomorphic three-form, respectively, that always exist on a Calabi-Yau threefold (and, in particular, on the $T^6/\mathbb{Z}_4$ orbifold we consider here). They can be defined locally by
\begin{equation}\label{2.Kähler} 
\Omega_3=dz^1\wedge dz^2\wedge dz^3,\quad J=i\sum_{i,\bar{j}}g_{i\bar{j}}dz^i\wedge d\bar{z}^{\bar{j}}.
\end{equation}
If two Lagrangian three-cycles are calibrated by different values of the phase $\varphi$, the corresponding D6-branes preserve different supersymmetries. Since the three-cycle wrapped by the O6-plane is automatically \textit{sLag}, the phase $\varphi$ is fixed to $\varphi=\varphi_\text{O6}$, and we need to search for the Lagrangian three-cycles which are calibrated by $\text{Re}(e^{i\varphi_\text{O6}}\Omega_3)$.

In terms of the $\vec{\theta}$, the angle $\varphi_\text{O6}$ is given by
\begin{equation}
\varphi_\text{O6}=-\frac{1}{2}\sum_i\theta_i\ (+\pi).
\end{equation}
The possible $+\pi$ in the previous equation arises from the freedom we have to choose either $\int (e^{i\varphi}\Omega_3) >0$ or $\int (e^{i\varphi+i\pi}\Omega_3) >0$ as the calibration condition, or, alternatively, the global sign of the three-cycle wrapped by the O6-plane.

$B_2 \times (A_1)^2 \times B_2$

Due to the factorisable structure of the $B_2 \times (A_1)^2 \times B_2$-lattice, any three-cycle with the wrapping numbers $(n^i,m^i)_{i=1,2,3}$ is automatically Lagrangian. The condition that a D6-brane wrapping such a three-cycle preserves the same supersymmetry as the O6-plane (i.e. is {\it sLag} with identical calibration) is 
\begin{equation}
\sum_i\phi_i=0\ \text{mod }2\pi
\end{equation}
where $\phi_i$ is an angle between the three-cycle and the O6-plane on the two-torus $T^2_{(i)}$. 
Fractional three-cycles with the same calibration arise when (fractions of) exceptional three-cycles are added through which a given bulk cycle passes with the one restriction on the relative prefactor discussed below equation~\eqref{Eq:factorisable-3-cycle}; for more details see \citep{Blumenhagen:2002gw}.

$A_3\times A_3$

With respect to the complex coordinates \eqref{A3A3compl_coord.}, 
the $Q$-invariant metric $g$ of equation~\eqref{3b} is given by
\begin{equation}\label{2.comlex_metric}
g_{i\bar{j}}=
\left(\begin{array}{ccc}
2R_1^2(1+a)&0&R_1R_2(e-c-i(c+2d+e))\\
0&-8u_2aR_1^2&0\\
R_1R_2(e-c+i(c+2d+e))&0&2R_2^2(1+b)\\
\end{array}\right) \, .
\end{equation}
The K\"ahler-form $J$ on the $A_3\times A_3$-lattice background is given by
\begin{equation}\label{2d}
J:=2R_1^2(1+a)\omega_1-2u_2aR_1^2\omega_2+2R_2^2(1+b)\omega_3+R_1R_2\left((c+2d+e)\omega_4+(c-e)\omega_5\right)
\end{equation}
with real $\mathbb{Z}_4$-invariant two-forms $\omega_i$:
\begin{equation}\label{2e}
\begin{split}
\omega_1&:=dx^{12}+dx^{23}\,,\\
\omega_2&:=dx^{14}-dx^{15}+dx^{16}-dx^{24}+dx^{25}-dx^{26}+dx^{34}-dx^{35}+dx^{36}\,,\\
\omega_3&:=dx^{45}+dx^{56}\,,\\
\omega_4&:=dx^{14}-dx^{16}+dx^{25}-dx^{34}+dx^{36}\,,\\
\omega_5&:=-dx^{15}+dx^{24}-dx^{26}+dx^{35}\,,
\end{split}
\end{equation}
where $dx^{ij} := dx^i\wedge dx^j$.

The K\"ahler two-form is negative under the orientifold projection, i.e. $\mathcal{R}J=-J$. This means that the $\mathcal{R}$-even part of $J$ has to vanish, which fixes part of the moduli, in addition to the fixing $u_1=0$ (i.e. $c+e=0$) mentioned in the previous subsection. The additional moduli fixing for the different orientifold projections is given in table \ref{tab:modulifixing}.

\begin{table}
\centering
\begin{tabular}{|c|c|c|c|c|}
\hline
\muc{5}{|c|}{\bf $A_3\times A_3$-orientifolds}\\
\hline\hline
Lattice & Involution & Invariant part of $J$ & Moduli fixing & $\varphi_{\text{O6}_i}$ \\ 
\hline 
\textbf{AAA} & $\mathcal{R}_1$ & $\omega_4$ & $d=0$ &0\\ 
\hline 
\textbf{BAA} & $\mathcal{R}_2$ & $\omega_4+\omega_5$ & $e=d$&$-\frac{\pi}{4}$ \\ 
\hline 
\textbf{AAB} & $\mathcal{R}_3$ & $\omega_4-\omega_5$ & $e=-d$&$-\frac{\pi}{4}$ \\ 
\hline
\textbf{BAB} & $\mathcal{R}_4$ & $\omega_4$& $d=0$&$\frac{\pi}{2}$ \\ 
\hline 
\end{tabular} 
\caption{Additional moduli fixing and calibration arising from the condition $\mathcal{R}J=-J$ for the different choices of $A_3 \times A_3$ lattice orientations.}
\label{tab:modulifixing}
\end{table}
While, in the factorisable case, i.e. $T^6=(T^2)^3$, all factorised three-cycles are Lagrangian, this does not hold true anymore on the orbifolds with some non-factorisable lattice. However, if we fix the moduli $c=d=e=0$, i.e. both $A_3$-tori are orthogonal to each other, and use the relation $dx^i(\pi_j)=\delta^i_j$, we can verify that any toroidal $\Z_2$-invariant three-cycle \eqref{2.frac.cond} is automatically Lagrangian.

The next step is to specify the {\it sLag} condition on our orientifolds $T^6/(\mathbbm{Z}_4\times\Omega\mathcal{R}_i)$ in order to find the supersymmetric three-cycles. The holomorphic three-form $\Omega_3=\text{Re}\Omega_3+i\text{Im}\Omega_3$ has the decomposition 
\begin{equation}\label{2.omega3_2} 
\begin{split}
\text{Re}(\Omega_3)&=\frac{1}{8\sqrt{u_2}}\bigl(\rho_1-u_2\rho_4\bigr)\,,\\
\text{Im}(\Omega_3)&=\frac{1}{8\sqrt{u_2}}\bigl(\rho_2+u_2\rho_3\bigr)\,,
\end{split}
\end{equation}
with real $\Z_4$-invariant three-forms
\begin{equation}\label{2.omega3_comp2} 
\begin{split}
\rho_1&:=-dx^{124}+dx^{125}+dx^{126}+2dx^{134}-2dx^{136}-dx^{234}-dx^{235}+dx^{236}\,,\\
\rho_2&:=-dx^{124}-dx^{125}+dx^{126}+2dx^{135}+dx^{234}-dx^{235}-dx^{236}\,,\\
\rho_3&:=dx^{145}-2dx^{146}+dx^{156}-dx^{245}+dx^{256}-dx^{345}+2dx^{346}-dx^{356}\,,\\
\rho_4&:=dx^{145}-dx^{156}+dx^{245}-2dx^{246}+dx^{256}-dx^{345}+dx^{356}\,.
\end{split}
\end{equation}
Because this case is not so interesting from the phenomenological point of view, we do not go into detail here and only give the supersymmetry parameters $\varphi_{\text{O6}_i}$ in table \ref{tab:modulifixing} for each involution. 

$A_3\times A_1\times B_2$

With respect to the complex coordinates $\eqref{A3A1B2complex_coord}$, the $A_3\times A_1\times B_2$-orbifold has the hermitian metric
 \begin{equation}\label{2.comlex_metric2}
g_{i\bar{j}}=
\left(\begin{array}{ccc}
2R_3^2(1+a)&0&\frac{R_2R_3}{\sqrt{2}}(b+i(b+2c))\\
0&-8u_2aR_3^2&0\\
\frac{R_2R_3}{\sqrt{2}}(b-i(b+2c))&0&2R_2^2\\
\end{array}\right)\,,
\end{equation}
which gives rise to the K\"ahler-form 
\begin{equation}\label{c6} 
J=2R_3^2(1+a)\omega_1+2R_2^2\omega_2-4R_3^2au_2\omega_3-R_3R_2(b+2c)\omega_4-R_3R_2b\omega_5
\end{equation}
with the following $\mathbb{Z}_4$-invariant two-forms:
\begin{equation}\label{c7} 
\begin{split}
\omega_1&:=dx^{12}+dx^{23}\,,\\
\omega_2&:=dx^{56}\,,\\
\omega_3&:=dx^{14}-dx^{24}+dx^{34}\,,\\
\omega_4&:=2dx^{15}-dx^{16}+dx^{26}-2dx^{35}+dx^{36}\,,\\
\omega_5&:=-dx^{16}+2dx^{25}-dx^{26}+dx^{36}\,.
\end{split}
\end{equation}
The anti-symmetry of the orientifold projection on $J$ leads here again to the fixing of some moduli, as shown in the table \ref{tab:T5}.
\begin{table}[!h] 
\centering
\begin{tabular}{|c|c|c|c|c|}
\hline 
\muc{5}{|c|}{\bf $A_3\times A_1\times B_2$-orientifolds}\\
\hline\hline
Lattice & Involution & Invariant part of $J$ & Moduli fixing & $\varphi_{\text{O}6_i}$ \\ 
\hline 
\textbf{AAB} & $\mathcal{R}_1$ & $\omega_4$ & $b=-2c$& 0 \\ 
\hline 
\textbf{AAA} & $\mathcal{R}_2$ & $\omega_4+\omega_5$ & $b=-c$ & $\frac{5\pi}{4}$ \\ 
\hline 
\textbf{ABA} & $\mathcal{R}_3$ & $\omega_4$ & $b=-2c$& $\frac{3\pi}{2}$ \\ 
\hline 
\textbf{ABB} & $\mathcal{R}_4$ & $\omega_4-\omega_5$ & $c=0$& $\frac{\pi}{4}$ \\ 
\hline
\end{tabular} 
\caption{Moduli fixing due to $\mathcal{R}J=-J$ and calibration for different choices of the anti-holomorphic involution on the $A_3 \times A_1 \times B_2$ lattice.}
\label{tab:T5}
\end{table}

As in the $A_3\times A_3$-case, we verify that any toroidal $\Z_2$-invariant three-cycle \eqref{2.fractional_condition} is automatically Lagrangian for the choice of fixing the moduli to $b=c=0$.

Using the complex coordinates $\eqref{A3A1B2complex_coord}$, we can write $\Omega_3$ in real coordinates
\begin{equation}\label{2.omega3} 
\begin{split}
\text{Re}(\Omega_3)&=\frac{1}{4\sqrt{u_2}}\bigl(\rho_1+u_1\rho_3-u_2\rho_4\bigr)\,,\\
\text{Im}(\Omega_3)&=\frac{1}{4\sqrt{u_2}}\bigl(\rho_2+u_2\rho_3+u_1\rho_4\bigr)\,,
\end{split}
\end{equation}
where the $\mathbb{Z}_4$-invariant three-forms are:
\begin{equation}\label{2.omega3_comp} 
\begin{split}
\rho_1&:=-dx^{125}+dx^{126}-dx^{136}+2dx^{135}-dx^{235}\,,\\
\rho_2&:=-dx^{125}+dx^{136}+dx^{235}-dx^{236}\,,\\
\rho_3&:=2dx^{145}-dx^{146}-dx^{246}-2dx^{345}+dx^{346}\,,\\
\rho_4&:=dx^{146}+2dx^{245}-dx^{246}-dx^{346}\,.
\end{split}
\end{equation}

The next step is to find the \textit{sLag}s on the orientifolds $T^6/(\mathbbm{Z}_4\times\Omega\mathcal{R}_i)$ with $A_3 \times A_1 \times B_2$ lattice.
The {\it sLag} condition 
for any three-cycle \eqref{2.torus_cycle} inherited from the torus can be expressed as follows,
\begin{subequations}\label{2.sLags_torus}
\begin{align}
&\text{Im}(\Omega_3)\mid_{\pi^{\text{torus}}}=\frac{1}{4\sqrt{u_2}} ([Q-P] +u_1 [\bar{P}-\bar{Q}] -u_2 [\bar{P}+\bar{Q}] )\, ,\\ 
&\text{Re}(\Omega_3)\mid_{\pi^{\text{torus}}}=\frac{1}{4\sqrt{u_2}} ([P+Q] - u_1 [\bar{P}+\bar{Q}] - u_2 [\bar{P}-\bar{Q}] )\, ,\ 
\end{align}
\end{subequations} 
where $P,\,Q,\,\bar{P},\,\bar{Q}$ are the bulk wrapping numbers defined in~\eqref{2.bulk_decomposition}. 
Using this decomposition of a (fraction of a) bulk three-cycle into real and imaginary part in dependence of the complex structure $u$
and the bulk wrapping numbers for the corresponding O6-planes we can calculate the supersymmetry parameters $\varphi_{\text{O}6_i}$ per lattice orientation displayed in table \ref{tab:T5}.

\subsection{Pairwise relations between choices of orientifold axes}\label{Sss:Relations}

We know that in the factorisable case {\it a priori} six choices of the anti-holomorphic involution are possible, but that there are pairwise relations between them so that only four are physically inequivalent. 
This means that different but equivalent orientifold projections give rise to the same global semi-realistic particle models as discussed in section~\ref{Sss:B2A1-2B2-antihol}. \\
The same observation can be made in the non-factorisable cases. 
In order to find these relations between the lattices, we compute all supersymmetric three-cycles which 
 do not overshoot the bulk RR tadpole cancellation condition in~\eqref{tadpole}. Together with the supersymmetry conditions~\eqref{2.sLag} this gives rise to the restriction on the toroidal and corresponding bulk wrapping numbers, e.g. for the \textbf{A}$_\textbf{a}$\textbf{AB}-lattice of the $A_3\times A_1\times B_2$ lattice, they are bounded by 
\begin{equation}\label{bulk_tadpole}
\begin{array}{cl}
A_1n^3-A_2n^3+2A_2m^3&\leqslant 8\,,\\
2B_3n^3-2B_2m^3&\leqslant 16\,,
\end{array}
\quad\Longleftrightarrow\quad
\begin{array}{rl}
P+Q&\leqslant 8\,,\\
\bar{Q}-\bar{P}&\leqslant16.
\end{array}
\end{equation}
The restrictions on the other lattice orientations take a similar form. The required O6-plane bulk wrapping numbers entering~\eqref{tadpole} are given in \eqref{2.O6-planes} for the $A_3\times A_3$ lattice, and in \eqref{2.O6-Planes} for the $A_3\times A_1\times B_2$ lattice.

$A_3\times A_3$ 

For this orbifold we found {\it a priori} four possible involutions. 
However, due to the relation $\mathcal{R}_4Q^3=-\mathcal{R}_1$ between the involutions $\mathcal{R}_1$ and $\mathcal{R}_4$ the \textbf{AAA}- and \textbf{BAB}-lattices 
give physically identical models on the $A_3\times A_3$-orientifold. 
But the relation between the corresponding complex structure values cannot be verified more precisely at this point. There are two possibilities
\begin{equation}\label{2.dualities2}
\text{or}\;\begin{array}{cc}
&\text{\textbf{AAA} dual to \textbf{BAB} and }\; u_2=u_2^\prime\,, \\
&\text{\textbf{AAA} dual to \textbf{BAB} and }\;u_2=\frac{1}{u_2^\prime}\,.
\end{array}
\end{equation}
Moreover, the orientifold projections $\mathcal{R}_2$ and $\mathcal{R}_3$ - corresponding to the {\bf BAA} and {\bf AAB} lattice orientation, respectively, according to table~\ref{tab:modulifixing} -
act on the three-cycles in the same way, and therefore the supersymmetric three-cycles on the corresponding orientifolds are the same. At first sight one might be tempted to identify these orientifolds, but the distinction of the length of the O6-planes, $\pi_{\text{O6}_2}=2\pi_{\text{O6}_3}$, gives in principle rise to different allowed ranks and lengths of bulk cycles in the 
RR tadpole cancellation conditions and consequently to more possible models for the \textbf{BAA}-orientifold.

In conclusion, by investigating the structure of bulk three-cycles and their RR tadpole cancellation conditions, we arrive at three physically inequivalent $A_3 \times A_3$-lattice orientations 
{\bf AAA}, {\bf BAA} and {\bf AAB}.

$A_3\times A_1\times B_2$

In this case there are {\it a priori} eight possible lattice orientations. 
The number of the corresponding fractional cycles not overshooting the bulk RR tadpole cancellation conditions 
 and the number of possible complex structure values $u_2$ are presented in the table \ref{tab:T6}.
\begin{table}[!h] 
\centering
\begin{tabular}{|c|c|c|c|}
\hline 
\muc{4}{|c|}{\bf $A_3\times A_1\times B_2$-orientifolds}\\
\hline\hline
Lattice & Orien. proj. & \# of frac. cycles & \# of $u_2$\\ 
\hline
\textbf{A}$_\textbf{a}$\textbf{AA} & $\mathcal{R}_2$ $(u_1=0)$ & 2126 & 96\\ 
\hline 
\textbf{A}$_\textbf{a}$\textbf{AB} & $\mathcal{R}_1$ $(u_1=0)$ & 2126 & 96 \\ 
\hline 
\textbf{A}$_\textbf{a}$\textbf{BA} & $\mathcal{R}_3$ $(u_1=0)$ & 5134 & 210\\ 
\hline 
\textbf{A}$_\textbf{a}$\textbf{BB} & $\mathcal{R}_4$ $(u_1=0)$ & 5134 & 210\\ 
\hline\hline
\textbf{A}$_\textbf{b}$\textbf{AA} & $\mathcal{R}_2$ $(u_1=\frac{1}{2})$ & 2410 & 118\\ 
\hline 
\textbf{A}$_\textbf{b}$\textbf{AB} & $\mathcal{R}_1$ $(u_1=\frac{1}{2})$ & 3646 & 140 \\ 
\hline 
\textbf{A}$_\textbf{b}$\textbf{BA} & $\mathcal{R}_3$ $(u_1=\frac{1}{2})$ & 3646 & 140\\ 
\hline 
\textbf{A}$_\textbf{b}$\textbf{BB} & $\mathcal{R}_4$ $(u_1=\frac{1}{2})$ & 2410 & 118\\ 
\hline
\end{tabular} 
\caption{The number of supersymmetric fractional cycles bounded by the bulk RR tadpole cancellation condition and the number of possible complex structure values $u_2$ for different choices of orientifold axes.}
\label{tab:T6}
\end{table}

Furthermore, we can easily verify the following pairwise relations between the different lattice orientations:
\begin{equation}\label{2.dualities}
\begin{split}
\text{\textbf{A}$_\textbf{a}$\textbf{AA} dual to \textbf{A}$_\textbf{a}$\textbf{AB} and }\; u_2=\frac{1}{2u_2^\prime}\,, \\
\text{\textbf{A}$_\textbf{a}$\textbf{BA} dual to \textbf{A}$_\textbf{a}$\textbf{BB} and }\; u_2=\frac{1}{2u_2^\prime}\,, \\
\text{\textbf{A}$_\textbf{b}$\textbf{AA} dual to \textbf{A}$_\textbf{b}$\textbf{BB} and }\; u_2=\frac{1}{4u_2^\prime}\,, \\
\text{\textbf{A}$_\textbf{b}$\textbf{BA} dual to \textbf{A}$_\textbf{b}$\textbf{AB} and }\; u_2=\frac{1}{4u_2^\prime}\,.
\end{split}
\end{equation}

Altogether, we have thus four physically inequivalent lattices (two with $u_1=0$ and two with $u_1=\frac{1}{2}$) and can restrict our further considerations to the lattice orientations \textbf{AAB} and \textbf{ABB}.

\subsection{Cross-check: D6-branes on top of O6-planes}\label{Sss:Crosscheck}
Non-factorisable $T^6/\mathbb{Z}_4$ orbifolds have been briefly studied in the past. For instance, in \cite{Blumenhagen:2004di} only D6-branes on top of the orientifold planes were considered, and only for a particular choice of the moduli in (\ref{3b}) and (\ref{1b}), i.e., the radii of all tori being equal and all the tori being orthogonal to each other. Our goal in this section is twofold. At first, we will reproduce and extend the CFT methods and results of \cite{Blumenhagen:2004di} to arbitrary values of the moduli, in particular providing the Kaluza-Klein and winding modes for generic D6-brane configuration. Secondly, we will compare this result with the geometric method described earlier on, i.e. by requiring that the 
{\it sLag} three-cycles wrapped by generic D6-branes satisfy the RR tapdole cancellation condition~(\ref{tadpole}) - in the case at hand with only two stacks of D6-branes. 

Let us start by reviewing the method used in \cite{Blumenhagen:2004di}. We denote a basis of the torus lattice by $\{e_i\}$. Let $\{e_i^*\}$ be a basis of the dual lattice such that $e_i\cdot e_j^*=\delta_{ij}$. If the lattice vectors $e_i$ form the $Q$-invariant metric $g_{ij}$, the dual vectors $\{e_i^*\}$ form the metric $g^*_{ij}=e_i^*\cdot e_j^*=g_{ij}^{-1}$. Note that the dual vectors transform under $Q^t$ and $\mathcal{R}_{m\;(m=1,2,3,4)}^t$.

\noindent In general, insertions of $\Omega\mathcal{R}_m Q^{2k}$ and $\Omega\mathcal{R}_m Q^{2k+1}$ in the Klein bottle trace (and strings starting on the $6_{2k}$ and $6_{2k+1}$ branes in the annulus and M\"obius strip) give different lattice contributions, so we need to compute both cases separately. In the rest of this section we will only consider the first case, but the second one is obtained analogously, by replacing $\mathcal{R}_m$ by $\mathcal{R}_mQ$ throughout the elaboration.

\noindent Let $\textbf{v}_i$ ($i=1,2,3$) be the lattice vectors that span the (fraction of the) bulk three-cycle wrapped by the O6-plane. They satisfy $\mathcal{R}_m\textbf{v}_i=\textbf{v}_i$. 
The winding modes are described by vectors $\textbf{w}_i$, $i=1,2,3$, satisfying $\mathcal{R}_m\textbf{w}_i=-\textbf{w}_i$. The momentum modes $\textbf{p}_i$ appearing in the Klein bottle amplitude in sectors where there are fixed tori correspond to vectors in the dual lattice invariant under $\mathcal{R}_m$. Using all these vectors we define the matrices

\begin{eqnarray}
 (M_{KB})_{ij}&:=&\textbf{p}_i\cdot \textbf{p}_j,\\
 (M_{A})_{ij}=(M_{MS})_{ij}&:=& \textbf{v}_i\cdot \textbf{v}_j,\\
 (W_{KB})_{ij}=(W_{MS})_{ij}&:=& \textbf{w}_i\cdot \textbf{w}_j.
\end{eqnarray}

\noindent The lattice mode contributions (in the corresponding sector) for the Klein bottle, annulus, and M\"obius strip are
\begin{eqnarray}
 KB&=&\frac{4^n}{(\det M_{KB}\det W_{KB})^{1/2}},\\
 A&=&\frac{\det M_A}{(\det g)^{1/2}},\\
 MS&=&\frac{4^n(\det M_{MS})^{1/2}}{(\det W_{MS})^{1/2}},
\end{eqnarray}
where $n=\dim (M)=\dim(W)$, ($n=3$ in the untwisted sector, which is the one we are interested in).

\noindent The tadpole cancellation condition is then given by
\begin{equation}
 KB+\frac{M^2\cdot A}{16}-\frac{M\cdot MS}{16}=0
\end{equation}
where $M$ is the number of identical branes (in the $\mathcal{R}_mQ$ case we will denote it by $N$).

$A_3\times A_3$

Let us start with the $A_3\times A_3$ lattice. For concreteness, we will focus on the orientation \textbf{AAB}. We will show that the number of supersymmetric D6-branes needed to cancel the 
(bulk) RR tadpole depends on the angle-moduli between the two $A_3$-tori.

Recall that for the orientifold \textbf{AAB} we have the involution $\mathcal{R}_3$ \eqref{2.r3}, which fixes the moduli to $c+e=0$ (as consequence of $u_1=0$) and $e+d=0$ (see table \ref{tab:modulifixing}). Using these constraints and the $Q$-invariant metric $g$, we obtain the following momentum modes \textbf{p}$_i$:
\begin{eqnarray}
\textbf{p}_1&=&2e_1^*-e_2^*,\\
 \textbf{p}_2&=&e_1^*-e_3^*,\\
 \textbf{p}_3&=&e_4^*+e_5^*-e_6^*.
\end{eqnarray}
The winding modes are:
\begin{eqnarray}
\textbf{w}_1&=&e_1+2e_2+e_3,\\
\textbf{w}_2&=&e_4-e_5,\\
\textbf{w}_3&=&e_4+e_6.
\end{eqnarray}
The $\textbf{v}_i$ vectors spanning the O6-planes are:
\begin{eqnarray}
\textbf{v}_1&=&e_1,\\
\textbf{v}_2&=&e_3,\\
\textbf{v}_3&=&e_4+e_5.
\end{eqnarray}

The determinants of the corresponding matrices $M_{KB}$, $M_{A}$ and $W_{KB}$ are
\begin{equation}
\begin{split}
\det M_{KB}&=-\frac{2}{a (1+a+b+ab-2e^2) R_1^4R_2^2}\,,\\
\det M_{A}&=-8 a (1+a+b+ab-2e^2) R_1^4R_2^2\,,\\
\det W_{KB}&=-32b(1+a+b+ab-2e^2) R_1^2R_2^4\,.
\end{split}
\end{equation}
 They give rise to the lattice mode contributions to the untwisted sector 
\begin{equation}
 KB=8\sqrt{\frac{a}{b}}\frac{R_1}{R_2},
 \quad A=2\sqrt{\frac{a}{b}}\frac{R_1}{R_2},\quad MS=32\sqrt{\frac{a}{b}}\frac{R_1}{R_2}.
\end{equation}
The untwisted RR tadpole cancellation condition is
\begin{equation}
 KB+\frac{M^2\cdot A}{16}-\frac{M\cdot MS}{16}=0\Rightarrow \left(M-8\right)^2=0.
\end{equation}
The $\mathcal{R}_3Q$-case leads to the analogous result $(N-8)^2=0$. Due to the action of the $\Z_2$ symmetry on the corresponding Chan-Paton factors, the gauge group will have rank 
$\frac{M}{2} \times \frac{N}{2}$, and the gauge group for generic values of moduli is $U(4)\times U(4)$ (with a possible gauge symmetry enhancement $U(4) \hookrightarrow USp/SO(8)$\footnote{This is a shorthand notation for either $USp(8)$ or $SO(8)$.}, for special choices of 
geometric moduli - in particular for the choice $c=d=e=0$ of the two $A_3$ lattices orthogonal to each other in~\eqref{3b} -, 
where determining the appropriate type of symmetry enhancement requires the development of CFT techniques at one-loop so far only available for $(T^2)^3$ factorisable backgrounds, see e.g.~\cite{Lust:2003ky,Akerblom:2007np,Blumenhagen:2007ip,Gmeiner:2009fb,Honecker:2011sm}).

The complementary purely geometric considerations are as follows:
The cycle wrapped by the O6-plane is $-2\gamma_1-2\bar{\gamma}_2$. For $e=0$, its contribution to the tadpole can be cancelled by a stack of $N_1=4$ branes wrapping the cycle $-\gamma_1$ and a second stack of $N_2=4$ branes wrapping $-\bar{\gamma}_2$, giving rise to a gauge group $USp/SO(8)\times USp/SO(8)$ in agreement with the CFT result. 

The results for the remaining orientifold projections are summarised in the table \ref{tab:branes}.
\begin{table}[h!]
\centering
\begin{tabular}{|c|c||c|c|c|}
\hline
\multicolumn{5}{|c|}{{\bf $A_3\times A_3$ orientifolds with D-branes on top of the O-plane}}\\
\hline
\hline 
\multicolumn{2}{|c||}{$A_3\times A_3$} & General & \multicolumn{2}{|c|}{$e=0$} \\ 
\hline 
Lattice & Involution & $M=N$ & $M=N$ & Gauge group \\ 
\hline \hline
\textbf{AAA}& $\mathcal{R}_1$& 8 & 8 & $USp/SO(8)\times USp/SO(8)$ \\ 
\hline 
\textbf{BAA}& $\mathcal{R}_2$&8 & 8 & $USp/SO(8)\times USp/SO(8)$ \\ 
\hline 
\textbf{AAB}& $\mathcal{R}_3$&8 & 8 & $USp/SO(8)\times USp/SO(8)$ \\ 
\hline 
\textbf{BAB}& $\mathcal{R}_4$&8 & 8 & $USp/SO(8)\times USp/SO(8)$ \\ 
\hline 
\end{tabular} 
\caption{IIA Orientifolds of $\Z_4$-orbifolds with the lattice of the type $A_3\times A_3$ with D-branes on O-plane. For generic values of the metric moduli, the gauge group is $U(\frac{M}{2}) \times U(\frac{N}{2})$.}
\label{tab:branes}
\end{table}

\vspace{2mm}

$A_3\times A_1 \times B_2$

Let us now consider the $A_3\times A_1 \times B_2$ lattice. For simplicity we will present the analysis of the case where the $B_2$-torus is orthogonal to $A_3\times A_1$. This means that the moduli $b,c$ vanish.\footnote{For an arbitrary choice of these moduli (up to the moduli fixing in table \ref{tab:T5}), we obtain the same results.} For the explicit computations, we will focus on the $\mathcal{R}_1$ involution (\ref{2.R1}).

\textbf{A}$_\textbf{a}$\textbf{AB}

The condition $u_1=0$ restricts the moduli $d$ and $a$. From the definition of the complex structure modulus (\ref{2.complex_structure}) we obtain that $d$ is fixed to 0 and $a$ has to be negative. \\
Since the dual basis vectors $e_i^*$ transform under the transposed orientifold action $\mathcal{R}_1^t$, the momentum modes $\textbf{p}_i$ are the eigenvectors of $\mathcal{R}_1^t$ to the eigenvalue $+1$:
\begin{equation}
\begin{split}
 \textbf{p}_1&=e_1^*-e_3^*,\\
 \textbf{p}_2&=2e_1^*-e_2^*+e_3^*,\\
 \textbf{p}_3&=e_5^*+e_6^*.
\end{split}
\end{equation}
The winding modes are 
\begin{equation}
\begin{split}
\textbf{w}_1&=e_1+2e_2+e_3,\\
\textbf{w}_2&=e_4,\\
\textbf{w}_3&=e_5-e_6,
\end{split}
\end{equation}
and the $\textbf{v}_i$ vectors describing the bulk part of a fractional cycle parallel to some O6-plane are
\begin{equation}
\begin{split}
\textbf{v}_1&=e_1,\\
\textbf{v}_2&=e_3,\\
\textbf{v}_3&=e_5+e_6.
\end{split}
\end{equation}
The determinants of the different matrices are 
\begin{equation}
\det M_{KB}:=-\frac{2}{a (1 + a) R_3^4R_2^2}\,,\quad
\det M_{A}:=-8 a (1 + a) R_3^4R_2^2\,,\quad
\det W_{KB}:=8 (1 + a) R_3^2R_2^2R_1^2\,.
\end{equation}
The lattice mode contributions to the untwisted sector are
\begin{equation}
 KB=16\sqrt{-a}\frac{R_3}{R_1},\quad A=4\sqrt{-a}\frac{R_3}{R_1},\quad MS=64\sqrt{-a}\frac{R_3}{R_1}.
\end{equation}
and the corresponding untwisted RR tadpole cancellation condition reads
\begin{equation}
 KB+\frac{M^2\cdot A}{16}-\frac{M\cdot MS}{16}=0\Rightarrow (M-8)^2=0.
\end{equation}
Thus, we get $M=8$ which agrees with \cite{Blumenhagen:2004di}.
In a similar way one shows that the tadpole condition from the $\mathcal{R}_1Q$-part gives rise to $N=16$.
The resulting gauge group is thus $U(4) \times U(8)$ or some rank preserving gauge enhancement to $SO(2M)$ or $USp(2M)$ (for one or both gauge factors).

The CFT calculation is complemented by purely geometric considerations as follows:
The cycle wrapped by the O6-planes is $\gamma_1+\gamma_2-2(\bar{\gamma}_1-\bar{\gamma}_2)$. Its contribution to the tadpole can be cancelled by a stack of $N_1=4$ branes wrapping the fractional cycle $\frac{1}{2}\gamma_1+\frac{1}{2}\gamma_2$ and a stack of $N_2=8$ branes wrapping $-\frac{1}{2}\bar{\gamma}_1+\frac{1}{2}\bar{\gamma}_2\pm\frac{1}{2}\bar{\gamma}_3\pm\frac{1}{2}\bar{\gamma}_4$ (any choice of sign for the exceptional part is allowed). Since the first stack is invariant under the orientifold action, the actual gauge group is $SO/USp(8)\times U(8)$. 
Our result - which agrees for both (CFT and cycle homology) methods - differs slightly from~\cite{Blumenhagen:2004di} in the fact that the first gauge factor experiences a gauge group enhancement, which they do not mention. 

\vspace{2mm}

\textbf{A}$_\textbf{b}$\textbf{AB}
 
In this case, from the condition $u_1=\frac{1}{2}$ we obtain 
\begin{equation}
\begin{split}
d=-\frac{aR_3}{R_1}\,,\quad
a(1+a\frac{R_3^2}{R_1^2})<0\,.
\end{split}
\end{equation}
The momentum modes $\textbf{p}_i$ are
\begin{equation}
\begin{split}
\textbf{p}_1&=e_1^*-e_3^*,\\
\textbf{p}_2&=2e_1^*-e_2^*+e_4^*,\\
\textbf{p}_3&=e_6^*.
\end{split}
\end{equation}
The winding modes $\textbf{w}_i$ and lattice vectors $\textbf{v}_i$ are
\begin{equation}
\begin{split}
\textbf{w}_1&=e_1+2e_2+e_3,\\
\textbf{w}_2&=e_2+e_4,\\
\textbf{w}_3&=e_5,
\end{split}
\end{equation}
\begin{equation}
\begin{split}
\textbf{v}_1&=e_1,\\
\textbf{v}_2&=e_3,\\
\textbf{v}_3&=e_5+2e_6.
\end{split}
\end{equation}
The determinants of the different matrices are thus given by 
\begin{equation}
\det M_{KB}=\frac{-2}{(a+a^2)R_2^2R_3^4}\,,\quad \det M_{A}=-8(1+a)aR_2^2R_3^4\,,\quad \det W_{KB}=8(1+a)(R_1^2+aR_3^2)R_2^2R_3^2\,,
\end{equation}
and the resulting untwisted RR tadpole cancellation condition reads
\begin{equation}
 KB+\frac{M^2\cdot A}{16}-\frac{M\cdot MS}{16}=0\Rightarrow (M-8)^2=0,
\end{equation}
i.e. we obtain $M=8$. In a similar manner, the $\mathcal{R}_1Q$-insertion gives rise to $N=8$. Therefore, the gauge group is $U(4)\times U(4)$ or some rank preserving gauge group enhancement thereof.

The complementary considerations in terms of cycle homologies are as follows:
The cycle wrapped by the O6-planes is $2\gamma_2-2(\bar{\gamma}_1-\bar{\gamma}_2)$. Its contribution to the tadpole can be cancelled, for instance, by a stack of $N_1=4$ branes wrapping the fractional cycle $\frac{1}{2}\gamma_1+\frac{1}{2}\gamma_2$ and a stack of $N_2=4$ branes wrapping $-\frac{1}{2}\gamma_1+\frac{1}{2}\gamma_2-\bar{\gamma}_1+\bar{\gamma}_2\pm(\frac{1}{2}\gamma_3-\bar{\gamma}_3)\pm(\frac{1}{2}\gamma_4-\bar{\gamma}_4)$ (any choice of sign for the exceptional part is allowed). Naively, this gives rise to the gauge group $U(4)\times U(4)$, but the first stack maps to itself under the orientifold action, which enhances the symmetry to $USp/SO(8)\times U(4)$.

The results for the remaining orientifold projections can be derived analogously and are summarised in the table \ref{tab:T6'}.
\begin{table}[h!]
\centering
\begin{tabular}{|c|c||c|c|c|c|c|c|}
\hline
\multicolumn{8}{|c|}{{\bf $A_3\times A_1\times B_2$ orientifolds with D-branes on top of the O-planes}}\\
\hline
\hline 
\multicolumn{2}{|c||}{$A_3\times A_1\times B_2$} &\multicolumn{3}{|c|}{$u_1=0$} & \multicolumn{3}{|c|}{$u_1=\frac{1}{2}$} \\ 
\hline 
Lattice & Involution & $M$ & $N$ & Gauge group & $M$ & $N$ & Gauge group\\ 
\hline \hline
\textbf{AAA}& $\mathcal{R}_2$ & 16 & 8 & $U(8)\times USp/SO(8)$ & 16 & 4 & $U(8)\times USp/SO(4)$\\ 
\hline 
\textbf{AAB}& $\mathcal{R}_1$ &8 & 16 & $USp/SO(8)\times U(8)$ & 8 & 8 & $USp/SO(8)\times U(4)$\\ 
\hline 
\textbf{ABA}& $\mathcal{R}_3$ &16 & 8 & $U(8)\times USp/SO(8)$ & 8 & 8 & $U(8)\times USp/SO(8)$\\ 
\hline 
\textbf{ABB}& $\mathcal{R}_4$ &8& 16 & $USp/SO(8)\times U(8)$ & 4 & 16 & $USp/SO(4)\times U(8)$\\ 
\hline 
\end{tabular} 
\caption{IIA Orientifolds of $\Z_4$-orbifolds with the lattice of the type $A_3\times A_1\times B_2$ with D-branes on top of the O-planes canceling the RR tadpoles.}
\label{tab:T6'}
\end{table}

The column $u_1=0$ agrees mostly with the result \cite{Blumenhagen:2004di}, while the results for $u_1=\frac{1}{2}$ are presented here for the first time. We can also see the relation between the lattices \eqref{2.dualities} if we interchange $M$ and $N$.

\section{Factorisation of non-factorisable orbifolds}\label{S:Factorisation}

In this chapter we will show that both non-factorisable orbifolds can be written in a factorised form. This identification between the lattices was already detected in 
\cite{Blaszczyk:2011hs} by consideration of factorisable orbifolds. Here we want to explain which conclusions this identification means from the point of view of non-factorisable orbifolds. We make the restriction that in the first non-factorisable orbifold background, both $A_3$ lattices are orthogonal to each other, and in the second non-factorisable orbifold background, the $A_3\times A_1$ lattice is orthogonal to the $B_2$ lattice. This fixes the moduli to $c=d=e=0$ and $b=c=0$, respectively. On both orbifolds we can find a real basis such that the non-factorisable structure of the lattice decomposes into three two-tori, but an additional $\Z_2$ shift symmetry
 appears. Any three-cycle, written now as a product of three one-cycles on each two-torus, is automatically $\Z_2$-invariant. Moreover, only $\Z_2$-invariant three-cycles on the non-factorisable lattice can be expressed as three-cycles with respect to the new basis. Therefore, the number of wrapping numbers is reduced from twelve ($A_3\times A_3$) or ten ($A_3\times A_1\times B_2$) to six in agreement with the naive expectation from the known factorisable orbifold backgrounds $(T^2)^3/\Z_N$ or $(T^2)^3/(\Z_N \times \Z_M)$.\footnote{Note, however, that due to the additional symmetry some of the new wrapping numbers can now also be half-integer.} \\
Because any three-cycle, written in the new basis, is $\Z_2$-invariant, it follows that it is automatically Lagrangian. Furthermore, it can be verified that the {\it sLag} condition can be expressed in the same form as for the usual factorisable orbifolds: the sum of the angles between the one-cycles wrapped by supersymmetric D6-branes on each two-torus and the ${\cal R}_i$-invariant O6-plane has to vanish (mod $2\pi$).

$A_3\times A_3$

In this case, we introduce new coordinates along the directions
\begin{equation}
\begin{split}
v_1&:=\pi_1+\pi_2\,,\quad v_2:=\pi_2+\pi_3\,,\quad v_3:=\pi_1+\pi_3\,,\\
v_4&:=\pi_4+\pi_6\,,\quad v_5:=\pi_4+\pi_5\,,\quad v_6:=\pi_5+\pi_6\,.
\end{split}
\end{equation}
With respect to this basis, the metric $g$ and the Coxeter element $Q$ take a factorised form,
\begin{gather}\label{metric_fact_1}
g=\text{diag}\,\left(2(1+a)R_1^2,\,2(1+a)R_1^2,\,-4aR_1^2,-4bR_2^2,\,
2(1+b)R_2^2,\,2(1+b)R_2^2\right)\,,\\
Q=\text{diag}\left(
\begin{array}{cc}
 0 & -1 \\
 1 & 0 \\
\end{array}
\right)\oplus\left(
\begin{array}{cc}
-1 & 0 \\
0 & -1 \\
\end{array}
\right)\oplus\left(
\begin{array}{cc}
0 & -1 \\
 1 & 0 \\
\end{array}
\right)\,,
\end{gather}
and also the orientifold projections $\mathcal{R}_i$ become factorised, e.g.:
\begin{gather}
\mathcal{R}_4=\text{diag}\left(
\begin{array}{cc}
 1 & 0 \\
 0 & -1 \\
\end{array}
\right)\oplus\left(
\begin{array}{cc}
1 & 0 \\
0 & -1 \\
\end{array}
\right)\oplus\left(
\begin{array}{cc}
1 & 0 \\
0 & -1 \\
\end{array}
\right)\,.
\end{gather}

\begin{figure}[h!]
 	\centering
 		\includegraphics[width=15.5cm]{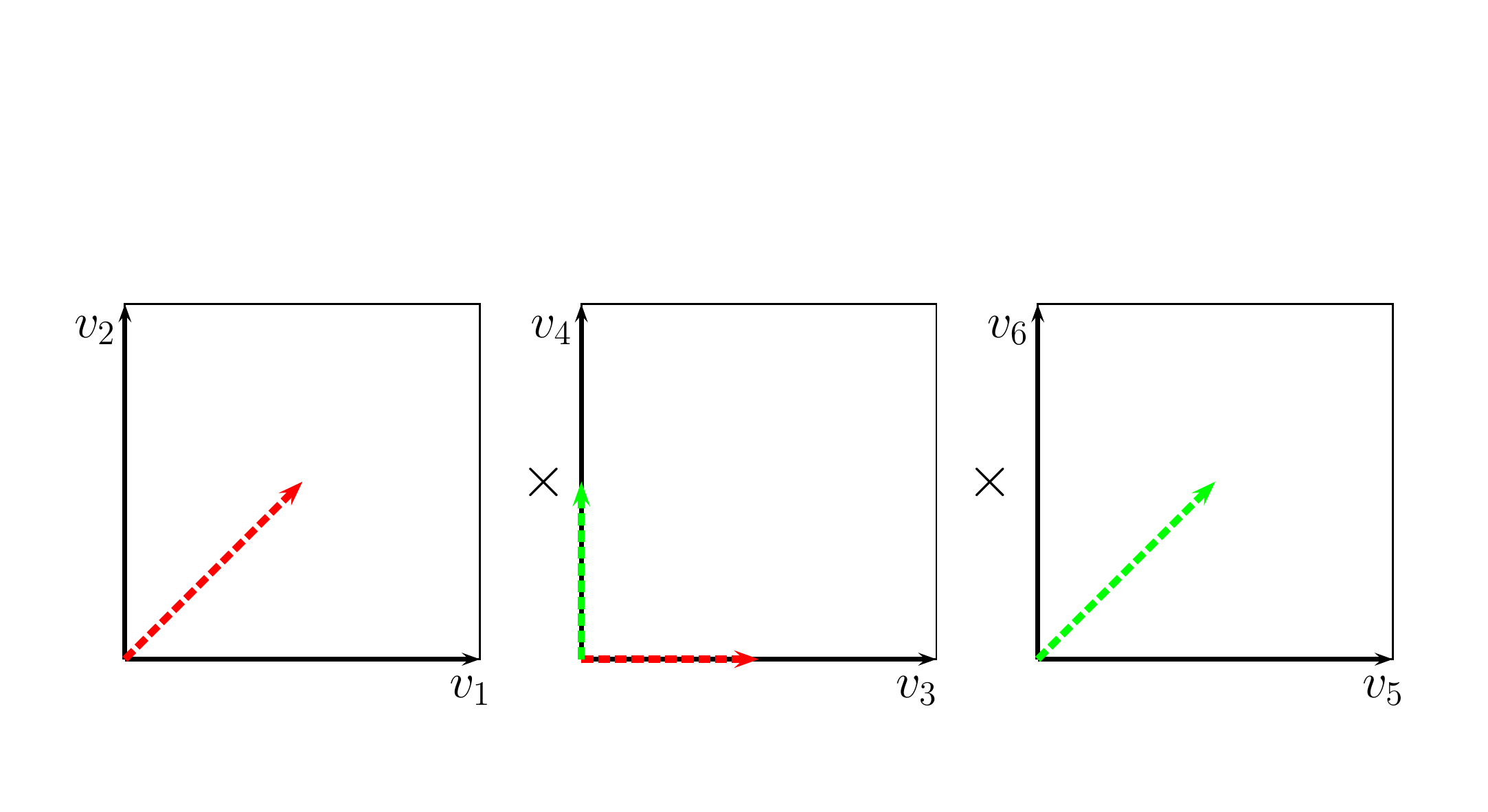}
 		\caption{Shift identifications of points on the factorised form of the $A_3\times A_3$-lattice by $\frac{v_1+v_2+v_3}{2}$ (red) and $\frac{v_4+v_5+v_6}{2}$ (green).}
 		\label{fig:Fig4}
 \end{figure}

But as already mentioned, the basis change gives rise to additional symmetries, which identify points of the factorised torus by the following shifts 
\begin{equation}\label{3_add.symmetry2}
\begin{split}
&p \simeq p+\frac{v_1+v_2+v_3}{2}\,,\\
&p \simeq p+\frac{v_4+v_5+v_6}{2} \, ,
\end{split}
\end{equation}
for any point $p$ on the torus, as depicted in figure~\ref{fig:Fig4}.

In the new coordinates, only six wrapping numbers occur,
\begin{equation}\label{3_reduced_wrap_numbers2}
\pi^{\text{torus}}:=(\tilde{n}^1v_1+\tilde{m}^1v_2)
\wedge(\tilde{n}^2v_3+\tilde{m}^2v_4)
\wedge(\tilde{n}^3\pi_5+\tilde{m}^3\pi_6)\,,
\end{equation}
in terms of which the bulk wrapping numbers defined in~\eqref{2.toroidal_3-cycle} and~\eqref{1f} can be rewritten as follows:
\begin{align}
\begin{array}{ll}
A_1q^1=-A_3q^1=-(\tilde{n}^1+\tilde{m}^1)\tilde{n}^2\tilde{n}^3\,,&
B_1m^1=-B_3m^1=\tilde{n}^1\tilde{m}^2(\tilde{n}^3+\tilde{m}^3)\,,\\
A_1r^1=-A_3r^1=-(\tilde{n}^1+\tilde{m}^1)\tilde{n}^2(\tilde{n}^3+\tilde{m}^3)\,,&
B_1n^1=-B_3n^1=(\tilde{n}^1+\tilde{m}^1)\tilde{m}^2(\tilde{n}^3+\tilde{m}^3)\,,
\\
A_1s^1=-A_3s^1=-(\tilde{n}^1+\tilde{m}^1)\tilde{n}^2\tilde{m}^3\,,&
B_1p^1=-B_3p^1=\tilde{m}^1\tilde{m}^2(\tilde{n}^3+\tilde{m}^3)\,,\\
A_2q^1=(\tilde{n}^1-\tilde{m}^1)\tilde{n}^2\tilde{n}^3\,,&
B_2m^1=\tilde{n}^1\tilde{m}^2(-\tilde{n}^3+\tilde{m}^3)\,,\\
A_2r^1=(\tilde{n}^1-\tilde{m}^1)\tilde{n}^2(\tilde{n}^3+\tilde{m}^3)\,,&
B_2n^1=(\tilde{n}^1+\tilde{m}^1)\tilde{m}^2(-\tilde{n}^3+\tilde{m}^3)\,,
\\
A_2s^1=(\tilde{n}^1-\tilde{m}^1)\tilde{n}^2\tilde{m}^3\,,&
B_2p^1=\tilde{m}^1\tilde{m}^2(-\tilde{n}^3+\tilde{m}^3)\,.
\end{array}
\end{align}
Note that due to the shift symmetry \eqref{3_add.symmetry2}, the wrapping numbers $\tilde{n}^2$ and $\tilde{m}^2$ can also have half-integer values.

On the factorised lattice, the geometric difference between the length of the O6-planes on the 
\textbf{AAB}- and \textbf{BAA}-orientations becomes clear: for the \textbf{AAB}-lattice, the ${\cal R}$-invariant
O6-plane is placed along the axes $v_1-v_2$ and $v_3$ on $T^2_{(1)} \times T^2_{(2)}$
and therefore it passes through the points which are identified by the shift symmetry \eqref{3_add.symmetry2}, but this is no longer the case for the \textbf{BAA}-lattice.

$A_3\times A_1\times B_2$
 
\begin{figure}[h]
 	\centering
 		\includegraphics[width=10.5cm]{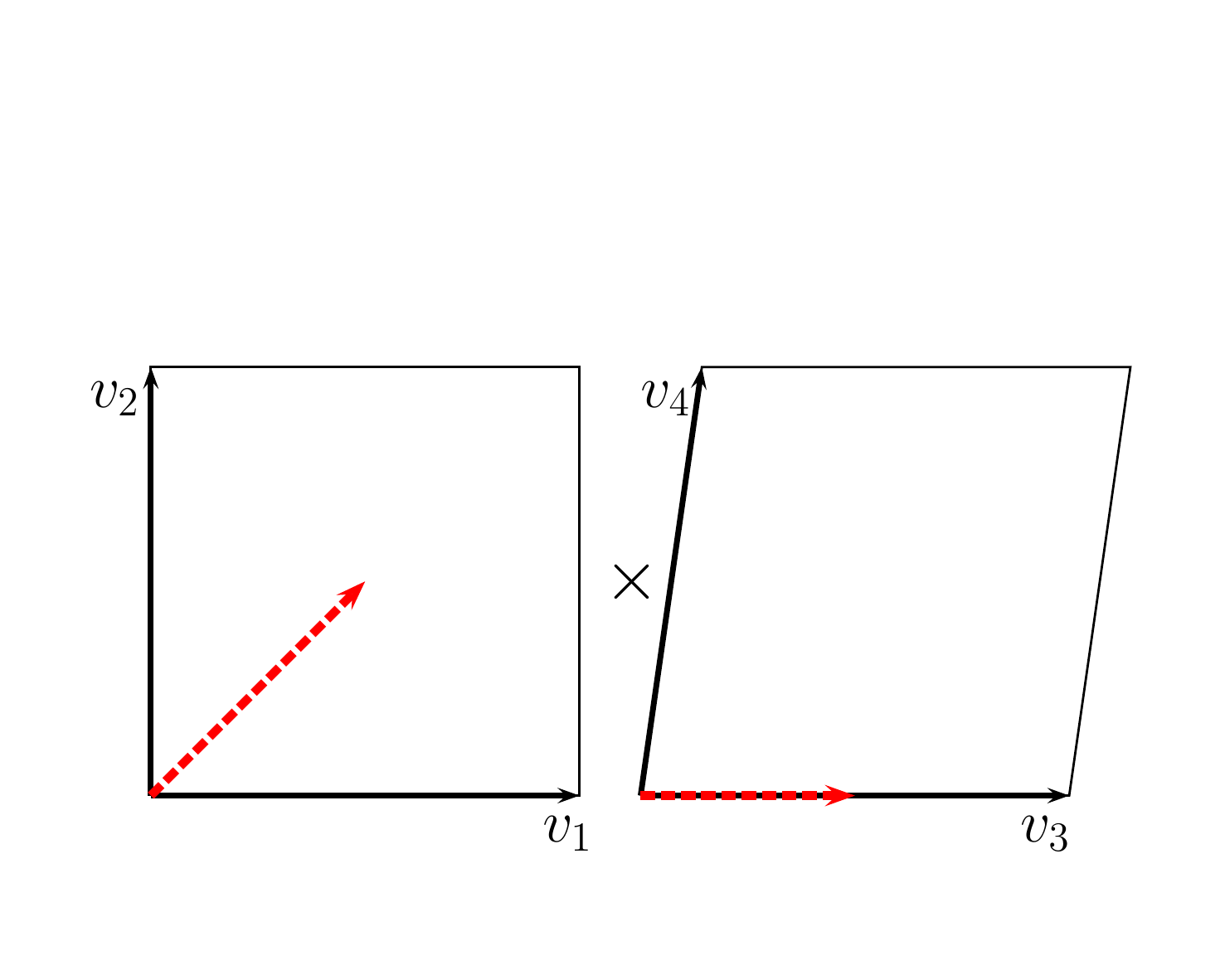}
 		\caption{Shift symmetry of the factorised form of the $A_3\times A_1$-lattice along $\frac{v_1+v_2+v_3}{2}$ (in red).}
 		\label{fig:Fig3}
 \end{figure}

Here it suffices to introduce new directions only on $A_3\times A_1$:
\begin{equation}\label{3.fact}
\begin{split}
v_1&:=\pi_1+\pi_2\,,\quad v_3:=\pi_1+\pi_3\,,\\
v_2&:=\pi_2+\pi_3\,,\quad v_4:=\pi_4\,.
\end{split}
\end{equation}
This basis factorises the torus, such that the metric becomes
\begin{equation}
g=\text{diag}\left(
\begin{array}{cc}
2(1+a)R_3^2& 0 \\
 0 & 2(1+a)R_3^2 \\
\end{array}
\right)\oplus\left(
\begin{array}{cc}
-4aR_3^2 & -4au_1R_3^2 \\
-4au_1R_3^2 & R_1^2 \\
\end{array}
\right)\oplus\left(
\begin{array}{cc}
2R_2^2 & -R_2^2 \\
 -R_2^2 & R_2^2 \\
\end{array}
\right)\,.
\end{equation}
For the Coxeter element $Q$ and the orientifold projections $\mathcal{R}_i$ we obtain
\begin{align}
Q=\text{diag}\left(
\begin{array}{cc}
 0 & -1 \\
 1 & 0 \\
\end{array}
\right)\oplus\left(
\begin{array}{cc}
-1 & 0 \\
0 & -1 \\
\end{array}
\right)\oplus\left(
\begin{array}{cc}
1 & -1 \\
 2 & -1 \\
\end{array}
\right)\,,\\
\mathcal{R}_1=\text{diag}\left(
\begin{array}{cc}
 0 & -1 \\
 -1 & 0 \\
\end{array}
\right)\oplus\left(
\begin{array}{cc}
1 & 2u_1 \\
0 & -1 \\
\end{array}
\right)\oplus\left(
\begin{array}{cc}
1 & -1 \\
0 & -1 \\
\end{array}
\right)\,,\\
\mathcal{R}_4=\text{diag}\left(
\begin{array}{cc}
 -1 & 0 \\
0 & 1 \\
\end{array}
\right)\oplus\left(
\begin{array}{cc}
1 & 2u_1 \\
0 & -1 \\
\end{array}
\right)\oplus\left(
\begin{array}{cc}
1 & -1 \\
0 & -1 \\
\end{array}
\right)\,.
\end{align} 
Also in this case the basis change gives rise to an additional shift symmetry displayed in figure \ref{fig:Fig3},
\begin{equation}\label{3_add.symmetry}
p\simeq p+\frac{v_1+v_2+v_3}{2}\quad \text{for any point $p$ on the torus}.
\end{equation}
Instead of ten wrapping numbers which we need to describe a fractional three-cycle in non-factorisable coordinates, in the $v$-basis the usual six wrapping numbers are sufficient:
\begin{equation}\label{3_reduced_wrap_numbers}
\pi^{\text{torus}}:=(\tilde{n}^1v_1+\tilde{m}^1v_2)
\wedge(\tilde{n}^2v_3+\tilde{m}^2v_4)
\wedge(m^3\pi_5+n^3\pi_6)\,.
\end{equation}
The relation between the non-factorised toroidal and bulk wrapping numbers in~\eqref{2.torus_cycle} and~\eqref{2.definition_of_A_and_B} and the new factorised ones is given by
\begin{align}
\begin{array}{ll}
A_1=-A_3=-(\tilde{n}^1+\tilde{m}^1)\tilde{n}^2\,,&
A_2=(\tilde{n}^1-\tilde{m}^1)\tilde{n}^2\,,\\
B_1=\tilde{n}^1\tilde{m}^2\,,&B_3=\tilde{m}^1\tilde{m}^2 = B_2 - B_1\,.
\end{array}
\end{align}
Notice that here, due to the shift symmetry \eqref{3_add.symmetry}, the wrapping number $\tilde{n}^2$ can be half-integer if $\tilde{m}^1+\tilde{n}^1$ is even.

\section{Concrete Pati-Salam Models on $A_3 \times A_1 \times B_2$}\label{S:ConcretePSModels}
In this chapter we present the construction of some local and global semi-realistic supersymmetric Pati-Salam (PS) models using intersecting D6-branes.\footnote{Notice that for generic configurations of gauge groups, {\it global} models do not only have to satisfy the RR tadpole cancellation conditions, but also the K-theory constraints, which are usually derived by scanning through all possible probe D-branes supporting $USp(2)$ gauge factors~\cite{Uranga:2000xp}. For PS models with all gauge groups (including `hidden' ones)
of even rank, the K-theory constraints are, however, trivially fulfilled.
 } 
In section~\ref{Sss:Relations} we have seen that there exist duality relations between the different orientifold projections ${\cal R}_i$ ($i=1,2,3,4$)
and that only four lattice orientations are independent. For the construction of particle spectra, we choose the \textbf{AAB}- and \textbf{ABB}-types lattices. 
In addition to the RR tadpole cancellation and {\it sLag} conditions, we require that the $U(4)_a$ stack is free from (anti-)symmetric representations.\footnote{The systematic computer search on the $A_3 \times A_1 \times B_2$ lattices actually showed that all two- and four-generation models of PS-type satisfy this condition without {\it a priori} imposing it.}
An intensive computer search showed that only global supersymmetric PS-models with two or four generations are possible. The supersymmetry conditions give rise to restrictions on the allowed values of the imaginary part $u_2$ of the complex structure modulus $\mathcal{U}$, and only a small number of values can be used for the construction of global PS-models without overshooting the bulk RR tadpole cancellation condition. 
The allowed values of $u_2$ are illustrated in the table \ref{tab:T7}. For some branes, some bulk wrapping numbers turn out to be zero, and thus 
the supersymmetry conditions do not fix the complex structure. If PS-models can be constructed just with such branes, we will list the arbitrariness in $\text{Im}(\mathcal{U})$ as $\infty$ in table~\ref{tab:T7}.
\begin{table}[h!]
\centering
\begin{tabular}{|c|c|c|}
\hline 
\muc{3}{|c|}{\bf $A_3\times A_1\times B_2$}\\
\hline\hline
 & Two generations & Four generations \\ 
\hline 
\textbf{A}$_\textbf{a}$\textbf{AB}& $\infty,1$ & $\infty,1$ \\ 
\hline 
\textbf{A}$_\textbf{b}$\textbf{AB} & $\infty$ & $\infty$ \\ 
\hline 
\textbf{A}$_\textbf{a}$\textbf{BB} & $\infty,1,3,\frac{1}{2}, \frac{1}{3}, \frac{1}{4}, \frac{1}{6}, \frac{1}{8}, \frac{1}{9}, \frac{1}{12}$ & $\infty,1,2, \frac{1}{2}, \frac{1}{3}, \frac{1}{4}, \frac{1}{6}, \frac{1}{9}, \frac{1}{12}$ \\ 
\hline 
\textbf{A}$_\textbf{b}$\textbf{BB}& $\infty,1,\frac{1}{4}$ & - \\ 
\hline 
\end{tabular}
\caption{The possible values of $\text{Im}(\mathcal{U})$ for global supersymmetric Pati-Salam models, with $\infty$ denoting that it can be choosen arbitrarily.}
\label{tab:T7}
\end{table}

\textbf{Two generations}

We start the search for global PS-models with the case of two particle generations and note some general properties of the models obtained. 
All models with $U(4)\times U(2)\times U(2)$-gauge group in the visible sector contain chiral particles which transform in the (anti-)symmetric representation of the $U(2)$ on the $b$ and/or $c$-stack. The spectrum of the models with gauge symmetry enhancement $U(1) \hookrightarrow USp/SO(2)$ on both the $b$- and $c$-stack, on the other hand, lacks the bifundamental representations in the $bc$-sector, i.e., there is no Higgs field in the chiral spectrum of such models. In table \ref{tab:T9}, we provide an explicit example of a PS-model with $U(4)\times U(2)\times USp/SO(2)$ gauge group in the visible sector. Its chiral spectrum is given in the table \ref{tab:T9s}. 

In appendix \ref{A:MoreModels}, we provide an explicit example for each other type of visible gauge group, i.e. $U(4)\times U(2)\times U(2)$ and $U(4)\times USp/SO(2)\times USp/SO(2)$.

\begin{table}[h!]
\bCentering
\resizebox{\linewidth}{!}{
\begin{tabular}{|c|c|c|}
\hline 
\muc{3}{|c|}{\bf D6-branes configuration for two generation PS-model on $A_3\times A_1\times B_2$-orientifold }\\
\hline\hline
Stack & $(m^1,n^1,p^1,q^1)\times(m^2,n^2,p^2,q^2)\times(m^3,n^3)$ & Homology cycle \\ 
\hline 
$U(4)_a$ & $(1,0,-1,0)\times(1,0,0,1)\times(-1,-2)$ & $\frac{1}{2}\gamma_{1}-\frac{1}{2}\gamma_2-\bar{\gamma}_1+\bar{\gamma}_2$ \\ 
\hline 
$U(2)_b$ & $(0,1,1,0)\times(0,0,0,1)\times(1,2)$ & $\bar{\gamma}_2-\bar{\gamma}_3+\bar{\gamma}_4$ \\ 
\hline 
$USp/SO(2)_c$ & $(1,1,0,0)\times(1,0,1,0)\times(-1,-2)$ & $\gamma_1+\gamma_3-\gamma_4$ \\ 
\hline \hline
$USp/SO(2)_{h_1}$ & $(1,1,0,0)\times(1,0,1,0)\times(-1,-2)$ & $\gamma_1-\gamma_3+\gamma_4$ \\ 
\hline
$U(2)_{h_2}$ & $(0,1,1,0)\times(0,0,0,1)\times(1,2)$& $\bar{\gamma}_2-\bar{\gamma}_3-\bar{\gamma}_4$ \\ 
\hline
\end{tabular} }
\caption{D6-branes for a two generation supersymmetric PS-model with \textbf{A}$_\textbf{a}$\textbf{BB}-lattice and $u_2=1/2$.}
\label{tab:T9}
\end{table}

\begin{table}[h!]
\centering
\begin{tabular}{|c|c|}
\hline
\muc{2}{|c|}{\bf Chiral spectrum of two generation PS-model on $A_3\times A_1\times B_2$-orientifold}\\
\hline\hline 
Sector & $SU(4)_a\times SU(2)_b\times USp/SO(2)_c\times USp/SO(2)_{h_1}\times SU(2)_{h_2}\times U(1)^3$ \\ 
\hline 
$ab$ & $(4,\bar{2},1,1,1)_{(1,-1,0)}$ \\ 
\hline 
$ab^\prime$ & $(4,2,1,1,1)_{(1,1,0)}$ \\ 
\hline 
$ac=ac^\prime$ & $2\times(\bar{4},1,2,1,1)_{(-1,0,0)}$ \\ 
\hline 
$bc=bc^\prime$ & $4\times(1,2,\bar{2},1,1)_{(0,1,0)}$ \\ 
\hline 
$ah_2$ & $(4,1,1,1,\bar{2})_{(1,0,-1)}$ \\ 
\hline 
$ah_2^\prime$ & $(4,1,1,1,2)_{(1,0,1)}$ \\ 
\hline 
\end{tabular} 
\caption{Chiral spectrum for the two generation $U(4)\times U(2)\times USp/SO(2)\times USp/SO(2)\times U(2)$-PS model with D6-brane configuration given in table~\protect\ref{tab:T9}.}
\label{tab:T9s}
\end{table}

\textbf{Four generations}

In a similar way, we can realise global supersymmetric PS-models with four generations with different gauge groups (without/with some enhancement of the `left' and `right' stack)
 in the visible sector. In general, only the $U(4)\times USp/SO(2)\times USp/SO(2)$-models do not contain chiral matter in the (anti-)symmetric representation on the $b$- and $c$- stacks. But these models also do not contain chiral states in the bifundamental representation of the $bc$-sector, which could serve as a (chiral) Higgs multiplet. 
In appendix \ref{A:MoreModels}, we collect several examples of global D6-brane configurations of such types together with their chiral spectrum.

\textbf{Three generations}

As mentioned above, there are no global supersymmetric Pati-Salam models with three generations. But it is possible to construct models where the bulk part of the tadpole vanishes and only the exceptional part remains. Such {\it local} models are only realisable on the \textbf{ABB}-orientifolds. More precisely, for the \textbf{A}$_\textbf{a}$\textbf{BB}-lattice only $u_2=3\,,\,\frac{1}{3}$ provides such models,
and the gauge group is enhanced on the $b$- and $c$-stacks. For the \textbf{A}$_\textbf{b}$\textbf{BB}-lattice, it is possible to construct {\it local} models with $u_2=\frac{3}{2}$ with symmetry enhancement on the $b$- and/or $c$-stack. A general property of all such models is the appearance of chiral particles which transform in the (anti-)symmetric representation of $U(4)_a$.

\section{Discussion and Conclusions}\label{S:Conclusions}

In this article, we explored the full three-cycle geometry of the non-factorisable \mbox{$T^6/\Z_4$} orbifold on the two possible root lattices $A_3 \times A_3$ and $A_3 \times A_1 \times B_2$ and compared it with the 
factorisable $B_2 \times (A_1)^2 \times B_2$ case. We found that, under the anti-holomorphic involution ${\cal R}$ of Type IIA/$\OR$ orientifolds, there exist {\it a priori} four different lattice orientations 
for $A_3\times A_3$, and eight for $A_3\times A_1\times B_2$. However, a closer look at the number of supersymmetric (i.e. {\it sLag}) fractional three-cycles bounded in their length by the RR tadpole cancellation conditions and by the allowed values for the complex structure moduli, which encode the relative angles between the different root lattices as well as between the generators within each $A_3$ lattice, reveals - in analogy to the factorisable cases~\cite{Gmeiner:2008xq,Honecker:2012qr,Ecker:2014hma} - the existence of several dualities, which lead to identical physics for different choices of lattice orientations under ${\cal R}$. More precisely, on $A_3\times A_3$ we found a duality relation between the {\bf AAA} and {\bf BAB} lattices, leaving three independent choices. For the $A_3\times A_1\times B_2$ lattice, there are four pairwise duality relations ({\bf A$_\textbf{a}$AA} and {\bf A$_\textbf{a}$AB}, {\bf A$_\textbf{a}$BA} and {\bf A$_\textbf{a}$BB}, {\bf A$_\textbf{b}$AA} and {\bf A$_\textbf{b}$BB}, {\bf A$_\textbf{b}$BA} and {\bf A$_\textbf{b}$AB}), leaving only four independent choices.

After computing the {\it sLag} three-cycles, in order to ascertain our results, we cross-checked agreement of the RR tadpole cancellation conditions among our new purely geometric derivation as well as via the `old' CFT method, in particular for the special D6-brane configurations on top of the O6-planes in~\cite{Blumenhagen:2004di}, compared to which we generalised the CFT results to arbitrary values of the
angles inside the $A_3$ lattices and the angle between the $A_3$ and the $A_1$ lattices. For the $A_3\times A_3$ lattices and the {\bf a}-lattices of $A_3\times A_1\times B_2$, our results derived in a twofold, mutually agreeing way (mostly) coincide with the results listed in~\cite{Blumenhagen:2004di}, while those corresponding to the ${\bf b}$-type lattices of $A_3 \times A_1 \times B_2$ are newly found here.

With the full list of allowed {\it sLag} three-cycles at hand, we proceeded to search for local and global semi-realistic Pati-Salam models on both types of lattices. The $A_3\times A_3$ lattices happened to be very restricted - e.g. by the small number of available three-cycles - and therefore unsuitable for model building. But the $A_3\times A_1\times B_2$ lattice with different orientations provided a very rich class of backgrounds with ample potential for model building. Although the search for global Pati-Salam models - to which we restricted ourselves here since the K-theory constraints are trivially fulfilled in that case - with three generations did not bear any fruit, many models with two and four generations were found.  From a qualitative point of view, our results are in agreement with~\cite{Nilles:2014owa} - which studied these non-factorisable orbifolds in the context of the heterotic string - in the sense that the $A_3\times A_1\times B_2$ lattice is the most promising non-factorisable one for model building.

Our first model searches typically lead to globally consistent models, where one or more stacks of D-branes wrap three-cycles invariant under the anti-holomorphic involution ${\cal R}$. While it is well known that in such cases an enhancement of the gauge group $U(N) \hookrightarrow USp(2N)$ or $SO(2N)$ occurs, the correct distinction can - with the generally available model building techniques to date - only be done by computing open string CFT amplitudes such as the gauge thresholds for factorisable backgrounds in~\cite{Forste:2010gw,Honecker:2012qr,Honecker:2013sww,Ecker:2014hma} and reading off the correct sign for the orientifold projection from the M\"obius strip contribution to either the one-loop beta function coefficient or to the RR tadpoles. Our finding in section~\ref{S:Factorisation}, that fractional three-cycles can be rewritten in a factorised form, suggests that the CFT methods can be straightforwardly generalised to the non-factorisable backgrounds of $T^6/\Z_4$ discussed here from a purely geometric viewpoint. By fully developing the corresponding CFT, not only the compete chiral plus vector-like matter spectrum can be determined, but also the low-energy effective action can (in principle) be derived. Identifying all probe D-branes supporting $USp(2)$ gauge groups by means of CFT techniques is furthermore necessary to determine the K-theory constraints for all future Standard Model or GUT model building.

Last but not least, it will be interesting to extend the methods for studying non-factorisable orbifolds to other point groups $\Z_{N \neq 4}$, and to incorporate closed string fluxes and study if particle physics models with (nearly) complete stabilisation of the closed string moduli are within reach.

\noindent
{\bf Acknowledgements:} 
This work is partially supported by the {\it Cluster of Excellence `Precision Physics, Fundamental Interactions and Structure of Matter' (PRISMA)} DGF no. EXC 1098,
the DFG research grant HO 4166/2-2, and the DFG Research Training Group {\it `Symmetry Breaking in Fundamental Interactions'} GRK 1581.


\appendix
\section{More Pati-Salam models}\label{A:MoreModels}

In this appendix, we provide additional explicit examples of global Pati-Salam models with two and four generations on the $A_3 \times A_1 \times B_2$ lattice. 

\subsection{2 generations}

In the main text, we presented a global model where one of the left- or right-symmetric groups of the Pati-Salam gauge group is provided by an enhanced $SO(2)_{L/R}$ or $USp(2)_{L/R}$ symmetry. We also found global models where none or both left- and/or right-symmetric groups are replaced by an enhanced gauge symmetry; we will show an example of each of type in the following.

Tables \ref{tab:T8} and \ref{tab:T8s} show the D6-branes and chiral spectrum, respectively, of a global Pati-Salam model with visible gauge group $U(4)\times U(2)\times U(2)$.
\begin{table}[h!]
\bCentering
\resizebox{\linewidth}{!}{
\begin{tabular}{|c|c|c|}
\hline 
\muc{3}{|c|}{\bf D6-branes configuration for a two generation PS-model on $A_3\times A_1\times B_2$-orientifold }\\
\hline\hline
Stack & $(m^1,n^1,p^1,q^1)\times(m^2,n^2,p^2,q^2)\times(m^3,n^3)$ & Homology cycle \\ 
\hline 
$U(4)_a$ & $(1,2,1,0)\times(0,0,0,1)\times(0,1)$ & $\bar{\gamma}_2+\bar{\gamma}_3$ \\ 
\hline 
$U(2)_b$ & $(1,1,0,0)\times(1,0,1,-1)\times(0,-1)$ & $\frac{1}{2}(\gamma_1+\gamma_2-\gamma_3-\gamma_4+\bar{\gamma}_1+\bar{\gamma}_2+\bar{\gamma}_3+\bar{\gamma}_4)$ \\ 
\hline 
$U(2)_c$ & $(1,1,0,0)\times(1,0,1,-1)\times(0,-1)$ & $\frac{1}{2}(\gamma_1+\gamma_2+\gamma_3+\gamma_4+\bar{\gamma}_1+\bar{\gamma}_2-\bar{\gamma}_3-\bar{\gamma}_4)$ \\ 
\hline \hline
$U(2)_{h_1}$ & $(0,1,1,0)\times(0,0,0,1)\times(1,2)$& $\bar{\gamma}_2-\bar{\gamma}_3+\bar{\gamma}_4$\\ 
\hline 
$USp/SO(4)_{h_2}$ & $(1,1,0,0)\times(1,0,1,0)\times(-1,-2)$ & $\gamma_1$\\
\hline
\end{tabular} }
\caption{D6-branes for a global two generation PS-model on the \textbf{A}$_\textbf{a}$\textbf{BB}-lattice with $u_2=1$.}
\label{tab:T8}
\end{table}
\begin{table}[h!]
\centering
\begin{tabular}{|c|c|}
\hline 
\muc{2}{|c|}{\bf Chiral spectrum of a two generation PS-model on $A_3\times A_1\times B_2$-orientifold}\\
\hline\hline 
Sector & $SU(4)_a\times SU(2)_b\times SU(2)_c\times SU(2)_{h_1}\times USp/SO(4)_{h_2}\times U(1)^4$ \\ 
\hline 
$ab$ & $2\times(4,\bar{2},1,1,1)_{(1,-1,0,0)}$ \\ 
\hline 
$ac^\prime$ & $2\times(\bar{4},1,\bar{2},1,1)_{(-1,0,-1,0)}$ \\ 
\hline 
$bc^\prime$ & $2\times(1,\bar{2},\bar{2},1,1)_{(0,-1,-1,0)}$ \\ 
\hline 
$bb^\prime$ & $2\times(1,3,1,1,1)_{(0,0,0,0)}$ \\ 
\hline 
$cc^\prime$ & $2\times(1,1,3,1,1)_{(0,0,0,0)}$ \\ 
\hline 
$bh_1^\prime$ & $2\times(1,\bar{2},1,\bar{2},1)_{(0,-1,0,-1)}$\\
\hline
$ch_1$ & $2\times(1,1,\bar{2},2,1)_{(0,0,-1,1)}$\\
\hline
$bh_2=bh^\prime_2$ & $(1,2,1,1,\bar{4})_{(0,1,0,0)}$\\
\hline
$ch_2=ch^\prime_2$ & $(1,1,2,1,\bar{4})_{(0,0,1,0)}$\\
\hline
\end{tabular} 
\caption{Chiral spectrum of the two generation $U(4)\times U(2)\times U(2)\times U(2)\times USp/SO(4)$-PS model with D-brane configuration displayed in table~\protect\ref{tab:T8}.}
\label{tab:T8s}
\end{table}

Tables \ref{tab:T10} and \ref{tab:T10s} show the D6-branes and chiral spectrum, respectively, of a global Pati-Salam model with visible gauge group $U(4)\times USp/SO(2)\times USp/SO(2)$.
\begin{table}[h!]
\bCentering
\resizebox{\linewidth}{!}{
\begin{tabular}{|c|c|c|}
\hline 
\muc{3}{|c|}{\bf D6-branes configuration for a two generation PS-model on $A_3\times A_1\times B_2$-orientifold }\\
\hline\hline
Stack & $(m^1,n^1,p^1,q^1)\times(m^2,n^2,p^2,q^2)\times(m^3,n^3)$ & Homology cycle \\ 
\hline 
$U(4)_a$ & $(1,2,1,0)\times(0,0,0,1)\times(0,1)$ & $\bar{\gamma}_2-\bar{\gamma}_4$ \\ 
\hline 
$USp/SO(2)_b$ & $(1,1,0,0)\times(1,0,1,0)\times(-1,-2)$ & $\gamma_1+\gamma_3+\gamma_4$ \\ 
\hline 
$USp/SO(2)_c$ & $(1,1,0,0)\times(1,0,1,0)\times(-1,-2)$ & $\gamma_1-\gamma_3-\gamma_4$ \\ 
\hline \hline
$U(4)_{h_1}$ & $(1,2,1,0)\times(0,0,0,1)\times(0,1)$ & $\bar{\gamma}_2+\bar{\gamma}_4$ \\ 
\hline
$USp/SO(2)_{h_2}$ & $(1,1,0,0)\times(1,0,1,0)\times(-1,-2)$ & $\gamma_1$\\
\hline
\end{tabular} }
\caption{\mbox{D6-branes for a global two generation PS-model on the \textbf{A}$_\textbf{a}$\textbf{BB}-lattice with non-fixed $u_2$.}}
\label{tab:T10}
\end{table}

\begin{table}[h!]
\centering
\begin{tabular}{|c|c|}
\hline 
\muc{2}{|c|}{\bf Chiral spectrum of a two generation PS-model on $A_3\times A_1\times B_2$-orientifold}\\
\hline\hline 
Sector & $SU(4)_a\times USp/SO(2)_b\times USp/SO(2)_c\times SU(4)_{h_1}\times USp/SO(2)_{h_2}\times U(1)^2$ \\ 
\hline 
$ab=ab^\prime$ & $2\times(4,\bar{2},1,1,1)_{(1,0)}$ \\ 
\hline 
$ac=ac^\prime$ & $2\times(\bar{4},1,2,1,1)_{(-1,0)}$ \\ 
\hline 
\end{tabular} 
\caption{Chiral spectrum of the two generation $U(4)\times USp/SO(2)\times USp/SO(2)\times USp/SO(2)\times U(4)$-PS model with D6-brane configuration displayed in table~\protect\ref{tab:T10}.}
\label{tab:T10s}
\end{table}

\subsection{4 generations}

As in the previous case, we also found four-generation global Pati-Salam models where none, one or both left- and/or right-symmetric groups are replaced by an enhanced symmetry $SO(2)_{L/R}$ or $Usp(2)_{L/R}$. Here we will show a concrete example for each of these three cases.

Tables \ref{tab:T14} and \ref{tab:T14s} show the D6-brane configuration and chiral spectrum, respectively, of a Pati-Salam model with visible sector $U(4)\times U(2)\times U(2)$.
\begin{table}[h!]
\bCentering
\resizebox{\linewidth}{!}{
\begin{tabular}{|c|c|c|}
\hline 
\muc{3}{|c|}{\bf D6-branes configuration for a four generation PS-model on $A_3\times A_1\times B_2$-orientifold }\\
\hline\hline
Stack & $(m^1,n^1,p^1,q^1)\times(m^2,n^2,p^2,q^2)\times(m^3,n^3)$ & Homology cycle \\ 
\hline 
$U(4)$ & $(0,1,1,0)\times(0,0,0,1)\times(1,2)$ & $\bar{\gamma}_2-\bar{\gamma}_3-\bar{\gamma}_4$ \\ 
\hline 
$U(2)$ & $(0,1,1,0)\times(1,0,1,-1)\times(0,1)$ & $\frac{1}{2}(\gamma_1-\gamma_2+\gamma_3+\gamma_4-\bar{\gamma}_1+\bar{\gamma}_2+\bar{\gamma}_3+\bar{\gamma}_4)$ \\ 
\hline 
$U(2)$ & $(1,1,0,0)\times(1,0,1,-1)\times(0,-1)$ & $\frac{1}{2}(\gamma_1+\gamma_2-\gamma_3-\gamma_4+\bar{\gamma}_1+\bar{\gamma}_2+\bar{\gamma}_3+\bar{\gamma}_4)$ \\ 
\hline \hline
$U(2)_{h_1}$ & $(0,1,1,0)\times(0,0,0,1)\times(1,2)$& $\bar{\gamma}_2-\bar{\gamma}_3+\bar{\gamma}_4$\\ 
\hline 
$USp/SO(4)_{h_2}$ & $(1,1,0,0)\times(1,0,1,0)\times(-1,-2)$ & $\gamma_1$\\
\hline
\end{tabular} }
\caption{D6-branes for a global four generation PS-model on the \textbf{A}$_\textbf{a}$\textbf{BB}-lattice with $u_2=1$.}
\label{tab:T14}
\end{table}
\begin{table}[h!]
\centering
\begin{tabular}{|c|c|}
\hline
\muc{2}{|c|}{\bf Chiral spectrum of a four generation PS-model on $A_3\times A_1\times B_2$-orientifold}\\
\hline\hline 
Sector & $SU(4)_a\times SU(2)_b\times SU(2)_c\times SU(2)_{h_1}\times USp/SO(4)_{h_2}\times U(1)^4$ \\ 
\hline 
$ab$ & $(4,\bar{2},1,1,1)_{(1,-1,0,0)}$ \\ 
\hline 
$ab^\prime$ & $3\times(4,2,1,1,1)_{(1,1,0,0)}$\\
\hline
$ac$ & $(\bar{4},1,2,1,1)_{(-1,0,1,0)} $\\
\hline
$ac^\prime$ & $3\times(\bar{4},1,\bar{2},1,1)_{(-1,0,-1,0)}$ \\ 
\hline 
$bc$ & $2\times(1,2,\bar{2},1,1)_{(0,1,-1,0)}$ \\ 
\hline 
$bb^\prime$ & $2\times(1,3,1,1,1)_{(0,0,0,0)}$ \\ 
\hline 
$cc^\prime$ & $2\times(1,1,3,1,1)_{(0,0,0,0)}$ \\ 
\hline 
$bh_1^\prime$ & $2\times(1,\bar{2},1,\bar{2},1)_{(0,-1,0,-1)}$\\
\hline
$ch_1$ & $2\times(1,1,\bar{2},2,1)_{(0,0,-1,1)}$\\
\hline
$bh_2=bh^\prime_2$ & $(1,2,1,1,\bar{4})_{(0,1,0,0)}$\\
\hline
$ch_2=ch^\prime_2$ & $(1,1,2,1,\bar{4})_{(0,0,1,0)}$\\
\hline
\end{tabular} 
\caption{Chiral spectrum of the four generation $U(4)\times U(2)\times U(2)\times USp/SO(4)\times U(2)$-PS model with D6-brane configuration displayed in table~'\protect\ref{tab:T14}.}
\label{tab:T14s}
\end{table}

Tables \ref{tab:T15} and \ref{tab:T15s} show the D6-brane configuration and chiral spectrum, respectively, of a Pati-Salam model with visible sector $U(4)\times U(2)\times USp/SO(2)$.
\begin{table}[h!]
\bCentering
\resizebox{\linewidth}{!}{
\begin{tabular}{|c|c|c|}
\hline 
\muc{3}{|c|}{\bf D6-branes configuration a for four generation PS-model on $A_3\times A_1\times B_2$-orientifold }\\
\hline\hline
Stack & $(m^1,n^1,p^1,q^1)\times(m^2,n^2,p^2,q^2)\times(m^3,n^3)$ & Homology cycle \\ 
\hline 
$U(4)_a$ & $(0,1,1,0)\times(0,0,0,1)\times(1,2)$ & $\bar{\gamma}_2+\bar{\gamma}_3+\bar{\gamma}_4$ \\ 
\hline 
$U(2)_b$ & $(0,1,1,0)\times(1,0,1,1)\times(0,1)$ & $\frac{1}{2}(\gamma_1-\gamma_2-\gamma_3-\gamma_4-\bar{\gamma}_1+\bar{\gamma}_2-\bar{\gamma}_3-\bar{\gamma}_4)$ \\ 
\hline 
$USp/SO(2)_c$ & $(1,1,0,0)\times(1,0,1,0)\times(-1,-2)$ & $\gamma_1+\gamma_3+\gamma_4$ \\ 
\hline \hline
$U(3)_{h_1}$ &$(0,1,1,0)\times(0,0,0,1)\times(1,2)$ & $\bar{\gamma}_2+\bar{\gamma}_3-\bar{\gamma}_4$\\ 
\hline
$USp/SO(2)_{h_2}$ &$(1,1,0,0)\times(1,0,1,0)\times(-1,-2)$ & $\gamma_1$ \\ 
\hline
\end{tabular} }
\caption{D6-branes for a global four generation PS-model on the \textbf{A}$_\textbf{a}$\textbf{BB}-lattice with $u_2=1$.}
\label{tab:T15}
\end{table}
\begin{table}[h!]
\centering
\begin{tabular}{|c|c|}
\hline 
\muc{2}{|c|}{\bf Chiral spectrum of a four generation PS-model on $A_3\times A_1\times B_2$-orientifold}\\
\hline\hline 
Sector & $SU(4)_a\times SU(2)_b\times USp/SO(2)_c\times SU(3)_{h_1}\times USp/SO(2)_{h_2}\times U(1)^3$ \\ 
\hline 
$ab$ & $(4,\bar{2},1,1,1)_{(1,-1,0)}$ \\ 
\hline 
$ab^\prime$ & $3\times(4,2,1,1,1)_{(1,1,0)}$ \\ 
\hline 
$ac=ac^\prime$ & $4\times(\bar{4},1,2,1,1)_{(-1,0,0)}$ \\ 
\hline 
$bc=bc^\prime$ & $(1,2,\bar{2},1,1)_{(0,1,0)}$ \\ 
\hline 
$bb^\prime$&$2\times(1,3,1,1,1)_{(0,0,0)}$\\
\hline
$bh_1$ & $(1,2,1,\bar{3},1)_{(0,1,-1)}$ \\ 
\hline 
$bh_1^\prime$ & $(1,2,1,3,1)_{(0,1,1)}$ \\ 
\hline 
\end{tabular} 
\caption{Chiral spectrum of the four generation $U(4)\times U(2)\times USp/SO(2)\times U(3)\times USp/SO(2)$-PS model with D6-brane configuration given in table~\protect\ref{tab:T15}.}
\label{tab:T15s}
\end{table}

Tables \ref{tab:T16} and \ref{tab:T16s} show the D6-brane configuration and chiral spectrum, respectively, of a Pati-Salam model with visible sector $U(4)\times USp/SO(2)\times USp/SO(2)$.
\begin{table}[h!]
\bCentering
\resizebox{\linewidth}{!}{
\begin{tabular}{|c|c|c|}
\hline 
\muc{3}{|c|}{\bf D6-branes configuration for a four generation PS-model on $A_3\times A_1\times B_2$-orientifold }\\
\hline\hline
Stack & $(m^1,n^1,p^1,q^1)\times(m^2,n^2,p^2,q^2)\times(m^3,n^3)$ & Homology cycle \\ 
\hline 
$U(4)_a$ & $(1,0,-1,0)\times(0,0,0,1)\times(-1,-2)$ & $-\bar{\gamma}_1+\bar{\gamma}_2+2\bar{\gamma}_4$ \\ 
\hline 
$USp/SO(2)_b$ & $(0,1,0,0)\times(1,0,1,0)\times(1,2)$ & $\frac{1}{2}\gamma_1+\frac{1}{2}\gamma_2-\gamma_4$ \\ 
\hline 
$USp/SO(2)_c$ & $(0,1,0,0)\times(1,0,1,0)\times(1,2)$ & $\frac{1}{2}\gamma_1+\frac{1}{2}\gamma_2+\gamma_4$ \\ 
\hline \hline
$USp/SO(4)_h$ & $(0,1,0,0)\times(1,0,1,0)\times(1,2)$ & $\frac{1}{2}\gamma_1+\frac{1}{2}\gamma_2$ \\ 
\hline
\end{tabular}} 
\caption{D6-branes for a global four generation PS-model one the \textbf{A}$_\textbf{a}$\textbf{AB}-lattice with non-fixed $u_2$.}
\label{tab:T16}
\end{table}
\begin{table}[h!]
\bCentering
\resizebox{\linewidth}{!}{
\begin{tabular}{|c|c|}
\hline 
\muc{2}{|c|}{\bf Chiral spectrum of a four generation PS-model on $A_3\times A_1\times B_2$-orientifold}\\
\hline\hline 
Sector & $SU(4)_a\times USp/SO(2)_b\times USp/SO(2)_c\times USp/SO(4)_{h}\times U(1)$ \\ 
\hline 
$ab=ab^\prime$ & $4\times(4,\bar{2},1,1)_{+1}$ \\ 
\hline 
$ac=ac^\prime$ & $4\times(\bar{4},1,2,1)_{-1}$ \\ 
\hline 
\end{tabular} }
\caption{Chiral spectrum of the four generation $U(4)\times USp/SO(2)\times USp/SO(2)\times USp/SO(4)$-PS model with D6-brane configuration displayed in table~\protect\ref{tab:T16}.}
\label{tab:T16s}
\end{table}
 
\vspace{10mm}

\clearpage


\addcontentsline{toc}{section}{References}
\bibliographystyle{ieeetr}
\bibliography{refs_NonFactorisable}

\end{document}